\documentclass{amsbook}

\usepackage{amssymb}

\newtheorem{prop}{Proposition}[chapter]
\newtheorem{thm}[prop]{Theorem}
\newtheorem{cor}[prop]{Corollary}
\newtheorem{lem}[prop]{Lemma}
\newtheorem{rem}[prop]{Remark}

\newcommand{\bbN}{{\mathbb{N}}}
\newcommand{\bbR}{{\mathbb{R}}}

\newcommand{\bbZ}{{\mathbb{Z}}}
\newcommand{\bbC}{{\mathbb{C}}}

\newcommand{\calD}{{\mathcal{D}}}

\newcommand{\calM}{{\mathcal{M}}}
\newcommand{\calE}{{\mathcal{E}}}
\newcommand{\calL}{{\mathcal{L}}}

\newcommand{\lam}{\lambda}
\newcommand{\sig}{\sigma}
\newcommand{\eps}{\epsilon}
\newcommand{\gam}{\gamma}
\newcommand{\ome}{\omega}
\newcommand{\Ome}{\Omega}
\newcommand{\al}{\alpha}
\newcommand{\del}{\delta}
\newcommand{\pa}{\partial}

\newcommand{\Del}{\Delta}

\newcommand{\sgn}{\operatorname{sgn}}
\newcommand{\tr}{\operatorname{Tr}}
\renewcommand{\ker}{\operatorname{Ker}}
\newcommand{\Div}{\operatorname{Div}}

\newcommand{\tl}{\operatorname{TL}}
\newcommand{\km}{\operatorname{KM}}
\newcommand{\sn}{\operatorname{sn}}

\newcommand{\rank}{\operatorname{rank}}
\newcommand{\res}{\operatornamewithlimits{res}}
\renewcommand{\dim}{\operatorname{dim}}
\renewcommand{\Re}{\operatornamewithlimits{Re}}
\renewcommand{\Im}{\operatornamewithlimits{Im}}

\newcommand{\ba}{\begin{eqnarray}}
\newcommand{\ea}{\end{eqnarray}}
\newcommand{\lb}{\label}
\newcommand{\bth}{{(3)}}

\newcommand{\btwo}{{(2)}}
\newcommand{\bze}{{(0)}}
\newcommand{\ti}{\tilde}
\newcommand{\bs}{\backslash}
\newcommand{\bi}{\bibitem}

\newcommand{\no}{\nonumber}
\newcommand{\uome}{{\underline{\ome}}}
\newcommand{\ual}{{\underline{\alpha}}}
\newcommand{\opi}{{\overline{\Pi}}}
\newcommand{\ul}{\underline}
\newcommand{\ua}{{\underline{A}}}
\newcommand{\uq}{{\underline{Q}}}
\newcommand{\un}{{\underline{n}}}
\newcommand{\um}{{\underline{m}}}
\newcommand{\uu}{{\underline{u}}}
\newcommand{\uv}{{\underline{v}}}
\newcommand{\uc}{{\underline{c}}}
\newcommand{\uut}{{\underline{\underline{t}}}}
\newcommand{\umu}{{\underline{\mu}}}
\newcommand{\tual}{{\ti{\underline{\alpha}}}}
\newcommand{\hua}{{ \underline{\hat{A} }}}
\newcommand{\hual}{{ \underline{\hat{\alpha} }}}

\newcommand{\ppm}{{\underset{(-)}{+}}}
\newcommand{\pmp}{{\underset{(+)}{-}}}
\newcommand{\infpm}{{\infty_+, \infty_-}}
\newcommand{\pze}{{P_0}}
\newcommand{\nze}{{n_0}}
\newcommand{\humu}{{ \hat{\underline{\mu} }}}
\newcommand{\uxi}{{\underline{\Xi}}}
\newcommand{\uU}{{\underline{U}}}
\newcommand{\uw}{{\underline{w}}}
\newcommand{\uzero}{{\underline{0}}}

\newcommand{\ukm}{{\underline{\km}}}

\newcommand{\uz}{{\underline{z}}}
\newcommand{\huxi}{{\hat{\underline{\Xi}}}}
\newcommand{\hmu}{{\hat{\mu}}}

\newcommand{\hatP}{\hat{P}}

\makeatletter
\def\@maketitle{%
  \cleardoublepage \thispagestyle{empty}%
  \begingroup \topskip\z@skip
  \null\vfil
  \begingroup
  \LARGE\bfseries \centering
  \openup\medskipamount
  \@title\par\vspace{24pt}%
  \begin{center}
    \ifx\@empty\@dedicatory\else{\rm\normalsize \@dedicatory\\[5mm]}\fi
  \end{center}
  \def\and{\par\medskip}\centering
  \mdseries\authors\par\bigskip
  \endgroup
  \vfil
  \begin{center}
    \ifx\@empty\@date\else\@date\fi
  \end{center}
  \vfil
  \ifx\@empty\addresses\else
   Authors' addresses:
   \@setaddresses
  \fi
  \vfill
  \newpage\thispagestyle{empty}
  \begin{center}
    \ifx\@empty\@subjclass\else{{\em 1991 Mathematics Subject
Classification.}\\
        \@subjclass\\[2mm]}\fi
    \ifx\@empty\@keywords\else{{\em Keywords and phrases.}\\
\@keywords\\[2mm]}\fi
    \ifx\@empty\@translators\else\vfil\@settranslators\fi
    \ifx\@empty\thankses\else\vfil\@setthanks\fi
    {\em Received by the editor July 2, 1995}
  \end{center}
  \vfil
  \@setabstract
  \endgroup}
\makeatother

\numberwithin{equation}{chapter}

\begin{document}
\pagenumbering{roman}

\title[Algebro-Geometric Solutions]{Algebro-Geometric
Quasi-Periodic
Finite-Gap Solutions of the Toda and Kac-van Moerbeke
Hierarchies}
\dedicatory{Dedicated to the memory of Henrik
H.\ Martens (1927--1993)}

\author{W.~Bulla}
\address{Institute for Theoretical Physics, Technical
University of Graz, A-8010 Graz, Austria.}
\email{bulla@itp.tu-graz.ac.at}
\author{F.~Gesztesy}
\address{Department of Mathematics, University of
Missouri, Columbia, MO 65211, USA.}
\email{fritz@math.missouri.edu}
\author{H.~Holden}
\address{Department of Mathematical Sciences,
Norwegian
University of Science and Technology, N-7034
Trondheim, Norway.}
\email{holden@math.ntnu.no}
\author{G.~Teschl}
\address{Institute for Theoretical Physics, Technical
University of Graz, A-8010 Graz, Austria and Department
of
Mathematics, University of Missouri, Columbia, MO 65211,
USA.}
\curraddr{Institut f\"ur Reine und Angewandte Mathematik,
RWTH Aachen, 52056 Aachen, Germany}
\email{gerald@iram.rwth-aachen.de}


\begin{abstract}
Combining algebro-geometric methods and factorization
techniques for
finite difference expressions we provide a complete and
self-contained treatment of all real-valued quasi-periodic
finite-gap
solutions of both the Toda and Kac-van Moerbeke hierarchies.

In order to obtain our principal new result, the
algebro-geometric
finite-gap solutions of the Kac-van Moerbeke hierarchy,
we employ particular
commutation methods in connection with Miura-type
transformations which
enable us to transfer whole classes of solutions (such as
finite-gap
solutions) from the Toda hierarchy to its modified
counterpart, the Kac-van
Moerbeke hierarchy, and vice versa.
\end{abstract}

\maketitle
\tableofcontents

\newpage
\pagenumbering{arabic}


\chapter{Introduction}
The primary goal of this exposition is to construct all
real-valued
algebro-geometric quasi-periodic finite-gap solutions
of the Kac-van
Moerbeke ($\km$) hierarchy of nonlinear evolution equations.

While there exists a direct method to construct the
finite-gap solutions
of the $\km$ hierarchy, we shall use an alternative route
that exploits the
close connection between the Toda and $\km$ hierarchies and
characterizes the
$\km$ hierarchy as the modified Toda hierarchy in precisely
the same manner
that connects the Korteweg-de Vries (KdV) and modified
Korteweg-de Vries
(mKdV) hierarchies or more generally, the Gel'fand-Dickey (GD)
hierarchy and
its modified counterpart, the Drinfeld-Sokolov (DS) hierarchy.
The deep
connection between these hierarchies of nonlinear evolution
equations and
their modified counterparts is based on Miura-type
transformations which
in turn rely on factorization techniques of the
associated Lax
differential, respectively difference expressions, as will
be indicated
below.  (Alternatively, one can use the discrete analog of
the formal
pseudo differential calculus in connection with the GD and
DS hierarchy
as in \cite{57}, Ch.~IV.)  Accordingly, our approach
consists of three
main parts:\\
(i) A thorough treatment of the Toda hierarchy.\\
(ii) The algebro-geometric approach to completely integrable
nonlinear
evolution equations.\\
(iii) A transfer of classes of solutions of the Toda
hierarchy to that of
the Kac-van Moerbeke hierarchy and vice versa.\\
Our major results then may be summarized as follows:\\
($\alpha$) Construction of an alternative approach to the
Toda
hierarchy, modeled after Al'ber \cite{6}, Jacobi \cite{46a},
McKean \cite{60},
and Mumford \cite{70a}, Sect. III a).1, particularly suited
to derive its
algebro-geometric quasi-periodic finite-gap solutions.
Derivation of an
intimate connection of this approach with spectral
properties
of the corresponding Lax operator.\\
($\beta$) A complete presentation of the algebro-geometric
approach to
the
Toda hierarchy which goes beyond results in the literature
and leads, in
particular, to an alternative theta function representation
of $b(n,t)$
(in Flaschka's variables \cite{34}, cf.~\eqref{6.64}).\\
($\gamma$) A complete derivation of all real-valued
algebro-geometric
quasi-periodic finite-gap solutions of the $\km$ hierarchy,
our principal new
result.

Before we describe the content of each chapter, and hence
($\alpha$)--($\gam$) in some detail, we shall comment on items
(i)--(iii) a bit further.

The (Abelian) Toda lattice ($\tl$) in its original variables
reads
\begin{multline}
\dfrac{d^2}{dt^2} Q(n,t) =\exp [Q(n-1, t)-Q(n,t)] -
\exp [ Q(n,t) -Q(n+1,
t)],\\ \quad (n,t) \in \bbZ \times \bbR
\lb{1.1}
\end{multline}
and similarly, the original Kac-van Moerbeke system, also
called the Volterra
system, in physical variables, is of the type
\begin{equation}
\dfrac{d}{dt} R(n,t) = \frac{1}{2} \{ \exp [-R(n-1,t)] -
\exp [-R (n+1,
t)]\},\quad (n,t) \in \bbZ\times \bbR.
\lb{1.2}
\end{equation}
In Flaschka's variables \cite{34} for \eqref{1.1} and
similarly for
\eqref{1.2},
\begin{align}
\begin{split}
a(n,t) & = \frac{\eps (n)}{2} \exp \{ [Q(n,t)
-Q(n+1,t)]/2\},\\
b(n,t) & = \dot Q (n,t) /2,
\quad \eps (n) \in \{+1, -1\},
\lb{1.3}
\end{split}\\
\rho (n,t) & = \frac{\eps (n)}{2} \exp [-R(n,t) /2],
\quad \eps (n) \in
\{+1,-1\} \lb{1.4}
\end{align}
one can rewrite \eqref{1.1} and \eqref{1.2} in the form
\begin{equation}
\tl_0 (a,b) = \binom{\dot a -a(b-b^+)}{\dot
b-2[(a^-)^2 -a^2 ]} =0
\lb{1.5}
\end{equation}
and
\begin{equation}
\km_0 (\rho) =\dot \rho -\rho [(\rho^+)^2 -(\rho^-)^2] =0,
\lb{1.6}
\end{equation}
the latter also known as Langmuir lattice.
Here ``$\dot{ \:\: }$'' denotes
$d/dt$ and we employed the notation
$f^\pm (n) =f(n\pm 1)$, $n\in \bbZ$ and regarded all
equations in
the multiplication algebra of sequences. Moreover,
introducing the shift
operators
\begin{equation}
(S^\pm f)(n) =f(n\pm 1) =f^\pm (n),\quad n\in\bbZ
\lb{1.7}
\end{equation}
in $\ell^\infty (\bbZ)$, the systems \eqref{1.5} and
\eqref{1.6} are
well-known to be equivalent to the Lax equations
\begin{align}
\dot L -[P_2, L] & =0
\lb{1.8}\\
\intertext{
and}
\dot M -[Q_2, M] & = 0.
\lb{1.9}
\end{align}
Here $L$ and $P_2$ are the difference expressions
\begin{equation}
L=aS^+ +a^- S^- -b, \quad P_2 =aS^+ -a^- S^-
\lb{1.10}
\end{equation}
defined on $\ell^\infty (\bbZ)$, and $M$ and $Q_2$ the
matrix-valued difference
expressions
\begin{align}
\begin{split}
M & = \begin{pmatrix}
0 & \rho_o^- S^- + \rho_e\\
\rho_o S^+ + \rho_e & 0
\end{pmatrix}, \\
Q_2 & = \begin{pmatrix}
\rho_e \rho_o S^+ - \rho_e^- \rho_o^- S^- & 0\\
0 & \rho_e^+ \rho_o S^+ - \rho_e \rho_o^- S^-
\end{pmatrix}
\lb{1.11}
\end{split}
\end{align}
defined on $\ell^\infty (\bbZ)\otimes \bbC^2$, with $\rho_e$
and $\rho_o$ the
``even'' and ``odd'' parts of $\rho$, that is,
\begin{equation}
\rho_e (n,t) =\rho (2n,t), \quad \rho_o (n,t)
=\rho (2n+1, t), \quad n\in
\bbZ,
\lb{1.12}
\end{equation}
assuming $a$, $b$, $\rho\in\ell^\infty (\bbZ)$.

Since the literature on the Toda lattice (even if one
considers only the infinite lattice $\bbZ$, our main
interest) is extensive, we will only
refer to a few standard monographs such as \cite{28},
\cite{72}, \cite{73},
\cite{80}.  In the case of the Kac-van Moerbeke system
we refer to \cite{15}, \cite{38}, \cite{47},
\cite{48}, \cite{51}, \cite{57}, \cite{58}, \cite{59},
\cite{68}, \cite{83} and the references therein.

While \eqref{1.5} and \eqref{1.6} describe the original
Toda and Kac-van
Moerbeke lattices, one can develop a systematic generalization
to Lax pairs
of the type $(L, P_{2g+2})$ and $(M, Q_{2g+2})$, where
$P_{2g+2}$
($Q_{2g+2}$) are (matrix-valued) difference expressions
of order $2g+2$
with certain polynomial coefficients in $a$, $b$ ($\rho$).
The associated
Lax equations
\begin{align}
\dot L -[P_{2g+2}, L] & = 0\lb{1.13}\\
\intertext{and}
\dot M -[Q_{2g+2}, M] & = 0
\lb{1.14}
\end{align}
(cf.~Chapter~\ref{s2}) are then equivalent to the $\tl_g$ and
$\km_g$ equations
denoted by
\begin{align}
\tl_g (a,b) & = 0 \lb{1.15}\\
\intertext{and}
\km_g (\rho) & = 0. \lb{1.16}
\end{align}
Varying $g\in \bbN_0$ then yields the corresponding
hierarchies of
nonlinear evolution equations for $(a,b)$ and $\rho$.

The special case of stationary $\tl_g$ and $\km_g$ equations,
characterized
by $\dot L =0$, $\dot M=0$ in \eqref{1.13}, \eqref{1.14}, or
equivalently, by commuting difference expressions of the type
\begin{align}
[P_{2g+1}, L] & = 0,\lb{1.17}\\
[Q_{2g+2}, M] & = 0,
\lb{1.18}
\end{align}
then yields a polynomial relationship between $L$ and
$P_{2g+2}$,
respectively $M$ and $Q_{2g+2}$.  In fact, \eqref{1.17}
and \eqref{1.18}
imply the following analogs of the Burchnall-Chaundy
polynomials familiar
from the theory of commuting ordinary differential
expressions \cite{16},
\cite{17}
\begin{alignat}{2}
P_{2g+2}^2 & = \prod_{m=0}^{2g+1} (L-E_m), && \quad
\{E_m \}_{0\leq m
\leq 2g+1} \subset \bbC, \lb{1.19}\\
Q_{2g+2}^2 & = \prod_{m=0}^{2g+1} (M^2 -e_m), &&
\quad \{e_m\}_{0 \leq m
\leq 2g+1} \subset \bbC.
\lb{1.20}
\end{alignat}
In particular, \eqref{1.19} and \eqref{1.20} yield
the following
hyperelliptic curves
\begin{align}
y^2 & = \prod_{m=0}^{2g+1} (z-E_m), \lb{1.21}\\
y^2 & = \prod_{m=0}^{2g+1} (w^2 -e_m),
\lb{1.22}
\end{align}
the fundamental ingredients of the algebro-geometric
approach for the $\tl$ and $\km$ hierarchies.  For spectral
theoretic reasons (see, e.g.,
Theorems~\ref{t4.2} and \ref{t8.2}) algebro-geometric
solutions $(a,b)$ and $\rho$ of
\eqref{1.15} and \eqref{1.16} are called $g$-gap solutions
following the
conventional terminology.

If $\theta$ denotes the Riemann theta function associated
with the curve
\eqref{1.21} (and a fixed homology basis), the ultimate
goal of the
algebro-geometric approach is then a $\theta$-function
representation of the
solutions $(a,b)$ and $\rho$ of the $\tl_r$ and
$\km_r$ equations,
\begin{equation}
\tl_r (a,b) =0, \quad \km_r(\rho) =0, \quad r\in\bbN_0
\lb{1.23}
\end{equation}
with $g$-gap initial conditions,
\begin{equation}
(a,b) =(a^{(0)}, b^{(0)}), \quad \rho
= \rho^\bze \text{ at } t=t_0,
\lb{1.24}
\end{equation}
where $(a^\bze, b^\bze)$ and $\rho^\bze$ are
stationary solutions of the
$\tl_g$ and $\km_g$ equations, that is,
\begin{alignat}{2}
& \tl_g (a^\bze, b^\bze) =0, && \quad  \dot a^\bze
=\dot b^\bze
=0,\notag\\
& \km_g (\rho^\bze) =0, && \quad  \dot \rho^\bze =0
\lb{1.25}
\end{alignat}
for some fixed $g\in\bbN_0$.

Next, we illustrate the close connection between the
Toda
hierarchy and its modified counterpart, the $\km$ hierarchy.
Introducing
the difference expressions
\begin{equation}
A=\rho_o S^+ + \rho_e, \quad A^*=\rho_o^- S^- + \rho_e
\lb{1.26}
\end{equation}
in $\ell^\infty (\bbZ)$ one infers that
\begin{align}
M & = \begin{pmatrix}
0 & A^*\\
A & 0
\end{pmatrix},
\lb{1.27}\\
M^2 & = \begin{pmatrix}
A^* A & 0\\
0 & A A^*
\end{pmatrix} =L_1 \oplus L_2 \; ,
\lb{1.28}
\end{align}
with
\begin{align}
L_1 & = A^* A, \quad L_2 = A A^*,
\lb{1.29}\\
L_k & =a_k S^+ +a_k^- S^- -b_k, \quad k=1,2,
\lb{1.30}
\end{align}
\begin{alignat}{2}
a_1 & = \rho_e \rho_o, && \quad  b_1 = -\rho^2_e -(\rho_o^-)^2,
\lb{1.31}\\
a_2 & = \rho_e^+ \rho_o, && \quad b_2 =-\rho_e^2 -\rho_o^2
\lb{1.32}
\end{alignat}
and
\begin{equation}
Q_{2g+2} =\begin{pmatrix}
P_{1, 2g+2} & 0\\
0 & P_{2, 2g+2}\end{pmatrix} =P_{1, 2g+2}\oplus P_{2, 2g+2}\; .
\lb{1.33}
\end{equation}
Here $P_{k, 2g+2}$ is constructed as in \eqref{1.10}
and \eqref{1.13}
with $(a,b)$ replaced by $(a_k, b_k)$, $k=1,2$,
respectively.  Relations
\eqref{1.28} and \eqref{1.33} then can be exploited to
prove the
implication
\begin{equation}
\km_g(\rho) =0 \Rightarrow \tl_g (a_k, b_k) =0, \quad k=1,2,
\lb{1.34}
\end{equation}
that is, a solution $\rho$ of the $\km_g$ equations
\eqref{1.16} yields two
solutions $(a_k, b_k)$, $k=1,2$ of the $\tl_g$ equations
\eqref{1.15}
related to one another by \eqref{1.31}, \eqref{1.32}, the
discrete analog
of Miura's transformation \cite{64}, familiar from the
(m)KdV hierarchy.
(According to a footnote in \cite{68}, the connection
between the $\km$
and $\tl$ lattices was mentioned by H\' enon in a letter
to Flaschka as
early as 1973.)  Incidentally, \eqref{1.28}--\eqref{1.32}
illustrate the
factorization of the Lax difference expression $L$
alluded to earlier.
The implication \eqref{1.34} was first systematically
studied by Adler
\cite{1} using factorization techniques.  Transformations
between the $\km$
and $\tl$ systems were studied earlier by Wadati \cite{83}
(see also
\cite{80}).  In a recent paper \cite{38} the converse
of \eqref{1.34} was
established.  More precisely, assuming the existence of
a solution $(a_1,
b_1)$ of the $\tl_g$ equations \eqref{1.15}, that is,
\begin{equation}
\tl_g (a_1, b_1)=0,
\lb{1.35}
\end{equation}
a solution $\rho$ of the $\km_g$ equations \eqref{1.16}
and another
solution $(a_2, b_2)$ of the $\tl_g$ equations
\eqref{1.15} are
constructed,
\begin{equation}
\km_g(\rho)=0,\quad \tl_g(a_2, b_2) =0
\lb{1.36}
\end{equation}
related to each other by the Miura-type transformation
\eqref{1.31},
\eqref{1.32} (we refer to Chapter~\ref{s7} for a detailed
discussion of these facts).
Equations \eqref{1.34} and especially \eqref{1.35}, \eqref{1.36}
yield the
possibility of transferring classes of solutions (such as
finite-gap
solutions) from the Toda hierarchy to the $\km$ hierarchy
and vice versa.

Having illustrated items~(i)--(iii) to some extent, we
finally turn to a description of the content of each
chapter.  In
Chapter~\ref{s2} we develop an alternative recursive
approach to the
Toda
hierarchy modeled after Al'ber \cite{6}.  In particular,
we recursively
compute the difference expressions $P_{2g+2}$
in \eqref{1.13}.
We chose to develop this approach in detail since it
most naturally leads
to the fundamental Burchnall-Chaundy polynomials and
hence to the
underlying hyperelliptic curves in connection with the
stationary Toda
hierarchy.  In addition it provides direct insight into
the spectral
properties of the underlying Lax operator as detailed
later in
Chapter~\ref{s4}.

Chapter~\ref{s3} is devoted to the algebro-geometric
approach to
integrate
nonlinear evolution equations and in particular to the
Baker-Akhiezer
(BA) function, the fundamental object of this approach.
Historically,
these techniques go back to the work of Baker \cite{8},
Burchnall and
Chaundy \cite{16}, \cite{17}, and Akhiezer \cite{5}.
The modern approach
was initiated by Its and Matveev \cite{45} in connection
with the
KdV equation and
further developed into a powerful machinery by Krichever
(see, e.g., the
review \cite{55}) and others.  We refer, in particular,
to the extensive
treatments in \cite{10}, \cite{24}, \cite{25}, \cite{26},
\cite{59},
\cite{69}, and \cite{72}.  In the special context of the
Toda equations
we refer to \cite{2}, \cite{19}, \cite{25}, \cite{26},
\cite{53},
\cite{55}, \cite{59}, \cite{63}, and \cite{67}.  Our own
presentation
starts with commuting difference expressions and their
associated
hyperelliptic curves and then develops the stationary
algebro-geometric
approach from first principles. In particular, we chose
to follow
Jacobi's classic representation of positive divisors of
degree $g$ of the
hyperelliptic curve \eqref{1.21} \cite{46a} which was
first applied to
the KdV case by Mumford \cite{70a}, Sect. III a).1 with
subsequent
extensions due to McKean \cite{60}. The reader will find
a meticulous
account which provides more details on the BA-function
than usually found
in the literature (see, e.g., Theorem~\ref{t3.4}).

Spectral theoretic properties and Green's functions of
self-adjoint
$\ell^2 (\bbZ)$ realizations $H$ of $L$ are the main
topic of
Chapter~\ref{s4}.  Assuming that $L$ is defined in
terms of stationary
solutions $(a,b)$ of the $\tl_g$ equations, we determine
the spectrum of
the Jacobi operator $H$ in Theorem~\ref{t4.2} and
provide a link between
the $2\times 2$ spectral matrix of $H$ and our recursive
approach to the
stationary Toda hierarchy (cf.
\eqref{4.32}--\eqref{4.36}).  The latter
result appears to be new and underscores the fundamental
importance of the
recursion formalism chosen in Chapter~\ref{s2}.

In Chapter~\ref{s5} we continue our stationary
algebro-geometric approach
to the Toda hierarchy and provide a detailed derivation
of the standard
$\theta$-function representation of all stationary
$\tl_g$ solutions
(cf.~Theorem~\ref{t5.2}).

In Chapter~\ref{s6} we finally complete the
algebro-geometric approach to
the Toda hierarchy.  In addition to a detailed
discussion of the
time-dependent BA-function in Theorem~\ref{t6.2} we
derive the well-known
(time-dependent) $\theta$-function representation of
the $\tl_r$
equations with $g$-gap initial conditions in
Theorem~\ref{t6.3}.  Our
detailed account in Chapter~\ref{s5} and \ref{s6}
also leads to an
alternative $\theta$-function representation of $b$ in
Corollaries~\ref{c5.6} and \ref{c6.5} which, much to
our surprise, seems
to have escaped notice in the literature thus far.

In Chapter~\ref{s7} we turn to the $\km$ hierarchy and
its connection with
the Toda hierarchy.  In addition to developing a
recursive approach to
the $\km$ hierarchy (which appears to be new) and in
particular to a
computation of $Q_{2g+2}$ in \eqref{1.14}, we describe
at length the
Miura-type transformation \eqref{1.31}, \eqref{1.32}
and especially the
transfer of solutions from the Toda to the $\km$
hierarchy in
Theorem~\ref{t7.2}.

In analogy to Chapter~\ref{s4}, Chapter~\ref{s8}
establishes spectral
properties of the self-adjoint realization $D$ of
$M$ in $\ell^2(\bbZ)
\otimes \bbC^2$ associated with finite-gap solutions
$\rho$ of the
$\km_g$ equations.  Theorem~\ref{t8.2}, in
particular, reduces the
spectral analysis of the Dirac-type difference
operator $D$ to that of
the Jacobi
operators $H_k$, the self-adjoint realizations of
$L_k$, $k=1,2$ in
$\ell^2 (\bbZ)$ (cf.~\eqref{1.27},
\eqref{1.29}--\eqref{1.32}) by using
factorization (commutation) methods indicated in
\eqref{1.28}.

Finally, in Chapter~\ref{s9} we complete the
principal objective of this exposition and derive
all real-valued algebro-geometric quasi-periodic
finite-gap solutions of the $\km$ hierarchy in
Theorems~\ref{t9.3} and
\ref{t9.5}.  Isospectral manifolds of finite-gap $\km$
solutions are briefly
considered in Remark~\ref{r9.4} and a brief outlook
on possible
applications of these completely integrable lattice
models ends this exposition.

For convenience of the reader, and for the sake of
being
self-contained, we added Appendix~\ref{app-a} which
summarizes basic
facts on hyperelliptic curves and their
$\theta$-functions and defines
the notation used in the main body of this exposition.
Appendix~\ref{app-b}
records the principal results of
Chapters~\ref{s3}--\ref{s5} in the
important special case of periodic rather than
quasi-periodic Jacobi
operators by explicitly invoking Floquet theory.
 Appendix~\ref{app-c}
finally records the simplest explicit examples
associated with
genus $g=0$ and~$1$.


\chapter[The Toda Hierarchy]{The Toda
Hierarchy, Recursion Relations, and Hyperelliptic
Cur\-ves} \lb{s2}
\setcounter{equation}{0}

In this chapter we review the construction of the Toda
hierarchy by using a recursive approach first advocated
by Al'ber \cite{6} and derive the Burchnall-Chaundy
polynomials in connection with the stationary Toda
hierarchy.  Our recursive approach to the Toda
hierarchy, though equivalent to the conventional one (see, e.g.,
\cite{57}, \cite{67},\cite{69}, \cite{73},\cite{77},\cite{78},
\cite{81}),
markedly differs from the standard treatment.  We have chosen to
present the formalism below since it most naturally yields the
Burchnall-Chaundy polynomials associated with the stationary Toda
hierarchy \linebreak and hence the underlying hyperelliptic
curves for
algebro-geometric quasi-periodic finite-gap solutions of the
Toda and
Kac-van Moerbeke hierarchies to be considered in
Chapters~\ref{s6} and
\ref{s9}.  Moreover, as shown in Chapter~\ref{s4}
(cf.~\eqref{4.32}--\eqref{4.36}), this recursive approach
provides a fundamental link to the spectral matrix of the
underlying Lax operator.

We start by introducing some notations. In the following
we denote by $\ell^p(M)$, where $1 \le p \le \infty$,
$M=\bbN$, $\bbN_0 =\bbN\cup \{0\}$, $\bbZ$, etc., the
usual space of $p$-summable respectively bounded (if
$p=\infty$) complex-valued sequences $f=\{f(m)\}_{m\in
M}$ and by
$\ell^p_{\bbR}(M)$ the corresponding restriction to real-valued
sequences. The
scalar product in the Hilbert space $\ell^2(M)$ will be
denoted by
\begin{equation}
(f,g) = \sum_{n \in M} \overline{f(n)} g(n), \quad \;
f,g \in \ell^2(M).
\lb{2.0}
\end{equation}
Since $\ell^\infty (\bbZ) \subseteq \ell^p (\bbZ)$ in a
natural way it suffices to make all further definitions
for $p=\infty$. In $\ell^\infty (\bbZ)$ we introduce the
shift operators
\begin{equation}
(S^\pm f)(n) =f(n\pm 1), \quad n \in \bbZ, \;
f\in \ell^\infty (\bbZ)
\lb{2.1}
\end{equation}
and in order to simplify notations we agree to use the
short cuts
\begin{align}
\begin{split}
& f^\pm = S^\pm f, \quad \text{ that is, } f^\pm (n) =
f(n\pm 1),\\ & (f+g) (n) =f(n) +g(n),  \quad  (fg) (n)
=f(n) g(n), \quad n \in \bbZ,
\; f,g\in\ell^\infty(\bbZ)
\lb{2.2}
\end{split}
\end{align}
whenever convenient. Moreover, if
$R: \: \ell^\infty(\bbZ) \to \ell^\infty(\bbZ)$
denotes a difference expression, let
\begin{equation}
R=\{R(m,n)\}_{m,n\in\bbZ}, \quad R(m,n) = (\del_m, R
\del_n)
\lb{2.3}
\end{equation}
denote its corresponding matrix representation with
respect to the standard basis
\begin{equation}
e_m =\{\del_m (n) \}_{n\in\bbZ}, \quad m\in\bbZ,\quad
\del_m(n) =\begin{cases}
1, & m=n\\ 0, & m \neq n
\end{cases}.
\lb{2.4}
\end{equation}
In connection with \eqref{2.3} we define the diagonal
and upper and lower triangular parts of $R$ as follows
\begin{alignat}{2}
R_0 & = \{ R_0 (m,n)\}_{m,n \in \bbZ}, \quad & R_0
(m,n) & =
\begin{cases} R(m,m), & m =n\\
0, & m \neq n
\end{cases}, \notag \\
R_\pm & =\{ R_\pm (m,n)\}_{m,n \in\bbZ}, & R_\pm (m,n)
& = \begin{cases} R(m,n), & \pm (n-m) > 0\\
0, & \text{ otherwise }
\end{cases} .
\lb{2.5}
\end{alignat}
Clearly,
\begin{equation}
R=R_+ + R_0 + R_-.
\lb{2.6}
\end{equation}

Given these notations one can now introduce the Toda
hierarchy.  Let
\begin{align}
\begin{split}
& a(t) =\{ a(n,t)\}_{n\in\bbZ} \in \ell^\infty(\bbZ),
\quad b(t) =\{b(n,t)\}_{n\in\bbZ} \in \ell^\infty(\bbZ),
\:\: t \in \bbR,\\ & 0 \neq a(n,.), \, b(n,.) \in C^1
(\bbR), \; n\in\bbZ
\lb{2.7}
\end{split}
\end{align}
and introduce the difference expressions $(L(t),
P_{2g+2} (t))$ (the Lax pair) in $\ell^\infty(\bbZ)$
\begin{align}
& L(t) = a(t) S^+ +a^- (t) S^- -b(t), \quad t\in\bbR,
\lb{2.8}\\
\begin{split}
& P_{2g+2} (t)=-L (t)^{g+1} + \sum_{j=0}^g [ g_j (t)
+2a(t) f_j (t) S^+] L(t)^{g-j} + f_{g+1} (t),\\
& \hspace{3in} g\in\bbN_0, \; t\in\bbR,
\lb{2.9}
\end{split}
\end{align}
where $\{f_j(n,t)\}_{0 \leq j \leq g+1}$ and $\{g_j
(n,t)\}_{0 \leq j \leq g}$ satisfy the recursion
relations
\begin{align}
\begin{split}
& f_0 = 1, \: g_0 = -c_1,\\
& 2 f_{j+1} +g_j + g^-_j +2b
f_j =0, \quad 0\leq j \leq g,\\
& g_{j+1} - g_{j+1}^- + 2[a^2 f^+_j -(a^-)^2f^-_j] +
b[g_j - g_j^-] =0, \quad 0\leq j \leq g-1.
\lb{2.10}
\end{split}
\end{align}
Note that $a$ enters in $f_j$ and $g_j$ only quadratically.
Then the Lax equation
\begin{equation}
\dfrac{d}{dt} L(t) -[P_{2g+2}(t), L(t)]=0, \quad
t\in\bbR
\lb{2.11}
\end{equation}
(here $[.\, ,\, .]$ denotes the commutator) is
equivalent to
\begin{align}
\begin{split}
\tl_g (a(t), b(t))_1 & = \dot a (t) +a(t) [g_g^+  (t)
+g_g (t) + f_{g+1}^+ (t) + f_{g+1} (t)\\ & + 2b^+ (t)
f_g^+ (t) ]=0,\\
\tl_g (a(t), b(t))_2 & = \dot b (t) +2 [b(t) (g_g (t)
+ f_{g+1} (t)) +a(t)^2 f_g^+ (t)\\
& -a^- (t)^2 f^-_g(t) +b(t)^2 f_g(t)]=0,\quad t\in\bbR.
\lb{2.12}
\end{split}
\end{align}
Varying $g\in \bbN_0$ yields the Toda hierarchy
\begin{equation}
\tl_g(a,b) =(\tl_g (a,b)_1, \tl_g (a,b)_2)^T =0, \quad
g\in\bbN_0.
\lb{2.13}
\end{equation}
Explicitly, one obtains from \eqref{2.10},
\begin{align}
\begin{split}
f_1 & = -b+c_1,\\
g_1 & = -2a^2 - c_2,\\
f_2 & = a^2 + (a^-)^2 +b^2 -c_1 b+c_2,\\
g_2 & = 2a^2 (b + b^+) - 2c_1 a^2 -c_3,\\
f_3 & = -(a^-)^2 (b^- +2b) - a^2 (b^+ + 2b) -b^3\\
&{} + c_1 (a^2 + (a^-)^2 +b^2) - c_2 b +c_3,\\
& \text{etc.}
\lb{2.14}
\end{split}
\end{align}
and hence from \eqref{2.12},
\begin{align}
\tl_0 (a,b) = & \binom{\dot a -a(b-b^+)}{\dot
b-2[(a^-)^2 -a^2 ]} =0,
\lb{2.15}\\
\tl_1 (a,b) = & \binom{\dot a -a[(a^+)^2 -(a^-)^2
+(b^+)^2 -b^2]}{\dot b -2a^2 (b^+ +b) +2(a^-)^2 (b+b^-)}
 + c_1 \binom{ -a(b-b^+)}{ -2 [(a^-)^2 -a^2]} =0,
\lb{2.16}\\ \no
\tl_2 (a,b) = & \binom {\scriptstyle
\dot a -a[b^3 -(b^+)^3 +2(a^-)^2 b-2(a^+)^2 b^+ +a^2
(b-b^+) -(a^+)^2 b^{++} -(a^-)^2 b^-] }
{\scriptstyle
\dot b-2(a^-)^2 [b^2 +bb^- +(b^-)^2 +(a^-)^2 +
(a^{--})^2 ] +2a^2 [b^2 +bb^+ +(b^+)^2 +a^2 +(a^+)^2] }\\
& + c_1 \binom
{\scriptstyle -a [(a^+)^2 -(a^-)^2 +(b^+)^2 -b^2] }
{\scriptstyle -2a^2 (b^+ +b) +2(a^-)^2 (b+b^-) }
+c_2 \binom {\scriptstyle -a(b-b^+)}{\scriptstyle - 2
[(a^-)^2 -a^2 ] } =0,
\lb{2.17}\\ \no & \text{etc.}
\end{align}
represent the first few equations of the Toda
hierarchy.  Here $c_\ell$ denote summation constants
which naturally arise by solving the resulting
difference equations for $g_{g+1, \ell}$ in
\eqref{2.10}. Throughout this exposition we will chose these
constants $c_\ell$ to be
real-valued. The corresponding homogeneous Toda equations
obtained by taking all summation constants
equal to zero, $c_\ell \equiv 0$, $\ell \in \bbN$,
are then denoted by
\begin{equation}
\widehat{\tl}_g (a,b):= \tl_g(a,b) \big|_{c_\ell
\equiv 0, \; 1\leq \ell \leq g} \lb{2.18}
\end{equation}
and similarly we denote by $\hat{P}_{2g+2} :=
P_{2g+2} (c_\ell \equiv 0)$, $\hat{f}_j := f_j
(c_\ell \equiv 0)$, $\hat{g}_j
:= g_j(c_\ell \equiv 0)$ the corresponding
homogeneous quantities. One verifies
\begin{equation}
P_{2g+2} =\sum_{m=0}^g c_{g-m} \hat{P}_{2m+2}, \quad c_0
=1.
\lb{2.29}
\end{equation}

Next we relate the homogeneous quantities  $\hat{f}_j$,
$\hat{g}_j$ to certain matrix elements of $L(t)^j$.

\begin{lem}\lb{l4.3}
The homogeneous coefficients $\{\hat{f}_j(t)\}_{0  \leq j
\leq g+1}$ and $\{\hat{g}_j(t)\}_{0 \leq j \leq g}$
satisfy
\begin{align}
\hat{f}_j(n,t) & = (\del_n, L(t)^j \del_n), \quad 0
\leq j \leq g+1, \; n \in
\bbZ, \lb{4.37}\\
\hat{g}_j(n,t) & = -2 a(n) (\del_{n+1}, L(t)^j \del_n),
\quad 0 \leq j \leq g,
\; n\in \bbZ, \lb{4.38}
\end{align}
where $\del_n =\{\del_{n,m}\}_{m\in\bbZ}$.
\end{lem}

\begin{proof}
We abbreviate
\begin{align}
\ti{f}_j (n) & = (\del_n, L^j \del_n), \quad \ti{g}_j (n) =
-2a(n) (\del_{n+1}, L^j
\del_n).
\lb{4.39}\\
\intertext{Then}
\ti{f}_{j+1} & = (L\del_n, L^j \del_n) =-\frac12
(\ti{g}_j + \ti{g}^-_j) -b \ti{f}_j
\lb{4.40}\\
\intertext{and similarly,}
\ti{g}_{j+1} & = -b \ti{g}_j -2a^2 \ti{f}_j^+ + \ti{h}_j =
-b^+ \ti{g}_j -2a^2 \ti{f}_j +\ti{h}_j^+,
\lb{4.41}\\
\intertext{where}
\ti{h}_j(n) & = -2a(n) a(n-1) (\del_{n+1}, L^j \del_{n-1}).
\lb{4.42}
\end{align}
Eliminating $\ti{h}_j$ in \eqref{4.41} results in
\begin{equation}
\ti{g}_{j+1} - \ti{g}^-_{j+1} =-2 [a^2 \ti{f}_j^+
-(a^-)^2 \ti{f}_j^-]
-b [\ti{g}_j -\ti{g}_j^-].
\lb{4.43}
\end{equation}
By inspection, \eqref{4.40} and \eqref{4.43} are
equivalent to \eqref{2.10}.  In order to determine which
solution of \eqref{2.10} has been found (i.e., determine
the summation constants $c_1, \ldots, c_g$) we
temporarily assign the weight one to $a(n)$ and $b(n)$,
$n\in\bbZ$. Then $\hat{f}_j$ and $\hat{g}_{j+1}$ have
weight $j$ and hence
\begin{equation}
c_0 =1, \; c_j =0, \quad 1 \leq j \leq g
\lb{4.44}
\end{equation}
completing the proof.
\end{proof}

Now we are in the position to reveal the connections
with the usual approach to the Toda equations. It
suffices to consider the homogeneous case.

\begin{lem}
The homogeneous Lax operator $\hat{P}_{2g+2}$ satisfies
\begin{equation}
\hat{P}_{2g+2} (t) =[L(t)^{g+1}]_+ -[L(t)^{g+1}]_-
\lb{2.19}
\end{equation}
(cf. the notation in \eqref{2.5}).
\end{lem}

\begin{proof}
We use induction on $g$. $g=0$ is trivial. Suppose \eqref{2.19}
holds for $g=0,\dots,g_0-1$.  By
\eqref{2.9} we have
\begin{equation}
\hat{P}_{2g_0+2}(t) = \hat{P}_{2g_0}(t) L(t) +[\hat{g}_{g_0}(t) +
2 a(t) \hat{f}_{g_0}(t) S^+] - \hat{f}_{g_0}(t) L(t) +
\hat{f}_{g_0+1}(t).
\end{equation}
In order to prove \eqref{2.19} one considers $(\del_m,
\hat{P}_{2g_0+2}(t) \del_n)$ and makes the case distinctions
$m<n-1,m=n-1,m=n,m=n+1,m>n+1$.  Explicitly, one verifies, for
instance, in the case $m=n$,
\begin{align}
& \quad (\del_m,\hat{P}_{2g_0+2} \del_n) \no \\
& \quad=(\del_n,\hat{P}_{2g_0}(a\del_{n-1}+a^-\del_{n+1}-b\del_n))
+\hat g_{g_0}(n)+b(n)\hat f_{g_0}(n)+\hat f_{g_0+1}(n) \no \\
&\quad=(\del_n,[(L^{g_0})_+ - (L^{g_0})_-](a\del_{n-1}
+a^-\del_{n+1}-b\del_n))
+\hat g_{g_0}(n)+b(n)\hat f_{g_0}(n)+\hat f_{g_0+1}(n) \no \\
&\quad=(\del_n,(L^{g_0})_+\; a^-\del_{n+1})
-(\del_n,(L^{g_0})_- \; a\del_{n-1})
+\hat g_{g_0}(n)+b(n)\hat f_{g_0}(n)+\hat f_{g_0+1}(n) \no \\
&\quad=a(n)(\del_n,L^{g_0}\del_{n+1})-a(n-1)
(\del_n,L^{g_0}\del_{n-1})
+\hat g_{g_0}(n)+b(n)\hat f_{g_0}(n)+\hat f_{g_0+1}(n) \no \\
&\quad=\frac12\hat g_{g_0}(n)+\frac12\hat g_{g_0}(n-1)
+b(n)\hat f_{g_0}(n)+\hat f_{g_0+1}(n)=0 \no
\\ & \text{} \label{tull}
\end{align}
using \eqref{2.10}, \eqref{4.37}, and \eqref{4.38}. Since obviously
\begin{equation}
(\del_n,[(L^{g_0+1})_+-(L^{g_0+1})_-]\del_n)=0
\end{equation}
by \eqref{2.5}, this settles the case $m=n$ in \eqref{2.19}.  The
remaining cases are settled one by
one in a similar fashion.
\end{proof}

Before we turn to a discussion of the stationary Toda
hierarchy we briefly sketch the main steps leading to
\eqref{2.9}--\eqref{2.12}.  If
$\ker (L(t) -z)$, $z\in\bbC$ denotes the two-dimensional
nullspace of $L(t) -z$ (in the algebraic sense as
opposed to the functional analytic one), we seek a
representation of
$P_{2g+2}(t)$ in
$\ker (L(t)-z)$ of the form
\begin{equation}
P_{2g+2}(t)\big|_{\ker(L(t)-z)} =2a(t) F_g(z,t) S^+ +
G_{g+1}(z,t),
\lb{2.20}
\end{equation}
where $F_g$ and $G_{g+1}$ are polynomials in $z$ of the
type
\begin{equation}
F_{g} (z,t) =\sum_{j=0}^g z^j f_{g-j}(t), \; G_{g+1}
(z,t) =-z^{g+1} + \sum_{j=0}^g z^j g_{g-j}(t) +
f_{g+1}(t),
\lb{2.21}
\end{equation}
with $f_{\ell}(t) = \{f_{\ell} (n,t)\}_{n\in\bbZ}
\in\ell^\infty(\bbZ)$, $g_{\ell} (t) = \{
g_{\ell}(n,t)\}_{n\in\bbZ}  \in\ell^\infty(\bbZ)$.
The Lax equation \eqref{2.11} restricted to
$\ker (L(t) -z)$ then yields
\begin{align}
\begin{split}
0 & = \{ \dot L -[P_{2g+2}, L]\} \big|_{\ker (L-z)}
=\{ \dot L+(L-z) P_{2g+2}\}\big|_{\ker (L-z)}\\
& = \big\{ a\big[ \dfrac{\dot a}{a} -\dfrac{\dot
a^-}{a^-} +2 (b^+ +z) F_g^+ -2(b+z) F_g +G_{g+1}^+
-G_{g+1}^- \big] S^+\\ & \quad + \big[ - \dot b +(b+z)
\dfrac{\dot a^-}{a^-} +2(a^-)^2 F_g^-\\ & \quad -2a^2 F_g^+ +(b+z)
(G_{g+1}^- -G_{g+1})\big]\big\}
\big|_{\ker (L-z)}.
\lb{2.22}
\end{split}
\end{align}
Hence one obtains
\begin{align}
& \dfrac{\dot a}{a} - \dfrac{\dot a^-}{a^-} = 2 (b^+ +z) F_g^+
-2(b+z)
F_g +  G_{g+1}^- -  G_{g+1}^+,
\lb{2.23}\\
& \dot b = (b+z) \dfrac{\dot a^-}{a^-} +2(a^-)^2 F_g^-
-2a^2 F_g^+ +(b+z) (G_{g+1}^- -G_{g+1}).
\lb{2.24}
\end{align}
Upon summing \eqref{2.23} (adding $G_{g+1} -G_{g+1}$ and
neglecting a trivial summation constant) one infers
\begin{equation}
\dot a =-a [2(b^+ +z) F_g^+ +G_{g+1}^+ +G_{g+1}], \quad
g\in\bbN_0.
\lb{2.25}
\end{equation}
Insertion of \eqref{2.25} into \eqref{2.24} then
implies
\begin{equation}
\dot b =-2 [(b+z)^2 F_g +(b+z) G_{g+1} +a^2 F_g^+
-(a^-)^2 F_g^-], \quad g\in\bbN_0.
\lb{2.26}
\end{equation}
Insertion of \eqref{2.21} into \eqref{2.25} and
\eqref{2.26} then produces the recursion relation
\eqref{2.10} (except for the relation involving $f_{g+1}$ which
serves
as a definition) and the result \eqref{2.12}. Relation
\eqref{2.20} then yields \eqref{2.9}.  We omit further
details and just record as an illustration a few of the
polynomials $F_g$ and
$G_{g+1}$,
\begin{align}
\begin{split}
F_0 & = 1 = \hat{F}_0,\\
G_1 & = -b-z = \hat{G}_1,\\
F_1 & = c_1 -b+z = c_1 \hat{F}_0 + \hat{F}_1,\\
G_2 & = c_1 (-b-z) +(a^-)^2 -a^2 +b^2 -z^2 = c_1
\hat{G}_1 + \hat{G}_2,\\ F_2 & = c_2 +c_1 (-b+z) +a^2
+(a^-)^2 +b^2 -bz+z^2= c_2 \hat{F}_0 + c_1
\hat{F}_1 +\hat{F}_2,\\
G_3 & = c_2 (-b-z) +c_1 ((a^-)^2 -a^2 +b^2 -z^2)
+a^2b^+ -(a^-)^2 b^-\\ & \quad -2 (a^-)^2 b-b^3 -2a^2 z
-z^3 = c_2
\hat{G}_1 + c_1  \hat{G}_2 +\hat{G}_3,\\
& \text{etc.}
\lb{2.27}
\end{split}
\end{align}

\begin{rem}\lb{r2.1}
Since by \eqref{2.10}, \eqref{2.21}, $a$ enters
quadratically in
$F_g$ and $G_{g+1}$, the Toda hierarchy \eqref{2.12}
(respectively \eqref{2.25}, \eqref{2.26}) is invariant
under the substitution
\begin{equation}
a(t) \to a_\eps (t) =\{\eps(n) a(n,t)\}_{n\in\bbZ},
\quad \eps (n) \in \{+1, -1\}, \; n\in\bbZ.
\lb{2.28}
\end{equation}
This result should be compared with (the last part of)
 Lemma~\ref{l3.1} and Lemma~\ref{l4.1}.
\end{rem}

Finally, we specialize to the stationary Toda
hierarchy characterized by
$\dot a =\dot b =0$ in \eqref{2.13} (respectively
\eqref{2.12}), or more precisely, by commuting
difference expressions
\begin{equation}
[P_{2g+2}, L]=0
\lb{2.30}
\end{equation}
of order $2g+2$ and $2$, respectively.
Equations \eqref{2.24} and \eqref{2.25} then yield
\begin{align}
& (b+z) (G_{g+1} -G_{g+1}^-) =2(a^-)^2 F^-_g -2a^2 F_g^+,
\lb{2.31}\\
& G_{g+1}^+ + G_{g+1} =-2 (b^+ +z) F_g^+.
\lb{2.32}
\end{align}
Because of \eqref{2.30} one computes
\begin{align}
\begin{split}
\big[P_{2g+2}\big|_{\ker (L-z)} \big]^2  =
&\big[(2aF_g S^+ +G_{g+1}) \big|_{\ker (L-z)} \big]^2\\
= & \big\{ 2a F_g [G_{g+1}^+ + G_{g+1} +2 (b^+ +z)
F_g^+]S^+ \\ & {}+ G_{g+1}^2 -4a^2 F_g F_g^+
\big\}\big|_{\ker (L-z)}\\
= &\{ G_{g+1}^2 -4a^2 F_g F_g^+\} \big|_{\ker (L-z)} =:
R_{2g+2}.
\lb{2.33}
\end{split}
\end{align}
A simple calculation, using \eqref{2.31} and
\eqref{2.32} then proves that $R_{2g+2}$ is a lattice
constant and hence a polynomial of degree
$2g+2$ with respect to $z$:
\begin{align}
\begin{split}
& \quad (b+z) (R_{2g+2} -R^-_{2g+2})\\
& = (b+z) \{ (G_{g+1} + G_{g+1}^-) (G_{g+1} -G_{g+1}^-)
-4F_g [a^2 F_g^+ -(a^-)^2 F_g^-] \}\\
& \overset{\text{\eqref{2.31}}}{=}
- [G_{g+1} +G_{g+1}^- +2 (b+z)
F_g] 2 [a^2 F_g^+ -(a^-)^2 F_g^-]
\overset{\text{\eqref{2.32}}}{=}
 0.
\lb{2.34}
\end{split}
\end{align}
Thus one infers
\begin{equation}
R_{2g+2}(z) =\prod_{m=0}^{2g+1} (z-E_m), \quad
\{E_m\}_{0 \leq m \leq 2g +1} \subset \bbC
\lb{2.35}
\end{equation}
and, since $z\in\bbC$ is arbitrary, obtains the
Burchnall-Chaundy polynomial (see \cite{16}, \cite{17}
in the case of  differential expressions) relating
$P_{2g+2}$ and $L$,
\begin{equation}
P_{2g+2}^2 =R_{2g+2} (L) =\prod_{m=0}^{2g+1} (L-E_m).
\lb{2.36}
\end{equation}
The resulting hyperelliptic curve $K_g$ of (arithmetic)
genus $g$ obtained upon compactification of the curve
\begin{equation}
y^2 =R_{2g+2}(z) =\prod_{m=0}^{2g+1} (z-E_m)
\lb{2.37}
\end{equation}
will be the basic ingredient in our algebro-geometric
treatment of the Toda and Kac-van Moerbeke hierarchies in
the remainder of this exposition.

The spectral theoretic content of the polynomials
$F_g$ and $G_{g+1}$ is clearly displayed in \eqref{4.8},
\eqref{4.19},\eqref{4.20} and especially in
\eqref{4.32}--\eqref{4.36}.


\chapter{The Stationary Baker-Akhiezer Function}
\lb{s3}
\setcounter{equation}{0}
\setcounter{prop}{0}

In this chapter we provide a major part of our thorough
review of the
algebro-geometric methods to construct quasi-periodic
finite-gap
solutions of the Toda hierarchy, a subject we shall
complete in
Chapter~\ref{s6}. As explained in the Introduction,
the origins of our
approach go back to a classic representation of
positive divisors of
degree $g$ of $K_g$ due to Jacobi \cite{46a} and
its application to
the KdV case by Mumford \cite{70a}, Sect. III a).1
and subsequently
McKean \cite{60}.

Although these finite-gap integration techniques
(especially in the
special case of spatially periodic solutions) have
been discussed in
several references on the subject, see, for instance,
\cite{2}, \cite{19},
\cite{25}, \cite{26}, \cite{53}, \cite{55},
\cite{59}, \cite{63},
\cite{67}, \cite{69}, \cite{80}, we have chosen to
give a detailed
account.  This decision is based both on the
necessity of this material
for our main Chapter~\ref{s9} on algebro-geometric
solutions of the
Kac-van Moerbeke hierarchy and on the fact
that we believe to be
able to offer a simpler and more streamlined approach
than the existing
ones.

As indicated at the end of Chapter~\ref{s2}
(cf.\ \eqref{2.30},
\eqref{2.35}--\eqref{2.37}), the stationary Toda
hierarchy is intimately
connected with pairs of commuting difference
expressions $(P_{2g+2}, L)$ of
orders $2g+2$ and $2$, respectively and
hyperelliptic curves $K_g$
obtained upon compactification of the curve
\begin{equation}
y^2 =R_{2g+2} (z) =\prod_{m=0}^{2g+1} (z-E_m)
\lb{3.1}
\end{equation}
described in detail in Appendix~\ref{app-a} (whose
results and notations
we shall freely use in the remainder of this
exposition).  Since we are
interested in real-valued Toda solutions and
especially in their
expressions in terms of the Riemann theta function
associated with $K_g$,
we shall make the assumption (cf.~\eqref{a.1})
\begin{equation}
\{E_m\}_{0 \leq m \leq 2g+1} \subset \bbR,
\quad E_0 < E_1 < \cdots <
E_{2g+1}, \; g\in\bbN_0.
\lb{3.2}
\end{equation}

For a fixed but arbitrary point $n_0$ in $\bbZ$
consider
\begin{equation}
\{\hat \mu_j (n_0)\}_{1\leq j \leq g} \subset K_g,
\; \ti\pi (\hat\mu_j
(n_0))=\mu_j (n_0) \in [E_{2j-1}, E_{2j}],
\quad 1 \leq j \leq g
\lb{3.3}
\end{equation}
and
\begin{align}
F_g(z,n_0) & =\prod_{j=1}^g [z-\mu_j (n_0)],
\lb{3.4}\\
G_{g+1} (z,n_0) & =
 -\sum_{j=1}^g R^{1/2}_{2g+2} (\hat\mu_j (n_0))
\prod^g_{\substack{ k=1\\ k\neq j}}
\dfrac{[z-\mu_k (n_0)]}{[\mu_j (n_0)
-\mu_k (n_0)]} -[z+b(n_0)] F_g
(z,n_0),
\lb{3.5}
\end{align}
where
\begin{equation}
b(n_0) =\sum_{j=1}^g \mu_j (n_0)
-\frac12 \sum_{m=0}^{2g+1} E_m
\lb{3.6}
\end{equation}
and
\begin{align}
\begin{split}
& R^{1/2}_{2g+2} (\hat\mu_j (n_0))
=\sig_j (n_0) R_{2g+2} (\mu_j
(n_0))^{1/2} =-G_{g+1} (\mu_j (n_0), n_0),\\
& \hat\mu_j (n_0) = (\mu_j (n_0),
-G_{g+1} (\mu_j (n_0), n_0)), \quad 1
\leq j \leq g.
\lb{3.7}
\end{split}
\end{align}
Given \eqref{3.3}--\eqref{3.7} we define
$F_g (z,n_0+1)$ by
(cf.~\eqref{2.33})
\begin{equation}
G_{g+1} (z,n_0)^2 -4a(n_0)^2 F_g (z,n_0) F_g(z,n_0+1)
=R_{2g+2}(z),
\lb{3.8}
\end{equation}
where the constant $a(n_0)^2 \neq 0$ has been
introduced in \eqref{3.8} in order to guarantee
that $F_g(z,n_0+1)$ is a monic polynomial in $z$
(i.e., its highest coefficient is normalized to
one).  Since
\begin{equation}
a(n_0)^2 F_g(z,n_0) F_g(z,n_0+1) \geq 0 \text{ for }
z=E_{2j-1}, E_{2j},
E_{2g+1}
\lb{3.9}
\end{equation}
the left-hand side of \eqref{3.9} has at least two
zeros in $[E_{2j-1},
E_{2j}]$.  Thus $F_g(z, n_0+1)$ is of the form
\begin{equation}
F_g (z,n_0+1)
=\prod_{j=1}^g [z-\mu_j (n_0+1)],\; \mu_j
(n_0+1) \in [E_{2j-1}, E_{2j}], \quad
1 \leq j \leq g.
\lb{3.10}
\end{equation}
Equation~\eqref{3.9} with $z=E_{2g+1}$ shows
$a(n_0)^2 \geq 0$ and hence
$a(n_0)^2 > 0$ and one computes from \eqref{3.5}
and \eqref{3.8} that
\begin{equation}
a(n_0)^2 =
\frac12 \sum_{j=1}^g R^{1/2}_{2g+2} (\hat \mu_j (n_0))
\prod^g_{\substack{k=1\\ k\neq j}}
[\mu_j (n_0) -\mu_k (n_0)]^{-1} - \frac14 [b(n_0)^2
+b^{(2)} (n_0)] > 0,
\lb{3.11}
\end{equation}
where we used the notation
\begin{equation}
b^{(k)} (n_0)  = \sum_{j=1}^g \mu_j (n_0)^k
- \frac12 \sum_{m=0}^{2g+1}
E_m^k, \quad k\in \bbN,
\lb{3.12}
\end{equation}
(thus $b(n_0) = b^{(1)} (n_0)$). Introducing
$\hat\mu_j(n_0 +1)$ by
\begin{equation}
\hat\mu_j (n_0+1) =(\mu_j (n_0 +1),
G_{g+1} (\mu_j (n_0+1), n_0)), \quad
1\leq j \leq g,
\lb{3.13}
\end{equation}
we have constructed the set
$\{\hat\mu_j (n_0+1)\}_{1\leq j \leq g}$ from
the set $\{\hat\mu_j (n_0)\}_{1\leq j \leq g}$
and get in addition,
\begin{align}
\begin{split}
G_{g+1} (z,n_0) &
= \sum_{j=1}^g R^{1/2}_{2g+2} (\hat\mu_j (n_0+1))
\prod^g_{\substack{ k=1\\ k\neq j}}
\dfrac{[z-\mu_k (n_0+1)]}
{[\mu_j (n_0+1) -\mu_k (n_0 +1)]}\\
& \quad - [z+ b(n_0 +1) ] F_g (z, n_0 +1),
\lb{3.14}
\end{split}
\end{align}
with
\begin{equation}
b(n_0+1) =\sum_{j=1}^g \mu_j (n_0 +1)
-\frac12 \sum_{m=0}^{2g+1} E_m
\lb{3.15}
\end{equation}
and
\begin{equation}
G_{g+1} (z, n_0+1) =-G_{g+1}(z,n_0)
-2[z+b(n_0+1)] F_g (z,n_0+1).
\lb{3.16}
\end{equation}
Since by \eqref{3.5} and \eqref{3.14}
$F_g (z, n_0)$ and $F_g (z, n_0+1)$
enter symmetrically in the expression for
$G_{g+1}(z, n_0)$, we can reverse this process,
that is, start with
$\{\hat\mu_j (n_0+1)\}_{1 \leq j \leq g}$ and
\eqref{3.14} and determine
$\{\hat\mu_j (n_0)\}_{1 \leq j \leq g}$.
Hence we obtain

\begin{lem} \lb{l3.1}
Given $\{\hat\mu_j (n_0)\}_{1\leq j \leq g}$
satisfying \eqref{3.3} we
can determine the numbers $\{\hat \mu_j (n)\}_{1 \leq j \leq g}$
for all $n\in\bbZ$
satisfying again \eqref{3.3}, that is,
\begin{equation}
\ti\pi (\hat \mu_j (n))
=\mu_j (n) \in [E_{2j-1}, E_{2j}], \quad 1 \leq j
\leq g, \; n\in \bbZ.
\lb{3.17}
\end{equation}
Moreover, we obtain two sequences
$\{a(n)\}_{n\in\bbZ}$,
$\{b(n)\}_{n\in\bbZ}\in \ell^\infty_\bbR(\bbZ)$
defined by
\begin{align}
a(n)^2 & =
\frac12 \sum_{j=1}^g R^{1/2}_{2g+2} (\hat \mu_j (n))
\prod^g_{\substack{ k=1\\ k\neq j}} [\mu_j (n)
-\mu_k (n)]^{-1}
- \frac14 [b(n)^2 +b^{(2)}(n)] > 0,
\lb{3.18}\\
b(n) & = \sum_{j=1}^g \mu_j (n)
-\frac12 \sum_{m=0}^{2g+1} E_m,
\lb{3.19}\\
b^{(k)} (n) & = \sum_{j=1}^g \mu_j (n)^k
-\frac12 \sum_{m=0}^{2g+1} E_m^k,
\quad k\in\bbN.
\end{align}
The sign of $a(n)$ is not determined by this
procedure and can be chosen
freely (cf.\ also Remark~\ref{r2.1} and
Lemma~\ref{l4.1}).
\end{lem}

\begin{proof}
It remains to prove the boundedness of $a$ and
$b$.  But this follows
immediately from \eqref{3.17}.
\end{proof}

\begin{rem}
While the trace formula \eqref{3.19} for $b(n)$ is a
standard result, the explicit representation \eqref{3.18}
of $a(n)$ in terms of the Dirichlet data $\{\hat \mu_j
(n)\}_{1 \leq j \leq g}$ appears to be new to the best of our
knowledge.
\end{rem}

We emphasize that \eqref{3.4}--\eqref{3.6}, \eqref{3.8}
--\eqref{3.12},
and \eqref{3.14}--\eqref{3.16}
still hold if $n_0$ is
replaced by $n$. In addition, we note for later use
\begin{align} \no
a(n)^2 F_g(z,n+1) - a(n-1)^2 F_g(z,n-1)
+ (b(n)+z)^2 F_g(z,n) \\
= -(b(n)+z) G_{g+1}(z,n).
\lb{3.20pr}
\end{align}

For reasons to become obvious in the next chapter
(cf.\ the spectral
properties described in Theorem~\ref{t4.2} in
connection with the
self-adjoint $\ell^2 (\bbZ)$ realization $H$
associated with $L=aS^+
+a^-S^- -b$) we shall call
$a=\{a(n)\}_{n\in\bbZ}$,
$b=\{b(n)\}_{n\in\bbZ}$ of the type \eqref{3.18},
\eqref{3.19} finite-gap
sequences (respectively $g$-gap sequences whenever we want to
emphasize the genus $g$
of the underlying curve $K_g$).

Next, we define a meromorphic function
$\phi(P,n)$ on $K_g$,
\begin{align}
\begin{split}
\phi(P,n) & = \dfrac{-G_{g+1} (\ti\pi (P), n)
+R^{1/2}_{2g+2} (P)}{2a (n)
F_g (\ti\pi(P), n)}
=\dfrac{-2a (n) F_g (\ti \pi (P), n+1)}{ G_{g+1} (\ti
\pi (P), n) + R^{1/2}_{2g+2} (P)},\\
P & =(z, \sig R_{2g+2}(z)^{1/2} )
=(\ti{\pi}(P), R^{1/2}_{2g+2} (P))
\lb{3.20}
\end{split}
\end{align}
and with the help of $\phi (P, n)$ another
meromorphic function $\psi (P,
n, n_0)$ on $K_g$, the stationary Baker-Akhiezer
(BA) function
\begin{equation}
\psi(P, n, n_0) = \begin{cases}
\prod_{m=n_0}^{n-1} \phi(P,m), & n \geq n_0 +1\\
1, & n=n_0\\
\prod_{m=n}^{n_0-1} \phi (P,m)^{-1}, & n\leq n_0 -1
\end{cases}.
\lb{3.21}
\end{equation}
As will turn out in the course of this chapter and
in Chapters~\ref{s5},
\ref{s6}, and \ref{s9}, these meromorphic functions
are the fundamental
ingredients of the finite-gap integration technique
and $\phi(., n)$, specifically,  is the central object in
our presentation of this material.

\begin{lem}\lb{l3.2}
The function $\phi (P, n)$ satisfies
the ``Riccati-type'' equation
\begin{equation}
a(n) \phi(P,n) +a(n-1) \phi(P,n-1)^{-1} =b(n) +
\ti\pi(P), \quad n
\in\bbZ
\lb{3.22}
\end{equation}
and the BA-function $\psi(P,n,n_0)$ satisfies the
Jacobi equation
\begin{multline}
a(n) \psi (P, n+1, n_0) +a(n-1) \psi (P, n-1, n_0)
=[b(n) + \ti \pi (P)]
\psi (P, n, n_0),\\ \qquad n, n_0 \in\bbZ.
\lb{3.23}
\end{multline}
\end{lem}

\begin{proof}
\eqref{3.16} implies
\begin{multline}
a(n) \phi(P,n) +a(n-1) \phi(P, n-1)^{-1}
=\tfrac12 [R^{1/2}_{2g+2} (P) -G_{g+1} (\ti \pi (P), n)]
F_g (\ti \pi (P), n)^{-1}\\
- \tfrac12 [R^{1/2}_{2g+2} (P) +G_{g+1} (\ti \pi (P), n-1)]
F_g (\ti \pi (P),n)^{-1} =b(n) +\ti \pi (P)
\lb{3.24}
\end{multline}
and \eqref{3.23} follows from \eqref{3.24} and
\begin{equation}
\phi(P,n) =
\psi (P, n+1, n_0) / \psi (P, n, n_0).
\lb{3.25}
\end{equation}
\end{proof}

We collect a few more useful relations which
follow from \eqref{3.8} and
\eqref{3.20}, \eqref{3.21},
\begin{align}
\phi(P,n) \phi(P^*, n) &
= F_g (\ti \pi (P), n+1)/ F_g (\ti \pi (P), n),
\lb{3.26}\\
\psi(P, n, n_0) \psi(P^*, n, n_0) &
= F_g (\ti \pi (P), n) / F_g (\ti \pi
(P), n_0), \lb{3.27}\\
\phi (P,n) -\phi(P^*, n)
&= R^{1/2}_{2g+2}(P) / [a(n) F_g (\ti \pi (P),
n)],
\lb{3.28}\\
\phi(P,n) + \phi(P^*, n) &
= -G_{g+1} (\ti \pi (P), n) / [a(n) F_g (\ti
\pi (P), n)].
\lb{3.29}
\end{align}

It will be convenient later on to denote by
$\phi_\pm (z,n)$, $\psi_\pm
(z,n,n_0)$ the chart expressions (branches)
of $\phi(P,n)$,
$\psi(P,n,n_0)$ in the charts
$(\Pi_\pm, \ti\pi)$ (see \eqref{a.13}).

In order to analyze $\phi$ and the BA-function
$\psi$ further, it is
convenient to express them in terms of the
Riemann theta function
associated with $K_g$. First we note that by
\eqref{3.20} and
\eqref{3.21}, the divisors $(\phi)$ of $\phi$
and $(\psi)$ of $\psi$ are
given by
\begin{align}
(\phi(., n)) & =
\calD_{\humu (n+1)} -\calD_{\humu (n)}
+\calD_{\infty_+} -\calD_{\infty_-}
\lb{3.30}\\
\intertext{and}
(\psi(.,n,n_0)) & = \calD_{\humu (n)} -
\calD_{\humu (n_0)} +(n-n_0)
(\calD_{\infty_+}
-\calD_{\infty_-})
\lb{3.31}
\end{align}
(cf.\ our notation established in
Appendix~\ref{app-a}).  By Abel's
theorem (cf.~\eqref{a.45}), \eqref{3.31} yields
\begin{align}
\begin{split}
\ual_{P_0} (\calD_{\hat\umu(n)}) &
= \ual_{P_0} (\calD_{\hat\umu (n_0)})
-(n-n_0) \underline{A}_{\infty_-} (\infty_+)\\
&= \ual_{P_0} (\calD_{\hat\umu (n_0)})
-2(n-n_0) \underline{A}_{P_0}
(\infty_+),
\lb{3.32}
\end{split}
\end{align}
where, for convenience only, from this point on
we agree to fix the base point $P_0$ as the
branch point $(E_0, 0)$,
\begin{equation}
P_0 = (E_0, 0).
\lb{3.33}
\end{equation}
Next we introduce the abbreviations,
\begin{align}
\uz (P,n) & = \hua_{P_0} (P)
-\hual_{P_0} (\calD_{\humu
(n)}) -  \huxi_{P_0} \in \bbC^g,
\lb{3.34}\\
\uz (n) & =\uz (\infty_+, n),
\lb{3.35}
\end{align}
where
\begin{equation}
\huxi_{P_0}
=(\hat\Xi_{P_0, 1}, \ldots, \hat\Xi_{P_0, g} ), \quad
\hat\Xi_{P_0, j}
= \left[ \dfrac{1+\tau_{j,j}}2
-\sum^g_{\substack{ k=1\\
k\neq j }} \int_{a_k} \hat A_{P_0,j} \ome_k\right]
\lb{3.36}
\end{equation}
denotes a representative of the vector of Riemann
constants $\ul{\Xi}_{P_0} =
\huxi_{P_0} \mod (L_g)$.  Since for any
$Q_0
\in K_g$,
\begin{equation}
\underline{A}_{Q_0} (.)  = \underline{A}_{P_0} (.)
-\underline{A}_{P_0}(Q_0), \quad
\underline\Xi_{Q_0}  = \underline\Xi_{P_0}
+ (g-1) \underline{A}_{P_0} (Q_0),
\lb{3.37}
\end{equation}
$\uz (P,n)$ is independent of the chosen base point
$P_0$. For later
purposes we recall that
\begin{equation}
2\uxi_{P_0} = \underline{0} \mod (L_g)
\lb{3.38}
\end{equation}
and, due to \eqref{3.32}, that
\begin{equation}
\uz(\infty_-, n) =\uz (\infty_+, n-1) \mod (\bbZ^g).
\lb{3.39}
\end{equation}
Next, consider the normal differential of the third
kind
$\ome^{(3)}_{\infty_+, \infty_-}$ which has simple poles
at $\infty_+$ and
$\infty_-$, corresponding residues $+1$ and $-1$,
vanishing $a$-periods,
and is holomorphic otherwise on $K_g$.  Hence we have
(cf.\ \eqref{a.39})
\begin{align}
& \ome^{(3)}_{\infty_+, \infty_-} =
\dfrac{\prod_{j=1}^g (\ti\pi -\lam_j)
d\ti\pi}{R_{2g+2}^{1/2}},
\quad \ome_{\infty_-, \infty_+}^{(3)}
=-\ome^{(3)}_{\infty_+, \infty_-},
\lb{3.40}\\
& \int_{a_j} \ome_{\infty_+, \infty_-}^{(3)} = 0,
\quad 1\leq j\leq g,
\lb{3.41}\\
& U_j^{(3)} = \frac{1}{2\pi i}
\int_{b_j} \ome_{\infty_+, \infty_-}^{(3)} =\hat
A_{\infty_-, j} (\infty_+) = 2 \hat A_{P_0, j}
(\infty_+), \quad 1 \leq j \leq g,
\lb{3.42}
\end{align}
where the numbers $\{ \lam_j \}_{1\leq j \leq g}$
are determined by the
normalization \eqref{3.41}.

Recalling that $\calD_{\humu (n)}$ are nonspecial
by \eqref{3.17} and
Lemma~\ref{la.2}, that is,
\begin{equation}
i(\calD_{\humu (n)} ) =0, \quad n\in\bbZ,
\lb{3.43}
\end{equation}
and that by (a special case of) Riemann's vanishing
theorem
\begin{equation}
\theta(\uz (P,n))=0 \text{ if and only if }
 P\in \{\hat \mu_j (n)\}_{1
\leq j \leq g},
\lb{3.44}
\end{equation}
the zeros and poles of $\phi$ and $\psi$ as
recorded in \eqref{3.30} and
\eqref{3.31} suggest consideration of the expressions
\begin{align}
& \dfrac{\theta (\uz (P, n+1))}
{\theta (\uz, (P,n))} \exp \left[
\int_{P_0}^P
\ome_{\infty_+, \infty_-}^{(3)}\right]
\lb{3.45}\\
\intertext{and}
& \dfrac{\theta(\uz (P,n))}
{\theta(\uz (P,n_0))} \exp \left[ (n-n_0)
\int_{P_0}^P \ome_{\infty_+, \infty_-}^{(3)}\right].
\lb{3.46}
\end{align}
Here we agree to use the same path of integration
from $P_0$ to $P$ on
$K_g$ in the Abel map $\underline{\hat{A}}_{P_0}(P)$
in $\uz(P,n)$ and in
the integral over $\ome_{\infty_+, \infty_-}^{(3)}$ in
the exponents of
\eqref{3.45} and \eqref{3.46}.  With this convention,
both expressions
\eqref{3.45}, \eqref{3.46} are well-defined on $K_g$
(due to
\eqref{3.41}, \eqref{3.42}, and \eqref{a.28}) and we
infer
\begin{align}
\phi(P,n) & = C(n) \dfrac{\theta (\uz (P,n+1))}
{\theta(\uz (P,n))} \exp
\left[ \int_{P_0}^P \ome_{\infty_+, \infty_-}^{(3)}\right],
\lb{3.47}\\
\psi (P, n, n_0) & = C(n, n_0) \dfrac{\theta(\uz (P,n))}
{\theta(\uz
(P,n_0))} \exp
\left[ (n-n_0) \int_{P_0}^P \ome_{\infty_+, \infty_-}^{(3)}
\right]
\lb{3.48}
\end{align}
since $\psi$ (and hence $\phi$) is determined up
to a constant by its
zeros and poles.  It remains to determine the
constants $C(n)$,
$C(n,n_0)$.  Since the original literature appears
to be vague at this
point we shall dwell on this a bit.  As a consequence
of \eqref{3.27} one
infers
\begin{equation}
\psi(\infty_+,n,n_0) \psi(\infty_-,n,n_0)=1
\lb{3.49}
\end{equation}
and hence \eqref{3.35} and \eqref{3.47} yield
\begin{equation}
C(n,n_0)^2 = \frac{\theta(\uz (n_0))\theta (\uz (n_0-1)) }
{ \theta(\uz (n))
\theta (\uz (n-1))}.
\lb{3.50}
\end{equation}
Because of
\begin{equation}
\phi(P,n) =\psi (P, n+1,n),
\lb{3.51}
\end{equation}
we get
\begin{equation}
C(n) = \left[ \frac{\theta(\uz (n-1)) }
{ \theta (\uz (n+1))} \right]^{1/2},
\lb{3.52}
\end{equation}
where the determination of the square root will be
given later (see
\eqref{3.69}).

Next we collect a few useful results.

\begin{lem}\lb{l3.3}
\newcounter{mycount}
(i).~$\ome_{\infty_+, \infty_-}^{(3)}$ satisfies
\begin{equation}
\Re \left( \int_{P_0}^P \ome_{\infty_+,
\infty_-}^{(3)} \right) \begin{cases}
=0, & \ti\pi (P) \in \bigcup_{j=0}^g [E_{2j}, E_{2j+1}]\\
<0, & \ti\pi(P) \in\bbR \bs \bigcup_{j=0}^g [E_{2j}
E_{2j+1}]\end{cases}, \quad P\in\Pi_+
\lb{3.53}
\end{equation}
(the sign in \eqref{3.53} being reversed for
$P\in\Pi_-$) and
\begin{equation}
\lam_j \in [E_{2j-1}, E_{2j}], \quad 1\leq j \leq g.
\lb{3.54}
\end{equation}
(ii).~Let $\calD \in \sig^g K_g$.  Then
\begin{align}
\begin{split}
& i \Im [\ul{\hat{\alpha}}_{P_0} (\calD) +\huxi_{P_0} ]
= \uzero \mod (L_g)\\
& \text{if and only if } \calD
=\calD_{\humu} \left(=\sum_{j=1}^g
\calD_{\hat\mu_j}\right), \; \humu
= (\hat\mu_1, \ldots, \hat \mu_g),\\
& \hspace*{3cm} \ti\pi (\hat \mu_j) \in [E_{2j-1}, E_{2j}],
\quad 1 \leq j \leq g.
\lb{3.55}
\end{split}
\end{align}
(iii).~Suppose $\hat\mu_j \in K_g$, $\ti\pi (\hat \mu_j)
=\mu_j \in[E_{2j-1},
E_{2j}]$, $1\leq j \leq g$.  Then
\begin{equation}
i \Im \left( \int_{P_0}^{\hat\mu} \ome_{\infty_+,
\infty_-}^{(3)} \right) =
-2\pi i \underline{\hat{A}}_{P_0} (\infty_+),
\quad 1 \leq j \leq g.
\lb{3.56}
\end{equation}
(iv).~One has
\begin{equation}
\theta(\uz (P,n)) \theta(\uz (P,n_0)) > 0
\text{ for } \ti\pi (P) < E_0
\text{ or } \ti\pi (P) > E_{2g+1},
\quad n, n_0 \in \bbZ,
\lb{3.57}
\end{equation}
in particular, $C(n)$ in \eqref{3.52} is
real-valued.
\end{lem}

\begin{proof}
(i). The normalization \eqref{3.41}, that is,
\begin{equation}
\int_{a_j} \ome_{\infty_+, \infty_-}^{(3)}
=-2 \int_{E_{2j-1}}^{E_{2j}}
\dfrac{\prod_{k=1}^g (z-\lam_k) \, dz}
{R_{2g+2} (z)^{1/2} } =0, \quad 1
\leq j \leq g,
\lb{3.58}
\end{equation}
immediately yields \eqref{3.54}.  In order to prove
\eqref{3.53} we
assume $P\in \Pi_+$ and choose as integration path
the lift of the
straight line from $E_0 +i\eps$ to $\ti\pi(P) +i\eps$
and then take
$\eps \downarrow 0$.  (Since we are required to stay
in the interior of
$\hat K_g$, whenever we are to intersect some
$b_j$-cycle we first go
around $a_j$ and then back on the other side of
$b_j$.  Since the parts
on $b_j$ cancel and due to our normalization
\eqref{3.58} the
$a_j$-periods of $\ome_{\infty_+, \infty_-}^{(3)}$ are
zero, this does not
alter the value of the integral in question.)  The
rest follows from
\eqref{3.54} and \eqref{a.5}.\\
(ii). First assume $\calD =\calD_{\humu}$ with
$\ti\pi (\hat\mu_j) \in
[E_{2j-1}, E_{2j}]$, $1 \leq j \leq g$.  Then,
using \eqref{a.5}, one can
show that
\begin{equation}
i\Im [\hat{A}_{P_0,j} (\hat \mu_k)]
=\dfrac12 \tau_{j,k} =i\Im
\left[ \int_{a_k} \hat{A}_{P_0, j} \ome_k\right]
\lb{3.59}
\end{equation}
by taking all integrals as lifts obtained from
limits of straight line
sequent of the type $\ti\pi (P_1) +i\eps$ to
$\ti\pi(P_2) +i\eps$ as
$\eps \downarrow 0$ for various points $P_1$, $P_2$.
Again, in order to
stay on $\hat K_g$, crossings of $b_j$-cycles can
be avoided by adding
contributions along $a_j$-cycles (which are
real-valued) as in the proof
of part~(i).  Equation \eqref{3.59} yields
\begin{equation}
i\Im [\hual_{P_0} (\calD_{\humu}) +
\ul{\hat{\Xi}}_{P_0}]= \uzero \mod (L_g).
\lb{3.60}
\end{equation}
Next consider $\ual_{P_0}$ as a holomorphic map
from $\sig^g K_g$ to
$J(K_g)$ and restrict $\ual_{P_0}$ to divisors
$\calD$ satisfying
\begin{equation}
\calD =\calD_{\humu}, \; \ti\pi (\humu_j )
=\mu_j \in [E_{2j-1}, E_{2j}],
\quad 1 \leq j \leq g.
\lb{3.61}
\end{equation}
Denote this restriction by $\tual_{P_0}$.  The
set of divisors $\calD$ in
\eqref{3.61} is a connected submanifold of
$\sig^gK_g$ since it is
isomorphic to $\times^g_{j=1} S^1$.  Moreover,
by the arguments leading to
\eqref{3.60}, the image of $\tual_{P_0}$ is a
subset of
$J=\{\underline{x} \in J(K_g) \big| i\Im (\underline{x}
+ \ul{\hat\Xi}_{P_0})
=\underline{0}\mod (L_g)\}$.  Since
\begin{equation}
\rank (d\ual_{P_0} (\calD_{\humu}))
=g-i(\calD_{\humu}) =g
\lb{3.62}
\end{equation}
as $\calD_{\humu}$ is nonspecial by \eqref{3.61}
and Lemma~\ref{la.2},
$\tual_{P_0}$ is invertible and hence its image is
all of $J$.  Thus
$\ual_{P_0}$ provides an isomorphism from
$\{ \calD_{\humu} \in \sig^g
K_g \big| \ti\pi (\hmu_j)
=\mu_j \in [E_{2j-1}, E_{2j}],\, 1 \leq j \leq
g\}$ onto $J$.\\
(iii). This is clear from \eqref{3.42}.\\
(iv). Since
\begin{equation}
i \Im [\uz (P,n)] = i \Im [\ul{\hat{A}}_{P_0} (P)]
=\underline{0} \mod (L_g),
\quad \ti\pi(P) < E_0, \; \ti\pi (P) > E_{2g+1},
\lb{3.63}
\end{equation}
a combination of \eqref{a.28}, \eqref{3.35},
\eqref{3.55}, and
\eqref{3.63} yields
\begin{equation}
\Im [\theta(\uz (P,n))\theta(\uz(P,n_0))]=0
 \text{ for } \ti\pi (P) < E_0,
\; \ti\pi (P) > E_{2g+1}, \; n, n_0 \in\bbZ.
\lb{3.64}
\end{equation}
But then \eqref{3.57} immediately follows by
considering $n=n_0$ and the
fact that all zeros of $\theta (\uz (P,n))$ occur
precisely at $P=\hat
\mu_j (n)$, $1\leq j \leq g$ for all $n\in\bbZ$.
\end{proof}

Lemma~\ref{l3.3} is the main ingredient for the
following
characterization of $\phi(P,n)$ and $\psi(P,n,n_0)$.

\begin{thm}\lb{t3.4}
(i).
\begin{align}
\phi(P,n) & = C(n) \dfrac{\theta(\uz (P,n+1))}
{\theta(\uz(P,n))} \exp
\left[ \int_{P_0}^P \ome_{\infty_+,
\infty_-}^{(3)} \right],
\lb{3.65}\\
\psi(P,n,n_0) & = C(n,n_0) \dfrac{\theta(\uz (P,n))}
{\theta (\uz
(P,n_0))} \exp
\left[ (n-n_0) \int_{P_0}^P \ome_{\infty_+,
\infty_-}^{(3)}\right],
\lb{3.66}
\end{align}
where $C(n)$, $C(n,n_0)$ are real-valued and
\begin{align}
C(n) & =C(n+1, n)
=[\theta(\uz(n-1))/\theta (\uz (n+1))]^{1/2},
\lb{3.67}\\
C(n, n_0) & = \begin{cases}
\prod_{m=n_0}^{n-1} C(m), & n\geq n_0 +1\\
1, & n=n_0 \\
\prod_{m=n}^{n_0-1} C(m)^{-1}, & n\leq n_0 -1
\end{cases}
\quad = \left[ \dfrac{\theta(\uz (n_0))
\theta(\uz (n_0 -1))}
{\theta (\uz (n)) \theta(\uz (n-1))} \right]^{1/2}.
\lb{3.68}
\end{align}
In addition, the sign of $C(n)$ is opposite
that of $a(n)$, that is,
\begin{equation}
\sgn [C(n)] =-\sgn [a(n)], \quad n\in\bbZ.
\lb{3.69}
\end{equation}
(ii). The function $\phi (P, n)$ and hence
$\psi (P, n,n_0)$ is real-valued for
all $P$ such that
$\pi (P) \in\bbR \bs \bigcup_{j=0}^g[E_{2j},
E_{2j+1}]$ and
all $n$, $n_0\in\bbZ$.\\
(iii). Let $\lam \in \bigcup_{j=0}^g [E_{2j},
E_{2j+1}]$.  Then
\begin{equation}
\psi_\pm (\lam, ., n_0) \in \ell^\infty(\bbZ).
\lb{3.70}
\end{equation}
(iv). Let $\lam \in\bbR \bs \bigcup_{j=0}^g [E_{2j},
E_{2j+1}]$. Then
there exist constants $M$, $K(\lam)>0$ such that
\begin{equation}
| \psi_\pm (\lam, n, n_0) | \leq M
e^{\mp (n-n_0) K (\lam)}.
\lb{3.71}
\end{equation}
Moreover,
\begin{equation}
\psi_\pm (\lam, . , n_0) \in \ell^2 ((N_0, \pm \infty)),
\quad N_0 \in\bbZ.
\lb{3.72}
\end{equation}
\end{thm}

\begin{proof}
(i). Equations \eqref{3.65}--\eqref{3.68} follow
from equations \eqref{3.21}, \eqref{3.47}--\eqref{3.52},
and Lemma~\ref{l3.3}~(iv)  except for the sign correlation
\eqref{3.69} of $C(n)$ and $a(n)$.  The  latter can be
inferred as follows.  Expanding $\phi(P,n)$ in \eqref{3.65}
near $P=\infty_+$ on $\Pi_+$ yields in the chart
$(\Pi_+ \bs \{ (0,R_{2g+2}(0)^{1/2}) \},
z=\zeta^{-1})$, integrating from
$P_0$ to $P$ along the lift of the corresponding
straight line segment
along the negative real axis,
\begin{align}
\begin{split}
\phi (P,n) &
\underset{\zeta\to 0}{=}
 C(n) \left[ \dfrac{\theta(\uz (n+1))}{\theta(\uz (n))}
+O(\zeta)\right]
\exp \{ \ln [\ti a \zeta +O(\zeta^2)]\}\\
& \underset{\zeta\to 0}{=}
C(n) \dfrac{\theta(\uz (n+1))}
{\theta(\uz (n))} \ti a \zeta +O(\zeta^2),
\lb{3.73}
\end{split}
\end{align}
where $\ti a < 0$ is an appropriate integration constant.
Inserting this
expansion into \eqref{3.22} (with $n$ replaced
by $n+1$) yields
\begin{equation}
a(n) \phi(P,n)^{-1} -\ti \pi (P)
\underset{\zeta\to 0}{=} O(1)
\underset{\zeta\to 0}{=}
\dfrac{a(n)}{C(n) \ti a} \dfrac{\theta (\uz(n))}
{\theta (\uz (n+1))}
\zeta^{-1} -\zeta^{-1} +O(1),
\lb{3.74}
\end{equation}
that is,
\begin{equation}
a(n) =\ti a C(n) \theta (\uz (n+1))\theta (\uz (n))^{-1},
\quad n\in\bbZ.
\lb{3.75}
\end{equation}
Since $[\theta (\uz (n+1))/ \theta (\uz (n))]>0$
by \eqref{3.57} we
obtain \eqref{3.69}.\\
(ii). This follows directly from \eqref{3.20} and
\eqref{3.21}, or
alternatively, by combining (i) and Lemma~\ref{l3.3}~(i),
(iii), (iv).\\
(iii), (iv). Relations \eqref{3.70}--\eqref{3.72} are an
immediate consequence
of the quasi-period\-icity and hence boundedness of
$\theta (\uz (P,n))$
with respect to $n\in\bbZ$ and Lemma~\ref{l3.3}~(i).
\end{proof}

Relation~\eqref{3.70} extends to all
$z\in\bbC\bs \bigcup_{j=0}^g
[E_{2j}, E_{2j+1}]$.  Since this is conveniently
proved by invoking Weyl
$m$-functions we postpone this fact to the next
chapter.


\chapter{Spectral Theory for Finite-Gap Jacobi
Operators} \lb{s4}
\setcounter{prop}{0}
\setcounter{equation}{0}

In this chapter we shortly digress into spectral
properties and Green's
functions of self-adjoint $\ell^2 (\bbZ)$ realizations
associated with
finite-gap difference expressions.

We start with a general difference expression $L$ of the
type
\begin{equation}
L=aS^+ +a^- S^- -b
\lb{4.1}
\end{equation}
assuming $a$, $b\in\ell^\infty_\bbR (\bbZ)$,
$a(n) \neq 0$, $n\in\bbZ$
implying that $L$ is in the limit point case near
$\pm\infty$.  Then the
following Jacobi operator $H$ in $\ell^2 (\bbZ)$
defined by
\begin{equation}
Hf =Lf, \quad f\in \calD (H) = \ell^2 (\bbZ)
\lb{4.2}
\end{equation}
is the unique self-adjoint realization associated
with $L$ in $\ell^2
(\bbZ)$.  The corresponding Green's function
$G(z, m, n)$ of $H$, defined
by
\begin{equation}
((H-z)^{-1} f)(m) =\sum_{n\in\bbZ} G(z,m,n) f(n),
\quad f\in\ell^2
(\bbZ), \; z\in\bbC \bs \sig (H),
\lb{4.3}
\end{equation}
with $\sig (H)$ denoting the spectrum of $H$, can
be expressed as
\begin{equation}
G(z,m,n) =W(f_- (z), f_+ (z))^{-1} \begin{cases}
f_- (z,m) f_+ (z,n), & m\leq n\\
f_+ (z,m) f_- (z,n), & m \geq n
\end{cases},\quad
z\in\bbC \bs \sig (H).
\lb{4.4}
\end{equation}
Here
\begin{equation}
f_\pm (z,.) \in\ell^2 ((N_0, \pm\infty)),
\quad z\in\bbC \bs \sig (H), \;
N_0 \in\bbZ
\lb{4.5}
\end{equation}
are weak solutions of
\begin{equation}
L\psi (z,n) =z\psi (z,n)
\lb{4.6}
\end{equation}
and
\begin{equation}
W(f,g)(n) =a(n) [f(n) g(n+1) -f(n+1) g(n)]
\lb{4.7}
\end{equation}
denotes the Wronskian of $f$ and $g$.

By general principles, $G(z, n_0, n_0)$,
$G(z, n_0+1, n_0+1)$, and $G(z, n_0, n_0+1)$ uniquely
determine both sequences $\{a(n)^2\}_{n\in\bbZ}$
and $\{b(n)\}_{n\in\bbZ}$.  These results are
standard and a consequence
of the $2\times 2$ Weyl $M$-matrix associated
with $H$,
\begin{equation}
M_{n_0} (z) =\begin{pmatrix}
G(z,n_0, n_0) & G(z, n_0, n_0+1)\\
G(z,n_0, n_0+1) & G(z, n_0+1, n_0+1)
\end{pmatrix}.
\lb{4.8}
\end{equation}
We recall the asymptotic behavior,
\begin{align}
\begin{split}
G(z,n,n) & \underset{z\to\infty}{=} -z^{-1} +O(z^{-2}),\\
G(z,n,n+1) & \underset{z\to\infty}{=}
-a (n) z^{-2} +O(z^{-3}), \quad n \in\bbZ.
\lb{4.9}
\end{split}
\end{align}
Moreover, the identity
\begin{equation}
[2a(n) G(z,n,n+1)-1]^2 =1+4a (n)^2 G(z,n,n)
G(z,n+1, n+1), \quad n \in\bbZ
\lb{4.10}
\end{equation}
proves that $G(z,n_0, n_0+1)$ is determined up
to a sign by $a(n_0)$,
$G(z,n_0, n_0)$, and $G(z, n_0+1, n_0+1)$.

Next, we define restrictions $H_{\pm, n_0}$ of
$H$ to
$\ell^2([n_0\pm1,\pm\infty))$ with a
Dirichlet boundary condition at $n_0$,
\begin{equation}
H_{\pm, n_0} f=Lf, \quad f\in\calD (H_{\pm, n_0})
=\{ f\in\ell^2 ([ n_0\pm1,
\pm\infty)) | f(n_0) =0\}.
\lb{4.11}
\end{equation}
Denoting by $G_{\pm, n_0}$ the corresponding
Green's functions of
$H_{\pm, n_0}$, the associated Weyl
$m$-functions on $[n_0, \pm\infty)$ then
read
\begin{align}
\begin{split}
m_{\pm, n_0} (z) & =
G_{\pm, n_0} (z,n_0 \pm 1, n_0 \pm 1) =(\del_{n_0 \pm
1}, (H_{\pm, n_0} -z)^{-1} \del_{n_0\pm1})\\
& = \begin{cases}
-a(n_0)^{-1} [f_+ (z,n_0+1)/ f_+ (z,n_0)]\\
-a(n_0-1)^{-1} [f_- (z,n_0-1) / f_- (z,n_0)]
\end{cases},
\lb{4.12}
\end{split}
\end{align}
with $f_\pm (n,z)$ as in \eqref{4.5} and $\del_m (n)
= \del_{m,n} = \begin{cases}
1, & m=n\\
0, & m\neq n
\end{cases}$.

The following is a simple but useful result
concerning the spectral
invariants of $H$.

\begin{lem} (See, e.g., \cite{28},~p.~141.)\lb{l4.1}
Suppose $a$, $b\in\ell^\infty_\bbR(\bbZ)$, $a(n) \neq 0$,
$n\in\bbZ$ and
introduce $a_\eps \in\ell^\infty_\bbR(\bbZ)$ by
\begin{equation}
a_\eps =\{\eps(n)a(n)\}_{n\in\bbZ},
\; \eps(n) \in\{+1, -1\},\quad
n\in\bbZ.
\lb{4.13}
\end{equation}
Define $H_\eps$ in $\ell^2 (\bbZ)$ as in
\eqref{4.2} with $L$ replaced by
$L_\eps =a_\eps S^+ +a_\eps^- S^- -b$.  Then
$H$ and $H_\eps$ are
unitarily equivalent, that is, there exists a
unitary operator $U_\eps$ in
$\ell^2 (\bbZ)$ such that
\begin{equation}
H=U_\eps H_\eps U_\eps^{-1}.
\lb{4.14}
\end{equation}
\end{lem}

\begin{proof}
$U_\eps$ is explicitly represented by the
diagonal matrix
\begin{equation}
U_\eps =(\ti \eps (n) \del_{m,n})_{m,n\in\bbZ},
\quad \ti\eps (n)
\in\{+1,
-1\}, \; \ti\eps (n) \ti\eps (n+1) =\eps (n),
\; n\in\bbZ.
\lb{4.15}
\end{equation}
\end{proof}

Next, let $c(z,n,n_0)$, $s(z,n,n_0)$, $z\in\bbC$ be
a fundamental system
of solutions of \eqref{4.6} satisfying
\begin{equation}
c(z,n_0,n_0) =s(z,n_0+1, n_0)=1,
\; c(z,n_0+1,n_0) =s(z,n_0,n_0)=0.
\lb{4.16}
\end{equation}
Returning to our special $g$-gap sequences
$a(n)$, $b(n)$ in
\eqref{3.18}, \eqref{3.19}, the branches
$\psi_\pm (z,n,n_0)$ of $\psi
(P,n,n_0)$ in \eqref{3.21} satisfy
\begin{equation}
\psi_\pm (z,n,n_0) = c(z,n,n_0)
+ \phi_\pm (z,n_0) s(z,n,n_0)
\lb{4.17}
\end{equation}
and
\begin{equation}
W(\psi_- (z,.,n_0), \psi_+ (z,.,n_0))
=a(n_0) [\phi_+ (z,n_0) -\phi_-(z,
n_0)],
\lb{4.18}
\end{equation}
with $\phi_\pm (z,n)$ the corresponding branches
of $\phi(P,n)$ in
\eqref{3.20}.  Taking into account
\eqref{3.26}--\eqref{3.29} then
enables one to further identify
\begin{align}
G(z,n,n) & =F_g (z,n) / R_{2g+2} (z)^{1/2},
\lb{4.19}\\
G(z,n,n+1) & = \{1-[G_{g+1} (z,n)
/ R_{2g+2} (z)^{1/2}]\} / [2a(n)],
\lb{4.20}
\end{align}
\begin{align}
\begin{split}
m_{+,n} (z) & = G_{+, n} (z, n+1, n+1)
=-\phi_+ (z,n) / a(n)\\
& = [G_{g+1} (z,n) -R_{2g+2} (z)^{1/2}]
/ [2a (n)^2 F_g (z,n)],
\lb{4.21}
\end{split}\\
\begin{split}
m_{-,n} (z) & = G_{-, n} (z, n-1, n-1)
=-1 / [a(n-1) \phi_- (z,n-1)]\\
& = [G_{g+1} (z,n-1) +R_{2g+2} (z)^{1/2}]
/ [2a (n-1)^2 F_g (z,n)].
\lb{4.22}
\end{split}
\end{align}
We note that $a(n_0)$ and $G_{g+1} (z,n_0)$
determine the sign of
$G(z,n_0, n_0+1)$ left open in \eqref{4.10}.

We conclude this chapter with a summary of
spectral properties of $H$,
$H_{\pm, n}$ in connection with the $g$-gap
sequences \eqref{3.18} and
\eqref{3.19}.  Let $\sig(.)$, $\sig_{ac}(.)$,
$\sig_{sc}(.)$, and $\sig_p
(.)$ denote the spectrum, absolutely continuous,
singularly continuous,
and point spectrum (set of eigenvalues),
respectively.  Then we have

\begin{thm} \lb{t4.2}
Suppose $a$, $b\in\ell_\bbR^\infty (\bbZ)$ are
$g$-gap sequences satisfying
\eqref{3.18}, \eqref{3.19} and let $H$,
$H_{\pm, n}$ be as in \eqref{4.2},
\eqref{4.11}.  Then
\begin{align}
\sig(H) & = \sig_{ac} (H)
=\bigcup_{j=0}^g [E_{2j}, E_{2j+1}],
\lb{4.23}\\
\sig_{sc}(H) & = \sig_p (H) =\emptyset
\lb{4.24}
\end{align}
and for all $n\in\bbZ$,
\begin{align}
\sig (H_{-, n} \oplus H_{+, n}) &
= \sig (H) \cup \{ \mu_j (n)\}_{1\leq j \leq
g},
\lb{4.25}\\
\sig_{ac} (H_{\pm, n} ) & = \sig (H),
\; \sig_{sc} (H_{\pm, n}) =\emptyset,
\lb{4.26}\\
\sig_p (H_{-, n} \oplus H_{+, n}) &
=\{ \mu_j (n)\}_{1\leq j \leq g} \cap \{
\bigcup_{k=1}^g (E_{2k-1}, E_{2k})\}.
\lb{4.27}
\end{align}
In addition, $\sig (H)$ has uniform spectral
multiplicity two whereas
$\sig(H_{\pm, n})$, $n\in\bbZ$ is simple.
\end{thm}

\begin{proof}
Consider the trace of the Weyl $M$-matrix
\eqref{4.8}, that is,
\begin{equation}
T(z) = G(z,n_0, n_0) +G(z, n_0 +1, n_0 +1).
\lb{4.28}
\end{equation}
Then $\bbC \bs \sig (H)$ coincides with the
holomorphy domain of $T$.
This identifies $\sig (H)$ in \eqref{4.23}.
 For any $\lam_0 \in\sig_p
(H)$ one must necessarily have
\begin{equation}
\lim\limits_{\eps \downarrow 0} [i\eps T(\lam_0
+ i\eps)] < 0.
\lb{4.29}
\end{equation}
The explicit structure of $G(z,n,n)$ in
\eqref{4.19} then proves the
impossibility of \eqref{4.29} and hence
$\sig_p (H) = \emptyset$.  In
order to
prove $\sig_{sc} (H) = \emptyset$ we apply
Theorem~XIII.20 of \cite{75} with
$D=\ell_0 (\bbZ)$ (the subspace of sequences
with at most
finitely many elements being nonzero), $p=2$
and $(a,b)=(E_{2j} +\eps,
E_{2j+1} -\eps)$, $\eps >0$.  Upon letting
$\eps \downarrow 0$ one infers
the spectrum to be purely absolutely continuous
on $[E_{2j}, E_{2j+1}]$.
This proves \eqref{4.23} and \eqref{4.24}.  Next,
define in $\ell^2
((-\infty,n_0-1]) \oplus \ell^2 ([n_0+1,\infty))$
\begin{equation}
H_n^D =H_{-, n} \oplus H_{+, n}.
\lb{4.30}
\end{equation}
Then the Green's function $G_n^D (z,m,m')$ of
$H_n^D$ is given by
\begin{multline}
G_n^D (z,m,m') =G(z,m,m') -G(z,n,n)^{-1} G(z,m,n)
G(z,n,m'),\\
z\in\bbC \bs \{\mu_j (n)\}_{1\leq j \leq g},
\; m,m' \in\bbZ \bs \{n_0\}.
\lb{4.31}
\end{multline}
This proves \eqref{4.25} and \eqref{4.27}
(cf.~\eqref{4.19}).  The
relation \eqref{4.26} can again be proved by
alluding to Theorem~XIII.20
of \cite{75}.  Finally, self-adjoint half-line
operators (like $H_{\pm, n}$)
always have simple spectra, and uniform spectral
multiplicity two of $H$
follows from the existence of the two linearly
independent branches
$\psi_\pm (\lam, n,n_0)$ for
$\lam\in \bigcup_{j=0}^g (E_{2j},
E_{2j+1})$.
\end{proof}

Alternatively, one can prove Theorem~\ref{t4.2}
directly, that is, by pure
ODE techniques by explicitly computing the
$2\times 2$ spectral matrix of
$H$ given the Weyl $M$-matrix \eqref{4.8}
(with entries \eqref{4.19},
\eqref{4.20}) and by calculating the spectral
function of $H_{\pm,n}$ given
the corresponding Weyl $m$-functions
$m_{\pm,n}$ in \eqref{4.21},
\eqref{4.22}.  This also settles the
multiplicity of the spectrum
following \cite{46} in the context of second-order
differential operators
in $L^2(\bbR)$ (see also Appendices~A--C of
\cite{37} for a short summary
of the relevant spectral theoretic results).
Here we only mention that
if
\begin{equation}
d\rho_{n_0} =(d\rho_{n_0,j,k})_{1\leq j, k\leq 2}
\lb{4.32}
\end{equation}
denotes the (self-adjoint) matrix-valued spectral
measure of $H$, related
to the Weyl $M$-matrix \eqref{4.8} via
\begin{equation}
M_{n_0} (z) =\int_\bbR \dfrac{d\rho_{n_0}
(\lam)}{z-\lam},
\lb{4.33}
\end{equation}
one explicitly obtains from \eqref{4.19},
\eqref{4.20} in the $g$-gap
context of Theorem~\ref{t4.2} that
\begin{align}
\dfrac{d\rho_{n_0, 1,1} (\lam)}{d\lam} &
= \begin{cases}
\dfrac{F_g (\lam, n_0)}
{\pi i R_{2g+2} (\lam)^{1/2} }, & \lam \in
\sig(H)\\
0, & \lam \in\bbR \bs \sig (H)
\end{cases},
\lb{4.34}\\
\dfrac{d\rho_{n_0, 1,2} (\lam)}{d\lam} & =
\dfrac{d\rho_{n_0, 2,1}(\lam)}{d\lam} =\begin{cases}
\dfrac{-G_{g+1} (\lam, n_0)}{2\pi ia(n_0)
R_{2g+2} (\lam)^{1/2}}, & \lam
\in \sig(H)\\
0, & \lam \in\bbR \bs \sig (H)
\end{cases},
\lb{4.35}\\
\dfrac{d\rho_{n_0, 2,2} (\lam)}{d\lam} &
= \begin{cases}
\dfrac{F_g (\lam, n_0+1)}
{\pi i R_{2g+2} (\lam)^{1/2}}, & \lam \in\sig
(H)\\
0, & \lam \in\bbR \bs \sig (H)
\end{cases}
\lb{4.36}
\end{align}
(cf.~\eqref{a.5} for our conventions on
$R_{2g+2}(\lam)^{1/2}$).

Finally we relate the polynomials
$F_g(z,n)$, $G_{g+1}(z,n)$ used
in Chapter~\ref{s3} to the homogeneous quantities
$\hat{F}_g(z,n)$, $\hat{G}_{g+1}(z,n)$
(cf.\ \eqref{2.18}). We introduce the
constants $c_j(\underline{E})$ via
\begin{equation}
R_{2g+2} (z)^{1/2} =
- z^{g+1} \sum_{j=0}^\infty c_j(\underline{E}) z^{-j},
\quad |z|> \| H \|, \lb{4.37a}
\end{equation}
where $\underline{E}=(E_0,\dots,E_{2g-1})$, implying
\begin{equation}
c_0(\underline{E}) = 1, \quad c_1(\underline{E}) =
-\frac{1}{2} \sum_{j=0}^{2g+1} E_j, \:\: \text{etc}.
\end{equation}

\begin{lem} \lb{l4.4}
Let $F_g(z,n)$, $G_{g+1}(z,n)$ be the
polynomials  defined in \eqref{3.4},
\eqref{3.5} and let $a(n)^2$, $b(n)$ be
defined as in \eqref{3.18},
\eqref{3.19}. Then we have
\begin{equation}
F_g(z,n) = \sum_{\ell=0}^g c_{g-\ell}(\underline{E})
\hat{F}_\ell(z,n), \quad
G_{g+1}(z,n) = \sum_{\ell=0}^g c_{g-\ell}(\underline{E})
\hat{G}_{\ell+1}(z,n).
\end{equation}
\end{lem}

\begin{proof}
 From \eqref{4.19}, \eqref{4.20} we infer for
$|z|>\| H \|$ using
Neumann's expansion for the
resolvent of $H$ and Lemma \ref{l4.3} that
\begin{eqnarray} \label{expfmu}
F_g(z,n) &=& -\frac{R_{2g+2} (z)^{1/2}}{z}
\sum_{\ell=0}^\infty \hat{f}_\ell(n)
z^{-\ell},\\
 G_{g+1}(z,n) &=& R_{2g+2} (z)^{1/2} \Big( 1
- \frac{1}{z} \sum_{\ell=0}^\infty
\hat{g}_\ell(n) z^{-\ell} \Big), \label{expgmu}
\end{eqnarray}
which, together with \eqref{4.37a}, completes
the proof.
\end{proof}

For interesting recent generalizations of almost
periodic Jacobi
operators to cases where formally $g\to\infty$, and
Cantor spectra and
solutions of infinite-dimensional Jacobi inversion
problems are involved,
we refer to \cite{7}, \cite{76}.


\chapter[Solutions of the Stationary Toda
Hierarchy]{Quasi-Periodic
Finite-Gap Solutions of the Stationary Toda
Hierarchy} \lb{s5}
\setcounter{prop}{0}
\setcounter{equation}{0}

Given the detailed preparations in Chapter~\ref{s3}
we are now ready to
derive the algebro-geometric finite-gap solutions of
the stationary Toda
hierarchy.

Starting with high-energy expansions for
$\ome_{\infty_+, \infty_-}^{(3)}$
and $\theta (\uz (P,n))$ we formulate

\begin{lem} \lb{l5.1}
Given the canonical charts
$(\Pi_\pm \bs \{ (0,R_{2g+2}(0)^{1/2}) \},
z=\zeta^{-1})$ we obtain the following
expansions for $P$ near $\infty_\pm$.\\
(i).
\begin{equation}
\exp \left[ \int_{P_0}^P \ome_{\infty_+,
\infty_-}^{(3)} \right]
\underset{\zeta\to 0}{=}
 (\ti a \zeta)^{\pm 1}
\left[ \sum_{\ell=0}^\infty \ti b_\ell (-\zeta)^\ell
\right]^{\pm 1}, \quad P \text{ near } \infty_\pm,
\lb{5.1}
\end{equation}
where $\ti a$, $\{\ti b_\ell\}_{\ell \in\bbN_0}$
only depend on $K_g$
(i.e., on $\{E_m\}_{0 \leq m \leq 2g+1}$) and
\begin{equation}
\ti a < 0
\lb{5.2}
\end{equation}
is an integration constant,
\begin{align}
\begin{split}
\ti b_0 & = 1,\\
\ti b_1 & = \sum_{j=1}^g \lam_j
-\frac12 \sum_{m=0}^{2g+1} E_m,\\
& \text{etc.}
\lb{5.3}
\end{split}
\end{align}
(ii).
\begin{equation}
\dfrac{\theta(\uz (P, n+1))}{\theta(\uz (P,n))}
\underset{\zeta\to 0}{=}
\dfrac{\theta\big(\uz \binom{n+1}{n}\big)}
{\theta \big(\uz
\binom{n}{n-1}\big)}
\sum_{\ell =0}^\infty \ti
\theta_{\pm,\ell} (n) \zeta^\ell, \quad P \text{
near } \infty_\pm,
\lb{5.4}
\end{equation}
where
\begin{align}
\begin{split}
\ti \theta_{\pm,0} (n) & = 1,\\
\ti \theta_{\pm, 1} (n) &
= \pm \sum_{j=1}^g c_j (g) \dfrac{\pa}{\pa w_j}
\ln \left. \left[ \dfrac{\theta \big(\uw
+\uz \binom{n+1}{n}\big)}{\theta
\big(\uw
+\uz
\binom{n}{n-1}\big)} \right] \right|_{\uw
= \uzero},\\
&\text{etc.},
\lb{5.5}
\end{split}
\end{align}
and $c_j (k)$ are defined in \eqref{a.20}.\\
(iii).
\begin{align}
\begin{split}
& \phi(P,n)
\underset{\zeta\to 0}{=}
 \ti a C(n) \dfrac{\theta (\uz (n+1))}
{\theta (\uz (n))} \zeta\\
&   + \ti a C(n)\dfrac{\theta (\uz (n+1))}
{\theta(\uz (n))}
\left\{ - \ti b_1 +\sum_{j=1}^g c_j (g) \dfrac{\pa}
{\pa w_j} \ln
\left. \left[
\dfrac{\theta(\uw +\uz (n+1))}
{\theta (\uw + \uz(n))} \right]
\right|_{\uw =\uzero} \right\} \zeta^2\\
&  + O(\zeta^3), \quad P \text{ near } \infty_+.
\lb{5.6}
\end{split}
\end{align}
\end{lem}

\begin{proof}
(i). Using the representation \eqref{3.40} for
$\ome_{\infty_+,
\infty_-}^{(3)}$ and Lemma~\ref{l3.3}~(i) one
readily finds
\eqref{5.1}--\eqref{5.3} as follows.  First one
expands
\begin{equation}
\ome_{\infty_+, \infty_-}^{(3)}
\underset{\zeta\to 0}{=} \pm
\zeta^{-1} \left\{
1+
\left[
\frac12 \sum_{m=0}^{2g+1} E_m -\sum_{j=1}^g \lam_j
\right] \zeta + 0
(\zeta^2) \right\} \, d\zeta,\quad
P \text{ near } \infty_\pm
\lb{5.7}
\end{equation}
in a sufficiently small neighborhood of $\infty_\pm$
and integrates
term by term.  The remaining contribution to the
integral in \eqref{5.1}
is then absorbed into the integration constant
$\ti a$ by integrating
along the lift of the straight line segment from
$E_0$ to $\ti \pi (\hatP)$ for some $\hatP$ near $\infty_\pm$
along the
negative real axis.\\
(ii). This follows from \eqref{3.34}, \eqref{3.35},
\eqref{3.39}, and from
\begin{equation}
(\underline{\hat{A}}_{P_0} \circ z^{-1})(\zeta)
\underset{\zeta\to 0}{=} \pm
\underline{c}(g) \zeta+ O(\zeta^2) \mod (L_g)
\quad\text{ near } \infty_\pm
\lb{5.8}
\end{equation}
(cf.~\eqref{a.25}).\\
Item~(iii) is obvious from (i), (ii), and \eqref{3.65}.
\end{proof}

Now we are in a position to derive the major
result of this chapter
expressing the $g$-gap sequences $a$, $b$ in
\eqref{3.18}, \eqref{3.19}
in terms of the $\theta$-function associated
with $K_g$.

\begin{thm}\lb{t5.2}
The stationary $\tl_g$ solutions, or equivalently,
the $g$-gap sequences
$a=\{ a(n)\}_{n\in\bbZ}$, $b=\{b(n)\}_{n\in\bbZ}$
in \eqref{3.18},
\eqref{3.19}, are given by
\begin{align}
a(n) & = \ti a [\theta (\uz (n-1) )
\theta (\uz (n+1))/ \theta (\uz
(n))^2 ] ^{1/2}, \quad n\in\bbZ,
\lb{5.9}\\
b(n) & = \sum_{j=1}^g \lam_j
-\frac12 \sum_{m=0}^{2g+1} E_m -\sum_{j=1}^g
c_j (g) \dfrac{\pa}{\pa w_j} \ln \left.
\left[ \dfrac{\theta (\uw +\uz
(n))}{\theta (\uw +\uz (n-1))} \right] \right|_{\uw
=\uzero}, \quad
n\in\bbZ.
\lb{5.10}
\end{align}
\end{thm}

\begin{proof}
Inserting expansion \eqref{5.6} near $\infty_+$ into
\eqref{3.22} yields
\begin{align}
\begin{split}
& 0  = a(n-1) \phi(P,n-1)^{-1}
-\ti{\pi} (P) -b(n) +O(\zeta)\\
& = \dfrac{a(n-1) \theta (\uz (n-1))}
{\ti a C(n-1) \theta (\uz (n))
\zeta} \Bigg\{ 1 +  \left[ \left. \ti b_1
- \sum_{j=1}^g c_j (g)
\dfrac{\pa}{\pa w_j} \ln \left( \dfrac{\theta (\uw
+\uz(n))}{\theta (\uw
+\uz (n-1))} \right) \right|_{\uw
=\uzero} \right]  \zeta\\
&  \quad + O(\zeta^2) \Bigg\} -\zeta^{-1}
-b(n) +O(\zeta),
\lb{5.11}
\end{split}
\end{align}
that is, one infers \eqref{3.75} (again) and
\eqref{5.10}.  Equation \eqref{5.9} is
clear from \eqref{3.67} and \eqref{3.75}.
\end{proof}

\begin{rem}\lb{r5.3}
Alternatively, one could have derived
\eqref{5.10} by evaluating the
integral
\begin{align}
\begin{split}
I & = \frac{1}{2\pi i} \int_{\pa \hat K_g}
\ti \pi (.) \, d \ln
[\theta(\uz (.,n))]\\
& = \sum_{j=1}^g \mu_j (n)
+\sum_{P\in\{\infty_\pm\}} \res_P \{ \ti\pi (.)
\, d\ln [\theta (\uz (.,n)]\}
\lb{5.12}
\end{split}
\end{align}
using the residue theorem.  Since on the other
hand a direct calculation
shows that
\begin{equation}
I =\sum_{j=1}^g \int_{a_j} \ti \pi \ome_j,
\lb{5.13}
\end{equation}
the trace relation \eqref{3.19} for $b(n)$
yields
\begin{equation}
b(n) =\sum_{j=1}^g \int_{a_j}
\ti \pi \ome_j -\frac12
\sum_{m=0}^{2g+1}
E_m -\sum_{j=1}^g c_j (g) \dfrac{\pa}
{\pa w_j} \ln \left[
\dfrac{\theta(\uw +\uz (n))}{\theta ( \uw
+ \uz (n-1))}
\right]\Bigg|_{\uw =\uzero}.
\lb{5.14}
\end{equation}
A comparison of \eqref{5.10} and \eqref{5.14}
then reveals that
\begin{equation}
\sum_{j=1}^g \int_{a_j} \ti \pi \ome_j
=\sum_{j=1}^g \lam_j.
\lb{5.15}
\end{equation}
\end{rem}

Next we shall show that the BA-function is
determined by the location of
its poles on $K_g \bs \{\infty_\pm\}$ and its
behavior near $\infty_\pm$.

\begin{lem} \lb{l5.4}
Let $\psi (.,n)$, $n\in\bbZ$ be meromorphic on
$K_g$ satisfying
\begin{align}
(\psi (.,n)) & \geq -\calD_{\humu(n_0)}
+ (n-n_0) (\calD_{\infty_+}
-\calD_{\infty_-}).
\lb{5.16}\\
\intertext{Define a divisor $\calD_0 (n)$ by}
(\psi (.,n)) & =\calD_0 (n) -\calD_{\humu (n_0)}
+(n-n_0) (\calD_{\infty_+}
-\calD_{\infty_-}).
\lb{5.17}
\end{align}
Then
\begin{equation}
\calD_0 (n) \in \sig^g K_g, \; \calD_0 (n) > 0,
\; \deg (\calD_0 (n)) =g.
\lb{5.18}
\end{equation}
Moreover, if $\calD_0 (n)$ is nonspecial for all
$n\in\bbZ$, that is, if
\begin{equation}
i(\calD_0 (n))=0, \quad n \in\bbZ
\lb{5.19}
\end{equation}
then $\psi(.,n)$ is unique up to a constant multiple
(which may depend on
$n$).
\end{lem}

\begin{proof}
By the Riemann-Roch theorem (see \eqref{a.44})
there exists at least one
such function $\psi(.,n)$.  If $\psi_j (.,n)$,
$j=1,2$ are two such
functions satisfying \eqref{5.17} with corresponding
divisors
$\calD_{0,j} (n)$, $j=1,2$, one infers
\begin{equation}
(\psi_1 (n) / \psi_2 (n)) = \calD_{0,1} (n)
-\calD_{0,2} (n).
\lb{5.20}
\end{equation}
Since $i(\calD_{0,2}(n))=0$, $\deg (\calD_{0,2}(n))
=g$ by hypothesis,
\eqref{a.44} yields $r(-\calD_{0,2}(n))=1$,
$n\in\bbZ$ and hence $\psi_1
(.,n) / \psi_2 (.,n)$ is a constant on $K_g$.
\end{proof}

One can use Lemma~\ref{l5.4} to obtain an
alternative proof of the fact
that $\phi$ and $\psi$ given by
\eqref{3.65}--\eqref{3.68} coincide with
the expressions \eqref{3.20}, \eqref{3.21} and
satisfy the Riccati and
Jacobi equations \eqref{3.22} and \eqref{3.23},
respectively.  We shall
use precisely this strategy in the $t$-dependent
context to be discussed
in the next chapter.

\begin{rem} \lb{r5.5}
In the special case where
$\mu_j (n_0)\in\{E_{2j-1}, E_{2j}\}$ for all
$1\leq j \leq g$ the following are equivalent.\\
(i).
\begin{equation}
\hat \mu_j (n_0+n) =\hat \mu_j (n_0-n)^*,
\quad 1 \leq j \leq g, \;
n\in\bbZ.
\lb{5.21}
\end{equation}
(ii).
\begin{equation}
\uz (P,n_0+n) =-\uz (P^*, n_0-n) \mod (L_g),
\quad n\in\bbZ.
\lb{5.22}
\end{equation}
(iii).
\begin{equation}
a(n_0 +n) =a(n_0-n+1), \; b(n_0 +n) =b(n_0 -n),
\quad n\in\bbZ.
\lb{5.23}
\end{equation}
\end{rem}

Next we derive an alternative and, to the best of
our knowledge, novel
$\theta$-function representation of $b(n)$.

\begin{cor} \lb{c5.6}
$b(n)$, $n\in\bbZ$ admits the representation
\begin{align}
b(n) = -E_0 +
\ti a \dfrac{\theta (\uz (n-1))\theta (\uz (P_0,
n+1))}{\theta (\uz (n)) \theta (\uz (P_0, n))}
+ \ti a \dfrac{\theta (\uz (n))
\theta (\uz ( P_0, n-1))}{\theta
(\uz (n-1))\theta (\uz (P_0,n))}.
\lb{5.24}
\end{align}
\end{cor}

\begin{proof}
It suffices to combine \eqref{3.22},
\eqref{3.65}, \eqref{3.67} (all at
$P=P_0$), and \eqref{5.9}.
\end{proof}

The results of this chapter, with the exception
of Corollary~\ref{c5.6},
are well-known (they are contained, e.g., in
Section~5 of \cite{67} which
is devoted to (stationary) $\theta$-function
representations of $\psi(.)$
and the coefficients $(a,b)$ of $L$).  Since
they represent a special
case of the time-dependent findings of
Chapter~\ref{s6} we postpone
further references to the original literature
to the next chapter.

Finally, for reasons of completeness, we also
mention the following
criterion for $\{a(n)\}_{n\in\bbZ}$,
$\{b(n)\}_{n\in\bbZ}$ to be periodic
of period $N\in\bbN$.

\begin{thm} (See \cite{55}, Ch.~2.) \lb{t5.7}
A necessary and sufficient condition for the
$g$-gap sequences
$\{a(n)\}_{n\in\bbZ}$, $\{ b(n)\}_{n\in\bbZ}$
in \eqref{3.18},
\eqref{3.19} to be periodic of period $N\in\bbN$
is that $R_{2g+2}(z)$ is
of the form
\begin{equation}
R_{2g+2} (z) Q(z)^2 =\Del (z)^2 -1,
\lb{5.25}
\end{equation}
where $Q(.)$ and $\Del(.)$ are polynomials.  The
period $N$ is given by
\begin{equation}
N=\deg(Q) +g+1.
\lb{5.26}
\end{equation}
\end{thm}

Since we are not using Theorem~\ref{t5.7} in our
main text we defer its
proof to Appendix~\ref{app-b}.

This completes our treatment of stationary
quasi-periodic finite-gap
sequences and we now turn to the $t$-dependent
case.


\chapter[The Time-Dependent Baker-Akhiezer Function]{Quasi-Periodic
Finite-Gap Solutions of the Toda Hierarchy and
the Time-Dependent Baker-Akhiezer Function}\lb{s6}
\setcounter{equation}{0}
\setcounter{prop}{0}

In this chapter we continue the construction of
$g$-gap sequences for the
Toda hierarchy and now treat the time-dependent case.

Our starting point will be a $g$-gap stationary
solution $(a^{(0)},
b^{(0)})$ of the type \eqref{5.9}, \eqref{5.10}, that is,
\begin{align}
a^\bze (n) & = \ti a [\theta (\uz (n-1))
\theta (\uz (n+1)) / \theta (\uz
(n))^2 ]^{1/2},\lb{6.1}\\
b^\bze (n) & = \sum_{j=1}^g \lam_j
-\frac12 \sum_{m=0}^{2g+1} E_m
-\sum_{j=1}^g c_j(g) \dfrac{\pa}{\pa w_j}
 \ln \left[ \dfrac{\theta (\uw
+ \uz (n))}{\theta (\uw + \uz (n-1))}
 \right] \Bigg|_{\uw = \uzero}
\lb{6.2}
\end{align}
satisfying \eqref{3.18}, \eqref{3.19},
\begin{align}
\begin{split}
a^\bze (n)^2 & =
\frac{1}{2} \sum_{j=1}^g R_{2g+2} (\hat \mu_j^\bze
(n))^{1/2} \prod^g_{\substack{ k=1\\ k\neq j }}
[ \mu_j^\bze (n)
-\mu_k^\bze (n)]^{-1}\\
& \quad -\frac{1}{4} [b^\bze (n)^2
+b^{\bze (2)} (n)]>0,
\lb{6.3}
\end{split}\\
\begin{split}
b^{\bze (k)} (n) &
=\sum_{j=1}^g \mu_j^\bze (n)^k -\frac12
\sum_{m=0}^{2g+1} E_m^k, \quad k\in\bbN,\\
b^\bze (n) & = b^{\bze (1)}
=\sum_{j=1}^g \mu_j^\bze (n) -\frac12
\sum_{m=0}^{2g+1} E_m,
\lb{6.4}
\end{split}
\end{align}
where
\begin{equation}
\ti \pi (\hat \mu_j^\bze (n))
=\mu_j^\bze (n) \in [E_{2j-1}, E_{2j} ],
\quad 1 \leq j \leq g, \; n\in\bbZ.
\lb{6.5}
\end{equation}
This $g$-gap stationary solution $(a^\bze,
b^\bze)$ represents the initial
condition for the following Toda flow (cf.~\eqref{2.12}),
\begin{equation}
\widetilde{TL}_r (a(t), b(t))=0, \quad (a(t_0), b(t_0))
=(a^\bze, b^\bze)
\lb{6.6}
\end{equation}
for some $r\in\bbN_0$, whose explicit solution we
seek below.  Explicitly, \eqref{6.6} amounts to (cf.\
\eqref{2.25}, \eqref{2.26})
\begin{align}
\dot a&=-a[2(b^++z)\ti F^+_r+\ti G^+_{r+1}+\ti G_{r+1}],
\quad a(t_0)=a^{(0)},  \\
\dot b&=-2[(b+z)^2\ti F_r+(b+z)\ti G_{r+1}
+a^2\ti F^+_r-(a^-)^2\ti F^-_r], \quad b(t_0)=b^{(0)}.
\end{align}

 From our treatment in Chapter~\ref{s3} we know that
$(a^\bze, b^\bze)$ is determined by the
band edges $\{ E_j \}_{0 \le j \le 2g+1}$ and the
Dirichlet eigenvalues $\{
\hat\mu_j^\bze(n_0) \}_{1 \le j \le g}$ at a
fixed point $n_0$. Hence we will
consider the following time evolution for the
Dirichlet eigenvalues
$\hat\mu_j(n_0,t)$
\begin{align}
\begin{split}
\dfrac{d}{dt} \mu_j (n_0,t) &
=-2\ti{F}_r (\mu_j (n_0,t), n_0,t)
\dfrac{R^{1/2}_{2g+2} (\hat\mu_j (n_0,t))}
{\prod^g_{\substack{ \ell =1\\
\ell\neq j}} [\mu_j (n_0,t) -\mu_\ell (n_0,t)]},\\
\hat \mu_j (n_0,t_0) & = \hat \mu_j^\bze (n_0),
\quad 1 \leq j \leq g, \;
(n_0,t) \in\bbZ\times\bbR
\lb{6.19}
\end{split}
\end{align}
(similar to those encountered in connection with
the KdV hierarchy,
see, e.g., \cite{6}, \cite{12}, \cite{13}). Here
$\ti{F}_r(z,n_0,t)$
has to be defined
using \eqref{2.21} (cf.\ \eqref{2.10}) and
$a(n_0,t)^2$, $b(n_0,t)$ have to
be expressed in terms of $\hat{\mu}_j(n_0,t)$
(this can be done by comparing the coefficients in \eqref{expfmu}).
In order to stress the fact that the  summation constants
$c_{\ell}$ in $\ti{F}_r$ and $F_g$ are different in general,
we decided
to indicate this by using the notation $\ti{c}_{\ell}$,
$\ti F_r$,
$\ti G_{r+1}$,
$\widetilde{TL}_r$, etc.

The appearance of the function $R^{1/2}_{2g+2}(.)$
in \eqref{6.19} indicates the natural way to interpret
this system as a
vector field on the
(complex) manifold $K_g \times \cdots \times K_g =K_g^g$.
Since we are
interested in real-valued solutions $(a(t), b(t))$ of
the Toda hierarchy, we
restrict this vector field to the submanifold
$\times_{j=1}^g \ti\pi^{-1} ([
E_{2j-1}, E_{2j}])$ which is isomorphic to the
torus $S^1 \times
\dots \times S^1 =T^g$. Standard theory for differential
equations on $C^\infty$
manifolds now implies the existence of a unique solution
$\{\hat\mu_j (n_0,t)\}_{1\leq j \leq g}$ satisfying the
initial condition $\hat
\mu_j (n_0,t_0) = \hat \mu_j^\bze (n_0)$.  An inspection
of \eqref{6.19} using
the charts \eqref{a.9}, \eqref{a.10} confirms that the
solution $\hat \mu_j
(n,t)$ changes sheets whenever it hits $E_{2j-1}$ or
 $E_{2j}$ and its projection
$\mu_j (n_0,t)=\ti\pi (\hat \mu_j(n_0,t))$ remains
trapped in
$[E_{2j-1}, E_{2j}]$ for all $t \in \bbR$. Thus up
to this point we have
\begin{equation}
\hat \mu_j (n_0,.) \in C^\infty (\bbR, K_g) \text{ and }
 \ti\pi (\hat \mu_j
(n_0,t))\in [E_{2j-1}, E_{2j}],
\quad 1\leq j \leq g,\; t \in \bbR,
\lb{6.20}
\end{equation}
and using \eqref{6.20} we can define polynomials
$F_g(z,n,t)$, $G_{g+1}(z,n,t)$ as in
Chapter~\ref{s3} (cf.\ \eqref{3.4}, \eqref{3.5}).

We start with calculating the time derivative of
$F_g(z,n_0,t)$. By virtue of \eqref{3.7} and
\eqref{6.19} we obtain
\begin{equation}
\dfrac{d}{dt} F_g (z,n_0,t)\big|_{z=\mu_j(n_0,t)} =
- 2 \ti{F}_r (\mu_j(n_0,t),n_0,t)
G_{g+1} (\mu_j(n_0,t),n_0,t), \quad 1 \le j \le g. \lb{deriv}
\end{equation}
Since two  polynomials of (at most) degree
$g-1$ coinciding at
$g$ points are
equal we infer
\begin{equation}
\dfrac{d}{dt} F_g (z,n_0,t)
=2[F_g (z,n_0,t) \ti{G}_{r+1} (z,n_0,t) -
\ti{F}_r (z,n_0,t) G_{g+1} (z,n_0,t)],
\lb{6.15pr}
\end{equation}
provided we can show that the right-hand side of
\eqref{6.15pr} is a
polynomial of degree at most
$g-1$.  It suffices to prove the special homogeneous case
where $\ti c_0=1$, $\ti c_\ell=0$,
$\ell\geq 1$. Dividing \eqref{6.15pr} by
$R_{2g+2}(z)^{1/2}$, using \eqref{expfmu}, \eqref{expgmu},
shows that our assertion is
equivalent to
\begin{align}
&-z^{-1}\sum_{j=0}^r \ti f_j(n_0,t)z^{-j}[-z^{r+1}
+\sum_{\ell=0}^r \ti g_{r-\ell}(n_0,t)z^\ell+\ti
f_{r+1}(n_0,t)]  \no \\
& -\sum_{\ell=0}^r\ti f_{r-\ell}(n_0,t)z^\ell[1
-z^{-1}\sum_{j=0}^\infty\ti
g_j(n_0,t)z^{-j}]=O(z^{-2}). \label{6.15a}
\end{align}
Since by inspection, the coefficient of $z^k$ for
$-1\leq k\leq r$ on the left-hand side of
\eqref{6.15a} turns out to be
\begin{align}
& \ti f_{r-k}-\ti f_{r-k}-[\ti g_{r-(k+1)}\ti f_0+
\ti g_{r-(k+2)}\ti f_1+\cdots+\ti g_0\ti
f_{r-(k+1)}]\no \\
& +[\ti f_{r-(k+1)}\ti g_0+\ti f_{r-(k+2)}\ti g_1+\cdots+
\ti f_0\ti g_{r-(k+1)}]=0,
\end{align}
this proves \eqref{6.15a} and hence \eqref{6.15pr}.

To obtain the time derivative of
 $G_{g+1}(z,n_0,t)$ we use
\begin{equation}
G_{g+1} (z,n_0,t)^2 -4a (n_0,t)^2 F_g (z,n_0,t)
F_g (z,n_0+1,t) =R_{2g+2}(z)
\lb{6.17}
\end{equation}
as in \eqref{3.8}. Again, evaluating the
$t$-derivative of \eqref{6.17} first at
$z=\mu_j(n_0,t)$, one obtains from \eqref{deriv}
\begin{align} \no
\dfrac{d}{dt} G_{g+1} (z,n_0,t) = &
4 a (n_0,t)^2 [F_g (z,n_0,t) \ti{F}_r (z,n_0+1, t)\\
& -\ti{F}_r (z,n_0,t) F_g (z,n_0+1, t)]
\lb{6.17pr}
\end{align}
for those $t \in \bbR$ such that
$\mu_j(n_0,t) \not\in \{E_{2j-1}, E_{2j} \}$,
provided the right-hand side of \eqref{6.17pr} is a
polynomial in $z$ of degree at
most $g-1$. Since
the exceptional set is discrete the identity will
then follow for all
$t \in \bbR$ by
continuity. Again it suffices to prove the special
homogeneous case $\ti c_0=1$, $\ti c_\ell=0$,
$\ell\geq 1$.  Thus we need to prove
\begin{align}
&\frac{F_g(z,n_0,t)}{R_{2g+2}(z)^{1/2}}\ti
F_r^+(z,n_0,t)-\frac{F^+_g(z,n_0,t)}
{R_{2g+2}(z)^{1/2}}\ti F_r(z,n_0,t) \no \\
&=z^{-1}\sum_{j=0}^\infty \ti
f_j^+(n_0,t)z^{-j}\sum_{\ell=0}^r\ti f_{r-\ell}(n_0,t)z^\ell
\lb{ekstra}\\
&  -z^{-1}\sum_{j=0}^\infty
\ti f_j(n_0,t)z^{-j}\sum_{\ell=0}^r\ti
f_{r-\ell}^+(n_0,t)z^\ell=O(z^{-2})  \no
\end{align}
using \eqref{expfmu}.  By inspection, the coefficient of
$z^k$ for $-1\leq k\leq r-1$ on the
left-hand side of \eqref{ekstra} indeed vanishes,
proving \eqref{6.17pr}.

Similarly, evaluating $(d/dt)F_g (z,n_0 +1,t)$ at
$z= \mu_j(n_0+1,t)$ (the zeros of $F_g
(z,n_0+1,t)$), we see that \eqref{6.15pr} also holds
with $n_0$ replaced by
$n_0+1$ and finally that
$\{\hat\mu_j (n_0+1,t)\}_{1\leq j \leq g}$ satisfies
\eqref{6.19} with initial condition
$\hat \mu_j (n_0+1,t_0) = \hat \mu_j^\bze
(n_0+1)$. Proceeding inductively we obtain this
result for all $n \ge n_0$ and
with a similar calculation (cf.\ Chapter~\ref{s3}) for all
$n \le n_0$.

Summarizing, we have constructed the set
$\{\hat\mu_j (n,t)\}_{1\leq j \leq g}$
for all $(n,t) \in \bbZ\times\bbR$ such that
\begin{align}
\begin{split}
\dfrac{d}{dt} \mu_j (n,t) &
=-2\ti{F}_r (\mu_j (n,t), n,t) \dfrac{R_{2g+2}
(\hat\mu_j (n,t))^{1/2}}{\prod^g_{\substack{ \ell
=1\\ \ell\neq j}} [\mu_j
(n,t) -\mu_\ell (n,t)]},\\
\hat \mu_j (n,t_0) & = \hat \mu_j^\bze (n),
\quad 1 \leq j \leq g, \;
(n,t) \in\bbZ\times\bbR,
\lb{6.16}
\end{split}
\end{align}
with
\begin{equation}
\hat{\mu}_j(n,.) \in C^\infty(\bbR,K_g) \text{ and } \ti\pi
(\hat{\mu}_j(n,t)) \in [E_{2j-1},E_{2j}],\quad 1 \leq j \leq g, \;
(n,t) \in\bbZ\times\bbR.
\end{equation}

As expected from the stationary $g$-gap outset in
\eqref{6.1}--\eqref{6.5},
the left-hand side in \eqref{6.16} is $t$-independent
for $r=g$, assuming the
same summation constants $c_\ell=c_\ell(E)$,
$1\leq \ell \leq g$ in $\ti{F}_r$
and $F_g$ (cf.\ Lemma \ref{l4.4}).

Furthermore,  we have corresponding polynomials
$F_g (z,n,t)$ and
$G_{g+1}(z,n,t)$
satisfying
\begin{align}
\dfrac{d}{dt} F_g (z,n,t) &=
2[F_g (z,n,t) \ti{G}_{r+1} (z,n,t) -
\ti{F}_r (z,n,t) G_{g+1} (z,n,t)],
\lb{6.15}\\
\dfrac{d}{dt} G_{g+1} (z,n,t) &=
4a (n,t)^2 [F_g (z,n,t) \ti{F}_r (z,n+1, t)
-\ti{F}_r (z,n,t) F_g (z,n+1, t)].
\lb{6.18}
\end{align}

In order to see that \eqref{6.19} was indeed the
correct choice one needs to
calculate $\dot{a}$ and $\dot{b}$. Differentiating
\eqref{6.17} involving
\eqref{6.15}, \eqref{6.18} and \eqref{3.16} yields
\eqref{2.25}.
Differentiating \eqref{3.16} using \eqref{6.15},
\eqref{6.18} and
\eqref{3.20pr} yields \eqref{2.26}. Thus
$\widetilde{TL}_r (a,b) =0$ and
\begin{equation}
\frac{d}{dt} L(t) - [\ti{P}_{2r+2}(t),L(t)] = 0, \quad t \in \bbR.
\end{equation}

In addition to the function $\phi (P,n,t)$
\begin{align}
\begin{split}
\phi (P, n, t) & =
\dfrac{ -G_{g+1} (\ti \pi (P), n,t)+R^{1/2}_{2g+2}
(P)}{2a (n,t) F_g (\ti\pi (P), n,t)}\\
& = \dfrac{-2a (n,t) F_g (\ti\pi (P), n+1, t)}
{G_{g+1} (\ti\pi (P), n,t)
+R^{1/2}_{2g+2}(P)},
\lb{6.7}
\end{split}
\end{align}
we now define the time-dependent BA-function
$\psi (P,n,n_0,t,t_0)$,
meromorphic on\\ $K_g \bs \{\infty_+, \infty_-\}$, by
\begin{multline}
\psi (P,n,n_0, t,t_0) =\\
\quad \exp \left\{ \int_{t_0}^t \, ds [2a (n_0, s) \ti{F}_r
(z,n_0,s) \phi (P, n_0, s)
+\ti{G}_{r+1} (z,n_0, s)]\right\} \times\\
\times \begin{cases}
\prod_{m=n_0}^{n-1} \phi (P,m,t), & n \geq n_0+1\\
1, & n=n_0\\
\prod_{m=n}^{n_0-1} \phi (P,m,t)^{-1}, & n\leq n_0 -1
\end{cases}.
\lb{6.8}
\end{multline}
Straightforward calculations then imply
\begin{align}
& a(n,t) \phi(P,n,t) +a(n-1,t) \phi (P,n-1,t)^{-1}
=b(n,t) + \ti\pi (P),
\; (n,t) \in\bbZ\times\bbR,
\lb{6.10}\\
\begin{split}
& \dfrac{d}{dt} \phi (P,n,t)
=-2 a (n,t) [\ti{F}_r (\ti\pi (P), n,t) \phi
(P,n,t)^2 +\ti{F}_r (\ti\pi (P), n+1, t)]\\
& \qquad +2 [b(n+1, t)
+\ti\pi (P)] \ti{F}_r (\ti\pi (P), n+1,t) \phi (P,n,t)\\
& \qquad + [\ti{G}_{r+1} (\ti \pi (P), n+1, t)
-\ti{G}_{r+1}
(\ti \pi (P), n,t)] \phi (P,n,t),
\quad (n,t) \in\bbZ\times \bbR
\lb{6.11}
\end{split}
\end{align}
and similarly,
\begin{align}
\begin{split}
& \quad a(n,t) \psi (P,n+1, n_0, t,t_0)
+a(n-1,t) \psi (P, n-1, n_0,
t,t_0)\\
& = [b(n,t) +\ti \pi (P)] \psi (P,n,n_0, t,t_0),
\quad (n,n_0, t,t_0)
\in\bbZ^2 \times \bbR^2,
\lb{6.12}
\end{split}
\end{align}
\begin{align}
\begin{split}
& \dfrac{d}{dt} \psi (P,n,n_0, t,t_0)
 = 2a (t,n) \ti{F}_r (\ti \pi (P), n,t)
\psi (P,n+1, n_0, t,t_0)\\
&  + \ti{G}_{r+1} (\ti \pi (P), n,t) \psi(P,n,n_0,t,t_0),
\quad (n,n_0,
t,t_0) \in\bbZ^2\times \bbR^2.
\lb{6.13}
\end{split}
\end{align}

The analogs of \eqref{3.8} (for all $n_0 \in \bbZ$)
and
\eqref{3.26}--\eqref{3.29} then extend to the
present $t$-dependent
situation.

Using (variants of) Lagrange's interpolation formula
and the trace
formula \eqref{3.19} for $b$, the flow \eqref{6.16}
is easily seen to be
linearized (straightened out) by the Abel map (i.e.,
$\frac{d}{dt}\,
\hual_{P_0} (\calD_{\humu (n,t)}) = \ti{\uU}_r$ for
some $\ti{\uU}_r \in\bbR^g$) for $r=0,1$.  Since this
approach gets fairly cumbersome
with increasing $r$, we omit further details at this
point and postpone a
proof of this fact until later (see Theorem \ref{t6.2})
when alternative and
more effective tools are available.

In order to express $\phi (P,n,t)$ and
$\psi(P,n,n_0,t,t_0)$ in
terms of the theta
function of $K_g$ we need a bit of notation.
Let $\ome_{\infty_\pm,
q}^{(2)}$ be the normalized Abelian dsk (i.e., with
vanishing $a$-periods)
with a single pole at $\infty_\pm$ of the form
\begin{equation}
\ome_{\infty_\pm,q}^{(2)} =[\zeta^{-2-q} +O(1)]
\, d\zeta \text{ near }
\infty_\pm, \quad q \in\bbN_0.
\lb{6.23}
\end{equation}
Given the summation constants $\ti{c}_1, \ldots,
\ti{c}_r$ in $\ti{F}_r$,
see \eqref{2.14}, \eqref{2.18}, \eqref{2.29}, and
\eqref{2.27}, we then
define
\begin{equation}
\ti{\Ome}_r^{(2)} =\sum_{q=0}^r (q+1) \ti{c}_{r-q}
(\ome_{\infty_+,q}^{(2)} - \ome_{\infty_-,q}^{(2)}),
\quad \ti{c}_0 =1.
\lb{6.24}
\end{equation}
Since the $\ome_{\infty_\pm,q}^{(2)}$ were supposed
to be normalized we
have
\begin{equation}
\int_{a_j} \ti{\Ome}_r^{(2)} =0, \quad 1\leq j \leq g.
\lb{6.25}
\end{equation}
Moreover, writing
\begin{equation}
\ome_j =\left( \sum_{m=0}^\infty d_{j,m}
 (\infty_\pm) \zeta^m \right) \,
d\zeta
=\pm \left( \sum_{m=0}^\infty d_{j,m} (\infty_+)
 \zeta^m \right) \, d
\zeta \text{ near } \infty_\pm,
\lb{6.26}
\end{equation}
relation \eqref{a.36} yields
\begin{equation}
\ti{U}_{r,j}^{(2)} = \frac{1}{2\pi i }\int_{b_j}
\ti{\Ome}_r^{(2)}
= 2 \sum_{q=0}^r \ti{c}_{r-q} d_{j,q}(\infty_+),
\quad 1\leq j \leq g.
\lb{6.27}
\end{equation}
We also will have to employ the following slight
generalization of
Lemma~\ref{l5.4}.

\begin{lem} \lb{l6.1}
Let $\psi(.,n,t)$, $(n,t) \in\bbZ\times \bbR$ be
meromorphic on $K_g \bs
\{\infty_+, \infty_-\}$ with essential singularities
at $\infty_\pm$ such
that $\ti \psi(.,n,t)$ defined by
\begin{equation}
\ti \psi (P,n,t) =\psi (P,n,t)
\exp \Big[ -(t-t_0) \int_{P_0}^P
\ti{\Ome}_r^{(2)}\Big]
\lb{6.28}
\end{equation}
is multivalued meromorphic on $K_g$ and its
divisor satisfies
\begin{equation}
(\ti \psi (.,n,t))\geq -\calD_{\humu(n_0, t_0)}
+(n-n_0)
(\calD_{\infty_+} -\calD_{\infty_-}).
\lb{6.29}
\end{equation}
Define a divisor $\calD_0 (n,t)$ by
\begin{equation}
(\ti \psi (.,n,t))=\calD_0 (n,t)
-\calD_{\humu (n_0, t_0)} +(n-n_0)
(\calD_{\infty_+} -\calD_{\infty_-}).
\lb{6.30}
\end{equation}
Then
\begin{equation}
\calD_0 (n,t) \in\sig^g K_g, \; \calD_0 (n,t) > 0,
\; \deg (\calD_0 (n,t))
=g.
\lb{6.31}
\end{equation}
Moreover, if $\calD_0 (n,t)$ is nonspecial for all
$(n,t) \in\bbZ\times
\bbR$, that is, if
\begin{equation}
i (\calD_0 (n,t) ) =0, \quad (n,t) \in\bbZ\times \bbR
\lb{6.32}
\end{equation}
then $\psi (.,n,t)$ is unique up to a constant
multiple (which may depend
on $n$ and $t$).
\end{lem}

Since the proof is analogous to that of Lemma~\ref{l5.4}
we omit further
details.

Given these preparations we obtain the following
characterization of
$\phi (P,n,t)$ and $\psi (P,n,n_0, t,t_0)$ in
\eqref{6.7} and \eqref{6.8}.

\begin{thm} \lb{t6.2}
Introduce
\begin{align}
\uz (P,n,t) & =\hua_{P_0}(P)
-\hual_{P_0} (\calD_{\humu(n_0,t_0)})
+(n-n_0)\uU^{(3)} +(t-t_0) \ti{\uU}^{(2)}_r
-\huxi_{P_0},\\
\uz (n,t) & = \uz (\infty_+, n,t).
\lb{6.57}
\end{align}
Then we have
\begin{align}
& \phi(P,n,t)  = C(n,t) \dfrac{\theta (\uz (P,n+1,t))}
{\theta (\uz
(P,n,t))} \exp \Big( \int_{P_0}^P \ome_{\infty_+,
\infty_-}^{(3)} \Big),
\lb{6.33}\\
\begin{split}
& \psi (P,n,n_0, t,t_0)
= C(n,n_0, t,t_0) \dfrac{\theta (\uz
(P,n,t))}{\theta (\uz (P,n_0, t_0))} \times\\
& \quad \times
\exp \Big[ (n-n_0) \int_{P_0}^P \ome_{\infty_+,
\infty_-}^{(3)}
+ (t-t_0) \int_{P_0}^P \ti{\Ome}_r^{(2)} \Big],
\lb{6.34}
\end{split}
\end{align}
where $C(n,t)$, $C(n,n_0, t,t_0)$ are real-valued,
\begin{align}
C(n,t) & = C(n+1, n,t,t)
=\left[ \dfrac{\theta (\uz (n-1,t))}{\theta (\uz
(n+1,t))} \right]^{1/2},
\lb{6.35}\\
C(n,n_0,t,t_0) & =
\left[ \dfrac{ \theta (\uz (n_0, t_0)) \theta (\uz
(n_0-1, t_0))}
{\theta (\uz (n,t))\theta (\uz (n-1, t))} \right]^{1/2},
\lb{6.36}
\end{align}
and the sign of $C(n,t)$ is opposite that
of $a(n,t)$, that is,
\begin{equation}
\sgn [C(n,t)] =-\sgn [a(n,t)],
\quad (n,t)\in\bbZ\times \bbR.
\lb{6.37}
\end{equation}
Moreover,
\begin{equation}
\hual_{P_0} (\calD_{\humu(n,t)})
=\hual_{P_0}(\calD_{\humu(n_0,t_0)})
-(n-n_0) \uU^{(3)} -(t-t_0)\ti{\uU}^{(2)}_r \mod (L_g)
\lb{6.55}
\end{equation}
and hence the flows \eqref{6.19} are linearized by the Abel map
\begin{equation}
\dfrac{d}{dt} \hual_{P_0} (\calD_{\humu (n,t)})
=-\ti{\uU}^{(2)}_r, \quad (n,t)
\in\bbZ\times \bbR.
\lb{6.56}
\end{equation}
\end{thm}

\begin{proof}
First of all note that \eqref{6.33} and \eqref{6.34} are
well defined due to
\eqref{3.41}, \eqref{3.42}, \eqref{6.25}, \eqref{6.27},
and \eqref{a.28}.

Denoting the right-hand side of \eqref{6.34} by
$\Psi (P,n,n_0,t,t_0)$,
our goal is to prove $\psi = \Psi$.  By inspection,
one verifies
\begin{equation}
\Psi (P,n,n_0,t,t_0)
=\Psi (P,n_0,n_0,t,t_0) \Psi (P,n,n_0,t,t).
\lb{6.38}
\end{equation}
Comparison of \eqref{3.21}, \eqref{3.30},
\eqref{3.65}--\eqref{3.67} and
\eqref{6.7}, \eqref{6.33}--\eqref{6.36} then yields
\begin{equation}
\psi (P,n+1,n,t,t) =\phi (P,n,t) =\Psi (P,n+1,n,t,t).
\lb{6.39}
\end{equation}
Moreover,
\begin{equation}
\psi (P,n,n_0,t,t) =\left.
\begin{cases}
\prod_{m=n_0}^{n-1} \phi (P,m,t), & n \geq n_0 +1\\
1, & n=n_0\\
\prod_{m=n}^{n_0-1} \phi (P,m,t)^{-1}, & n\leq n_0 -1
\end{cases} \right\} =\Psi (P,n,n_0,t,t).
\lb{6.40}
\end{equation}
By \eqref{6.38} it remains to identify
\begin{equation}
\psi (P,n_0, n_0, t,t_0) =\Psi (P,n_0, n_0, t,t_0).
\lb{6.41}
\end{equation}
This is a bit more involved.  We start by noting
that \eqref{6.7},
\eqref{6.8}, and \eqref{6.15} imply
\begin{align}
\begin{split}
& \psi (P,n_0, n_0, t,t_0)
= \exp \left\{ \int_{t_0}^t \!\! ds [2a(n_0,s)
\ti{F}_r (z,n_0,s) \phi (P,n_0,s)
+\ti{G}_{r+1} (z,n_0, s)] \right\}\\
& = \exp \left\{ \int_{t_0}^t \!\!\! ds
\left[\ti{F}_r (z,n_0, s) \dfrac{R_{2g+2}
(P)^{1/2}\! -G_{g+1} (z,n_0,s)}{F_g (z,n_0,s)}
+\ti{G}_{r+1} (z,n_0,s)\right]
\right\}.
\lb{6.42}
\end{split}
\end{align}
In order to spot the zeros and poles of $\psi$
on $K_g \bs \{\infty_+,
\infty_-\}$ we need to expand the integrand in
\eqref{6.42} near its
singularities (the zeros $\mu_j (n_0, s)$ of
$F_g (z,n_0,s)$).  Using
\eqref{6.16} one obtains
\begin{align}
\begin{split}
\psi (P,n_0,n_0,t,t_0) = \exp \left\{ \int_{t_0}^t
\, ds \left[\dfrac{
\frac{d}{ds} \mu_j (n_0,s)}{\mu_j (n_0,s) -\ti\pi(P)}
+O(1)\right]\right\} \hspace*{2cm}\\
\qquad =
\begin{cases}
[ \mu_j (n_0, t) -\ti \pi (P) ] O(1) & \text{ for
$P$ near }
\hat\mu_j (n_0, t) \neq \hat \mu_j (n_0, t_0)\\
O(1) &\text{ for $P$ near } \hat\mu_j (n_0, t) =
\hat\mu_j (n_0, t_0)\\  {}
[ \mu_j (n_0, t_0) -\hat\pi (P) ]^{-1} O(1)
& \text{ for $P$ near }
\hat\mu_j (n_0, t_0) \neq \hat\mu_j (n_0, t)
\end{cases},
\lb{6.43}
\end{split}
\end{align}
with $O(1) \neq 0$. Hence all zeros and all
poles of $\psi(P,n_0, n_0, t,t_0)$ on $K_g \bs \{ \infty_+,
\infty_-\}$ are simple and the poles coincide with those of $\Psi
(P,n_0,n_0,t,t_0)$.  Next we need to identify the essential
singularities of $\psi (P,n_0,n_0,t,t_0)$ at $\infty_\pm$.
For this purpose we use \eqref{6.10} and
rewrite \eqref{6.42} in the form
\begin{align}
\begin{split}
& \psi(P,n_0,n_0,t,t_0)
= \exp \left\{ \int_{t_0}^t \, ds \left[
\dfrac12
\, \dfrac{\frac{d}{ds}F_g (z,n_0,s)}{F_g (z,n_0,s)}
+R_{2g+2} (P)^{1/2}
\dfrac{\ti{F}_r(z,n_0,s)}
{F_g (z,n_0,s)} \right] \right\}\\
& = \left[\frac{F_g (z,n_0,t)}
{F_g (z,n_0,t_0)} \right]^{1/2} \exp \left\{
R_{2g+2}(P)^{1/2}
\int_{t_0}^t \, ds
\dfrac{\ti{F}_r (z,n_0,s)}{F_g (z,n_0,s)} \right\}.
\lb{6.44}
\end{split}
\end{align}
We claim that
\begin{equation}
R_{2g+2} (P)^{1/2} \ti{F}_r (z,n,t) / F_g (z,n,t)
=\mp \sum_{q=0}^r \ti{c}_{r-q}
\zeta^{-q-2}
+O(1) \text{ for } P \text{ near } \infty_\pm.
\lb{6.45}
\end{equation}
By \eqref{2.29}, in order to prove \eqref{6.45},
it suffices to prove the
homogeneous case $\ti{c}_0 =1$, $\ti{c}_q =0$,
$1\leq q \leq r$.  Using
\eqref{4.19}, we may rewrite \eqref{6.45}
in the form
\begin{align}
\begin{split}
\ti{F}_r (z,n,t) / z^{r+1} &
= z^{-1} \sum_{q=0}^r \hat{f}_{r-q} (n,t)
z^{q-r}\\ & \underset{z\to\infty}{=}
-G(z,n,n,t) +O(z^{-r-1}).
\lb{6.46}
\end{split}
\end{align}
Since
\begin{equation}
G(z,n,n,t) =(\del_n, (H(t)-z)^{-1}\del_n),
\lb{6.47}
\end{equation}
the Neumann expansion for $(H(t) -z)^{-1}$ then
shows that \eqref{6.46}
is equivalent to
\begin{equation}
z^{-1} \sum_{q=0}^r
\hat{f}_{r-q} (n,t) z^{q-r} \underset{z\to\infty}{=}
z^{-1} \sum_{q=0}^r (\del_n, H(t)^q \del_n) z^{-q}
+O(z^{-r-2}).
\lb{6.48}
\end{equation}
But \eqref{6.48} is proven in \eqref{4.37} of
Lemma~\ref{l4.3}.  Given
\eqref{6.45} we can apply Lemma~\ref{l6.1} to
conclude \eqref{6.41} since
$\calD_{\humu(n,t)}$ is nonspecial for all
$(n,t) \in\bbZ\times \bbR$
by \eqref{6.20} and Lemma~\ref{la.2}.  This yields
\eqref{6.33}, \eqref{6.34}, and \eqref{6.55}.
Equations \eqref{6.35}--\eqref{6.37}
are then proved as in Theorem~\ref{t3.4}~(i).
\end{proof}

For connections between complete  integrability and
linearizations on the Jacobian of a curve under
very general assumptions (i.e., for generalized Toda
systems in the sense of Lie Algebras) we
refer, for instance, to \cite{3}, \cite{4}.
Finally, we conclude with the $\theta$-function
representation for the $t$-dependent $g$-gap
solutions of the Toda
hierarchy.

\begin{thm} \lb{t6.3}
The solutions
$\{a(n,t)\}_{(n,t) \in\bbZ\times \bbR}$,
 $\{b(n,t)\}_{(n,t)
\in\bbZ\times \bbR}$ of the $\widetilde{TL}_r$ equations
\eqref{6.6} with $g$-gap
initial conditions $\{a^{(0)} (n) \}_{n\in\bbZ}$,
$\{b^\bze
(n)\}_{n\in\bbZ}$ in \eqref{6.1}--\eqref{6.4} are
given by
\begin{align}
\begin{split}
a(n,t)^2 = &
\frac12 \sum_{j=1}^g R_{2g+2} (\hat \mu_j (n,t))^{1/2}
\prod^g_{\substack{ k=1\\ k\neq j}} [\mu_j (n,t)
-\mu_k (n,t)]^{-1}
\\ & {}- \frac14 [b(n,t) +b^{(2)} (n,t) ] > 0,
\lb{6.58}\end{split}\\
\begin{split}
b^{(k)} (n,t) = & \sum_{j=1}^g \mu_j (n,t)^k
-\frac12 \sum^{2g+1}_{m=0}
E_m^k, \quad k\in\bbN,\\
b(n,t) = & b^{(1)} (n,t)
=\sum_{j=1}^g \mu_j (n,t) -\frac12
\sum_{m=0}^{2g+1} E_m,
\lb{6.59}
\end{split}
\end{align}
where $\{\hat\mu_j (n,t)\}_{1\leq j \leq g}$
solves \eqref{6.19}.
Moreover, we have
\begin{align}
a(n,t) & =
\ti a [ \theta (\uz (n-1,t))\theta (\uz (n+1, t)) / \theta
(\uz (n,t))^2]^{1/2}, \quad (n,t) \in\bbZ\times \bbR,
\lb{6.60}\\
\begin{split}
b(n,t) & = \sum_{j=1}^g \lam_j
-\frac12 \sum_{m=0}^{2g+1} E_m
-\sum_{j=1}^g c_j (g) \dfrac{\pa}{\pa w_j}
\ln \left[\dfrac{\theta (\uw
+\uz (n,t))}{\theta(\uw +
\uz (n-1,t))} \right] \Bigg|_{\uw =
\uzero},\\
& \hspace*{6cm} (n,t) \in\bbZ\times \bbR,
\lb{6.61}
\end{split}
\end{align}
with $\ti a <0$ introduced in \eqref{5.1},
\eqref{5.2}.
\end{thm}

\begin{proof}
Equations \eqref{6.58} and \eqref{6.59} are obtained
in precisely the same way
as \eqref{3.18} and \eqref{3.19} taking into
account \eqref{6.7},
\eqref{6.10}, \eqref{6.11}, and \eqref{6.17}
(for $n_0 \in \bbZ$).
Expressions \eqref{6.60} and \eqref{6.61} then follow
as in Theorem~\ref{t5.2}.
\end{proof}

\begin{rem} \lb{r6.4}
(i). Since in the special case $r=0$, that is, for the
original Toda lattice
equations, $\uU^{(2)}_0$ simplifies to
\begin{equation}
\uU^\btwo_0 =2\uc (g)
\lb{6.62}
\end{equation}
due to \eqref{6.26}, \eqref{6.27}, and \eqref{a.25},
the expression for
$b(n,t)$ in \eqref{6.61} can be rewritten in the
familiar form
\begin{equation}
b(n,t) =\sum_{j=1}^g \lam_j
-\frac12 \sum_{m=0}^{2g+1} E_m -\frac12
\dfrac{d}{dt} \ln \left[ \dfrac{\theta (\uz (n,t))}
{\theta(\uz
(n-1,t))}\right],\quad (n,t) \in\bbZ\times \bbR,\; r=0.
\lb{6.63}
\end{equation}
(ii). Furthermore, expanding equation~\eqref{6.13}
around $\infty_\pm$ (still for $r=0$) shows that
\begin{equation}
\int_{P_0}^P \Omega^{(2)}_0 = \mp \left[
\frac{1}{\zeta} + \ti{b}_1
+ O(\zeta) \right] \quad \text{near } \infty_\pm, \quad r=0,
\lb{expconst}
\end{equation}
where $\ti{b}_1$ is defined in \eqref{5.3}.
Conversely, proving \eqref{6.13} as in the KdV case
(by expanding both sides in \eqref{6.13} around
$\infty_\pm$ and using
Lemma~\ref{l6.1}) turns out to be equivalent to
proving \eqref{expconst}. However, since we are not aware
of an independent proof of \eqref{expconst}, we chose a
different strategy in the proof of Theorem~\ref{t6.2}.
\end{rem}

Moreover, in analogy to Corollary~\ref{c5.6},
$b(n,t)$ admits the
alternative $\theta$-function representation which,
to the best of our
knowledge, has thus far not been noted in the
literature.

\begin{cor} \lb{c6.5}
$b(n,t)$ admits the representation
\begin{align} \no
b(n,t) = & -E_0 +
\ti a \dfrac{ \theta (\uz (n-1,t))\theta(\uz
(P_0,n+1,t))}{\theta(\uz (n,t))\theta (\uz (P_0,n,t))}\\
& + \ti a \dfrac{ \theta (\uz (n,t))
\theta (\uz (P_0, n-1,t))}{\theta
(\uz (n-1,t))\theta (\uz (P_0(n,t))},
\quad (n,t) \in\bbZ\times \bbR.
\lb{6.64}
\end{align}
\end{cor}

Since the proof is identical to that of
Corollary~\ref{c5.6} we omit
further details.

The $\theta$-function representation of the
BA-function $\psi
(.,n,n_0,t,t_0)$ in \eqref{6.34} (except for
a determination of
$C(n,n_0,t,t_0)$, a minor point) can be found,
for instance, in \cite{2},
\cite{24}, \cite{25}, \cite{55} in the special
 case, $r=0$, that is, for the
original Toda lattice.  Similarly, the
$\theta$-function representations
for $(a,b)$ in \eqref{6.60}, \eqref{6.61} have
first been derived by
Krichever \cite{53}, again in the case $r=0$.
This result is reproduced
(assuming periodicity of $(a,b)$ with respect
to $n$) with more details,
for instance, in \cite{24}, \cite{25}, \cite{55}, and,
by somewhat different
methods, derived in \cite{2}.  The periodic case
was originally treated
by Date and Tanaka \cite{19} (and is reproduced
with more details in
\cite{80}, Ch.~4).

In accordance with the modern $\tau$-function
formulation of completely
integrable systems (see, e.g., \cite{11},
\cite{77}, \cite{78}, \cite{81}
and the references therein), the results of
this chapter clearly
illustrate the possibility of simultaneously
treating the entire Toda
hierarchy by introducing infinitely many time
variables $\uut = (t_0,
t_1, t_2, \ldots)$ and hence $a(n,\uut)$,
$b(n,\uut)$, $\theta(n,\uut))$,
etc., where the $r$-th coordinate $t_r$ in
$\uut$ is associated with the
homogeneous $\tl_r$ system.


\chapter[The Kac-van Moerbeke Hierarchy]{The
Kac-van Moerbeke Hierarchy and its
Relation to the Toda Hierarchy} \lb{s7}
\setcounter{prop}{0}
\setcounter{equation}{0}

This chapter is devoted to the Kac-van Moerbeke
hierarchy and its
connection with the Toda hierarchy.  Using a
commutation (sometimes also
called a supersymmetric) approach one can show
that the Kac-van Moerbeke
($\km$) hierarchy is a modified Toda hierarchy
precisely in the way that the
modified Korteweg-de Vries (mKdV) hierarchy is
related to the
Korteweg-de vries (KdV) hierarchy or more
generally, the Drinfeld-Sokolov
(DS) hierarchy is a modified version of the
Gel'fand-Dickey (GD) hierarchy.
The connection between these hierarchies and
their modified counterparts
is based on (suitable generalizations of) Miura-type
transformations
which in turn are based on factorizations of
differential (respectively
difference) expressions.  The literature on this
subject is too extensive
to be quoted here in full.  The interested reader can
consult, for instance,
\cite{35}, \cite{36}, \cite{39}, \cite{40}, \cite{41},
\cite{80}, Ch.~3,
\cite{83} and the references therein.  In the present
case of the Toda
and modified Toda, respectively Kac-van Moerbeke hierarchies,
the (discrete)
analog of Miura's transformation in connection with
factorization
methods was first systematically employed by Adler
\cite{1} and further
developed in \cite{38}.  In particular, the approach
presented in this
chapter is essentially modeled after \cite{38} where
further details on
the $\tl$ and $\km$ system can be found.  For an alternative
approach to the
modified Toda hierarchy we refer to \cite{57}, Ch.~4.

Let
\begin{equation}
\rho (t) =\{\rho (n,t)\}_{n\in\bbZ}
\in\ell_\bbR^\infty (\bbZ), \quad 0
\neq \rho (n,.) \in C^1 (\bbR), \; n\in\bbZ
\lb{7.1}
\end{equation}
and define the ``even'' and ``odd'' parts
of $\rho(t)$ by
\begin{equation}
\rho_e (n,t) =\rho(2n,t), \; \rho_o (n,t)
=\rho (2n+1,t),\quad (n,t)
\in\bbZ\times \bbR.
\lb{7.2}
\end{equation}
Next we consider the difference expressions in
$\ell^\infty (\bbZ)$
(respectively the bounded operators in $\ell^2 (\bbZ)$)
\begin{equation}
A(t) =\rho_o(t) S^+
+ \rho_e(t), \; A(t)^* =\rho_o^-(t) S^-
+\rho_e(t),
\lb{7.3}
\end{equation}
which enable us to define matrix-valued difference
expressions $(M(t),
Q_{2g+2}(t))$ (the Lax pair) in
$\ell^\infty (\bbZ) \otimes \bbC^2$ as
follows,
\begin{align}
M(t) & = \begin{pmatrix}
0 & A(t)^*\\
A(t) & 0
\end{pmatrix}, \quad t\in\bbR,
\lb{7.4}\\
Q_{2g+2} (t) & = \begin{pmatrix}
P_{1, 2g+2} (t) & 0\\
0 & P_{2,2g+2} (t)
\end{pmatrix} =
P_{1,2g+2} (t) \oplus P_{2,2g+2} (t), \quad
g\in\bbN_0,\; t\in\bbR.
\lb{7.5}
\end{align}
Here $P_{k,2g+2} (t)$, $k=1,2$ are defined as
in \eqref{2.9} respectively
\eqref{2.20}, that is,
\begin{align}
& P_{k, 2g+2} (t) = -L_k (t)^{g+1}
+\sum_{j=0}^g [ g_{k,j} (t) +2a_k
(t) f_{k,j}(t) S^+] L_k (t)^{g-j} +f_{k,g+1}(t),
\lb{7.6}\\
& P_{k,2g+2} (t) \Big|_{\ker (L_k (t) -z)}
=2a_k(t) F_{k,g}
(z,t) S^+ + G_{k,g+1} (z,t), \quad k =1,2
\lb{7.7}
\end{align}
and $\{f_{k,g,j}(n,t)\}_{0\leq j \leq g}$,
$\{g_{k,g+1,j}(n,t)\}_{0\leq j
\leq g+1}$ and $F_{k,g}(z,t)$, $G_{k,g+1} (z,t)$
are defined as
in \eqref{2.10} and \eqref{2.21} with
\begin{align}
a_1(t)  = \rho_e(t) \rho_o(t), \qquad b_1 (t)
=-\rho_e(t)^2 -\rho_o^-
(t)^2,
\lb{7.8}\\
a_2 (t) = \rho_e^+ (t) \rho_o(t), \qquad b_2(t)
=-\rho_e(t)^2
-\rho_o(t)^2,
\lb{7.9}\\
L_k(t) =a_k (t) S^+ +a_k^- (t) S^- -b_k (t),
\qquad k =1,2.
\lb{7.10}
\end{align}
One verifies the factorization,
\begin{equation}
L_1(t) =A(t)^* A(t), \quad L_2 (t) =A(t) A(t)^*.
\lb{7.11}
\end{equation}
The corresponding Lax equation for the $\km$ system
then reads
\begin{equation}
\dfrac{d}{dt}M(t) -[Q_{2g+2} (t), M(t)]=0,
\quad t\in\bbR
\lb{7.12}
\end{equation}
and as in the Toda context \eqref{2.11},
varying $g\in\bbN_0$ yields the
$\km$ hierarchy which we denote by
\begin{equation}
\km_g (\rho) =0, \quad g\in\bbN_0.
\lb{7.13}
\end{equation}
As in Chapter~\ref{s2} (cf.~\eqref{2.18}) we use the symbol
``$\hat{\:\:}$'' to distinguish between inhomogeneous and
homogeneous $\km$ equations,  that is,
\begin{equation}
\widehat{KM}_g (\rho): = \km_g (\rho) \big|_{c_\ell \equiv 0,
\; 1\leq \ell \leq g} \; ,
\lb{7.14}
\end{equation}
with $c_\ell$ the summation constants of
Chapter~\ref{s2}.  In order to
obtain explicit expressions for the $\km$
equations \eqref{7.13} we
proceed as follows.  First we note that
\begin{align}
\begin{split}
\ker (M(t) -w) & = \ker (M(t)^2 -w^2)\\
& = \ker (L_1 (t)-z) \oplus \ker (L_2 (t) -z),
\quad w^2 =z.
\lb{7.15}
\end{split}
\end{align}
(We shall only use \eqref{7.15} in an algebraic sense
as in \eqref{2.20}. However, using the methods in
the proof of Theorems~2.1 and
2.3 of \cite{40}, \eqref{7.15} is easily seen to be
valid in a functional
analytic sense as well.)  Since
\begin{equation}
\dot M- [Q_{2g+2}, M] =\begin{pmatrix}
0 & \dot A^* -P_{1,2g+2} A^* +A^* P_{2,2g+2}\\
\dot A -P_{2,2g+2} A+AP_{1,2g+2} & 0
\end{pmatrix},
\lb{7.16}
\end{equation}
relations
\eqref{7.15} and \eqref{7.3}
yield after some computations,
\begin{align}
\begin{split}
\dot \rho_e & = 2\rho_e \rho_o^2 (F^+_{1,g}
- F_{2,g} ) -\rho_e
(G_{1,g+1} -G_{2,g+1}),\\
\dot \rho_o & = -2 \rho_o (\rho_e^+)^2 (F_{1,g}^+
-F_{2,g}^+)
+\rho_o (G^+_{1,g+1} -G_{2,g+1}),\quad g\in\bbN_0.
\lb{7.17}
\end{split}
\end{align}
(Here ``$\dot{\text{ }}$'' $= d/dt$.) Given $F_{k,g}$,
$G_{k,g+1}$ from Chapter~\ref{s2}  with
$(a_k,b_k)$, $k=1,2$ defined in \eqref{7.8},
\eqref{7.9}, equations
\eqref{7.17} yield the hierarchy of $\km$ equations.
In particular, introducing
\begin{align}
\begin{split}
\ukm_g (\rho) & = (\km_g (\rho)_e, \; \km_g(\rho)_o)^T\\
& := \begin{pmatrix}
\dot \rho_e -2\rho_e \rho_o^2(F^+_{1,g} -F_{2,g}) +\rho_e (
G_{1,g+1} -G_{2,g+1})\\
\dot \rho_o +2\rho_o(\rho_e^+)^2 (F_{1,g}^+ -F_{2,g}^+ )
-\rho_o (G_{1,g+1}^+ -G_{2,g+1})
\end{pmatrix},
\lb{7.18}
\end{split}
\end{align}
one obtains the $\km$ equations \eqref{7.13} by taking
into account
\eqref{7.2} in \eqref{7.18}.  Explicitly, identifying
\begin{equation}
\km_g(\rho)(n,t)=\begin{cases}\km_g(\rho)_e(\frac{n}{2},t),
&\text{$n$ even}, \\
\km_g(\rho)_o(\frac{n-1}{2},t),&\text{$n$ odd},
\end{cases} \lb{7.18a}
\end{equation}
one infers
from \eqref{2.27},
\eqref{7.2}, \eqref{7.13},  \eqref{7.18}, and \eqref{7.18a},
\begin{align}
\km_0(\rho) & = \dot \rho -\rho [(\rho^+)^2
-(\rho^-)^2] =0,
\lb{7.19}\\
\begin{split}
\km_1(\rho) & = \dot \rho -\rho [(\rho^+)^4
-(\rho^-)^4 +(\rho^{++})^2
(\rho^+)^2 +(\rho^+)^2 \rho^2
-\rho^2 (\rho^-)^2\\
& \quad -(\rho^-)^2 (\rho^{--})^2]
+c_1 (-\rho) [(\rho^+)^2
-(\rho^-)^2]=0,\\
& \text{etc.}
\lb{7.20}
\end{split}
\end{align}

\begin{rem}\lb{r7.1}
In analogy to Remark~\ref{r2.1} one infers that
$\rho_e$ and $\rho_o$
enter $F_g$, $G_{g+1}$ quadratically so that
the $\km$ hierarchy
\eqref{7.13} is invariant under the substitution
\begin{equation}
\rho (t) \to \rho_\eps (t)
=\{ \eps(n) \rho(n,t)\}_{n\in\bbZ}, \quad
\eps(n) \in\{+1,-1\},\; n\in\bbZ.
\lb{7.21}
\end{equation}
This result should be compared with Lemma~\ref{l8.1}.
\end{rem}

The Miura-type relation between the $\tl$ and $\km$
hierarchies, alluded to at
the beginning of this chapter, is now obtained as
follows.  The connection
between $P_{k,2g+2}(t)$, $k=1,2$ and $Q_{2g+2}(t)$
is clear
from \eqref{7.5}, the corresponding connection
between $L_k(t)$, $k=1,2$
and $M(t)$ is provided by the elementary observation
\begin{equation}
M(t)^2 =\begin{pmatrix}
A(t)^*A(t) & 0\\
0 & A(t) A(t)^*
\end{pmatrix} =L_1(t) \oplus L_2(t),\quad t\in\bbR.
\lb{7.22}
\end{equation}
Moreover, recalling the notation employed in
\eqref{2.12}, \eqref{2.13},
that is,
\begin{equation}
\tl_g (a,b) =(\tl_g(a,b)_1, \tl_g(a,b)_2)^T,
\lb{7.23}
\end{equation}
one can verify that
\begin{equation}
\tl_g (a_k,b_k) =W_k \ukm_g(\rho),\quad k =1,2,
\lb{7.24}
\end{equation}
where $W_k(t)$ denote the matrix-valued
difference expressions
\begin{equation}
W_1(t) = \begin{pmatrix}
\rho_o(t) & \rho_e(t)\\
-2\rho_e(t) & -2\rho_o^- (t) S^-
\end{pmatrix}, \; W_2(t) =\begin{pmatrix}
\rho_o(t) S^+ & \rho_e^+ (t)\\
-2\rho_e(t) & -2\rho_o(t)
\end{pmatrix}.
\lb{7.25}
\end{equation}
Relation \eqref{7.24} is the analog of Miura's
identity \cite{64} between
the KdV and mKdV hierarchy (which extends to GD
and DS systems, see,
e.g., \cite{35}, \cite{36}, \cite{39}, \cite{40},
\cite{41} and the
references therein).  In particular, as
systematically studied by Adler
\cite{1}, the identity \eqref{7.24} yields the
implication
\begin{equation}
\km_g (\rho) =0 \Rightarrow \tl_g (a_k, b_k)=0,
\quad k=1,2
\lb{7.26}
\end{equation}
(since $\km_g(\rho)
=0 \Leftrightarrow \ukm_g(\rho) =0$),
that is, given
a solution $\rho$ of the $\km_g$ equation
\eqref{7.13} (respectively
\eqref{7.17}), one obtains two solutions,
$(a_1, b_1)$ and $(a_2, b_2)$, of
the $\tl_g$ equations \eqref{2.12} related
to each other by the
Miura-type transformations \eqref{7.8},
\eqref{7.9}.  As mentioned
briefly in the Introduction, a connection
between the $\km$ and Toda systems
was known to H\' enon in 1973.  Moreover,
transformations between $\km$ and
Toda lattices were investigated in some
detail by Wadati \cite{83} (see
also \cite{80}). B\"acklund transformations for
the Toda hierarchy and connections with the $\km$
system based on factorization techniques were
also studied by Knill \cite{51}, \cite{51a}.

The main result in \cite{38} describes a method
 to reverse the
implication in \eqref{7.26}, that is, starting from
 a solution, say $(a_1,
b_1)$ of the $\tl_g$ equation \eqref{2.12}, one
constructs a solution
$\rho$ of the $\km_g$ equation \eqref{7.17} and
 another $\tl_g$
solution $(a_2, b_2)$ of \eqref{2.12} related
to each other by the
Miura-type transformation \eqref{7.8},
\eqref{7.9}.  We now recall this
construction.

\begin{thm} \lb{t7.2} \cite{38}
Assume $(a_1, b_1)$ satisfies \eqref{2.7},
$a_1(n,t) <0$, $b_1 (n,t) <
0$, $(n,t) \in\bbZ\times \bbR$, and
$\widetilde{TL}_r (a_1,b_1) =0$ for some
$r\in\bbN_0$.  Suppose the associated
self-adjoint realization $H_1(t)$
in $\ell^2 (\bbZ)$ of $L_1 (t) =a_1(t) S^+
+ a_1^- (t) S^- -b_1(t)$ is
nonnegative for some (and hence for all)
$t\in\bbR$, $H_1 (t) \geq 0$, and
$0< \psi_{1,\pm}(n,.)\in C^1 (\bbR)$,
$n\in\bbZ$ are positive weak
solutions of
\begin{equation}
H_1(t) \psi_{1,\pm} (t) =0,
\quad \dot \psi_{1,\pm} (t) =\ti P_{1,2r+2}
(t) \psi_{1,\pm} (t), \quad t\in\bbR.
\lb{7.27}
\end{equation}
Define for $(n,t) \in\bbZ\times \bbR$,
\begin{align}
\psi_{1,\sig} (n,t) &
= \frac{1-\sig (t)}{2} \psi_{1,-} (n,t) +
\frac{1+\sig (t)}{2} \psi_{1,+} (n,t),
\lb{7.28}\\
\rho_{e,\sig} (n,t) &
= -[-a_1 (n,t) \psi_{1,\sig} (n+1,t) /
\psi_{1,\sig} (n,t) ]^{1/2},
\lb{7.29}\\
\rho_{o,\sig}(n,t) &
= [-a_1(n,t) \psi_{1,\sig} (n,t) / \psi_{1,\sig}
(n+1,t)]^{1/2},
\lb{7.30}\\
\rho_\sig (n,t) &= \begin{cases}
\rho_{e,\sig} (m,t), & n=2m\\
\rho_{o,\sig} (m,t), & n=2m+1
\end{cases},
\lb{7.31}\\
a_{2,\sig} (n,t) &
= \rho_{e,\sig}(n+1,t) \rho_{o,\sig} (n,t),
\lb{7.32}\\
b_{2,\sig} (n,t) &
= -\rho_{e,\sig}(n,t)^2 -\rho_{o,\sig}(n,t)^2,
\lb{7.33}
\end{align}
where $\sig : \bbR \to [-1,1]$,
$\sig \in C^1 (\bbR)$.  Then $\rho_\sig
(t)$, $a_{2,\sig}(t)$, $b_{2,\sig}(t)
\in\ell_\bbR^\infty (\bbZ)$,
$t\in\bbR$, $\rho_\sig (n,t) \neq 0$,
$a_{2,\sig}(n,t) < 0$, $(n,t)
\in\bbZ\times \bbR$ and
\begin{align}
\begin{split}
& \widetilde{KM}_r (\rho_\sig)=0,
\quad \widetilde{TL}_r (a_{2, \sig},
b_{2,\sig})=0\\
& \text{if and only if } \dot \sig =0
 \text{ or } W(\psi_{1,-}, \psi_{1,+}
) =0.
\lb{7.34}
\end{split}
\end{align}
\end{thm}

The proof of Theorem~\ref{t7.2}, according to
\cite{38}, can be reduced
to identities of the form
\begin{align}
\begin{split}
\widetilde{KM}_r (\rho_\sig)(2n,t) &
= - \frac14 \dot \sig (t) \rho_{e,\sig}
(n,t)^{-1} \psi_{1,\sig} (n,t)^{-2} W(\psi_{1,-},
\psi_{1,+}), \\
& \hspace*{6cm} t\in\bbR,
\lb{7.35}\end{split}\\
\begin{split}
\widetilde{KM}_r (\rho_\sig) (2n+1,t) &
= \frac14 \dot \sig (t)
\rho_{o,\sig}(n,t)^{-1}
\psi_{1,\sig} (n+1, t)^{-2} W(\psi_{1,-},
\psi_{1,+}), \\
& \hspace*{6cm} t\in\bbR,
\lb{7.36}
\end{split}
\end{align}
and similarly,
\begin{multline}
\widetilde{TL}_r (a_{2,\sig}, b_{2,\sig}) (n,t)
= \frac14 \dot \sig (t)
W(\psi_{1,-}, \psi_{1,+})\times\\
\times \begin{pmatrix}
\psi_{1,\sig} (n+1, t)^{-2} \left[
-\frac{\rho_{o,\sig} (n,t)}{\rho_{e,\sig} (n+1,t)} +
\frac{\rho_{e,\sig}(n+1,t)}{\rho_{o,\sig} (n,t)} \right] \\
2 [\psi_{1,\sig}(n,t)^{-2}
-\psi_{1,\sig} (n+1, t)^{-2}]
\end{pmatrix}, \quad (n,t) \in\bbZ\times \bbR.
\lb{7.37}
\end{multline}
(To be precise, Theorem~\ref{t7.2} is proved
in \cite{38} in the case
$r=0$, i.e., for the original Toda and $\km$ system.
However, as shown in
\cite{35} and \cite{39} in the case of the
(m)KdV and GD, DS
contexts, the proof directly extends to the
entire hierarchy,
$r\in\bbN_0$.)

Since the cases $\sig=\pm1$ with $\psi_{1,\pm}$
 the branches of the
BA-function associated with the finite-gap
operator $L_1$ will be the
most important ones for us in the following,
we shall identify
$\psi_{1,\pm 1} =\psi_{1, \pm}$, $\rho_{\pm 1}
=\rho_\pm$, $a_{2, \pm 1}
=a_{2, \pm}$, $b_{2,\pm 1} =b_{2, \pm}$, $L_{2, \pm}
=L_{2, \pm}$, $H_{2,
\pm 1} = H_{2, \pm}$, etc.\ for notational
convenience throughout the
rest of this exposition.  Moreover, in the special
case where $H_1$ is
critical (cf.\ Remark~\ref{r7.3}~(ii) below)
and hence $\psi_{1,+}
=\psi_{1,-} \equiv \psi_{1,0}$, we shall write,
$\rho_o$, $a_{2,0}$,
$b_{2,0}$, $L_{2,0}$, $H_{2,0}$, etc.

Theorem~\ref{t7.2} has been used in \cite{38} to
derive the soliton
solutions for the $\km$ system given the corresponding
solitons for the Toda
system.  In our final Chapter~\ref{s9} we shall use
Theorem~\ref{t7.2} to
derive the main objective of this exposition, viz.,
the algebro-geometric
quasi-periodic finite-gap solutions of the
$\km$ hierarchy, given the
corresponding results of Chapters~\ref{s5} and
\ref{s6} for the Toda
hierarchy.

We conclude this chapter with a series of remarks
further illustrating
Theorem~\ref{t7.2}.

\begin{rem} \lb{r7.3}
(i). Due to the Lax equation \eqref{2.11},
$H_1(t)$ is well-known to be
unitarily equivalent to $H_1(0)$ for all
$t\in\bbR$.  (Since
$H_1(t)$ are bounded self-adjoint operators
strongly continuous
with respect to $t\in\bbR$, the Dyson expansion
of the
associated unitary propagator converges in the
uniform operator
topology, see, e.g., Theorem~X.69 in \cite{74}.)
Moreover, the
Wronskian $W(\psi_{1,-}, \psi_{1,+})$ is independent
of $(n,t)
\in\bbZ\times \bbR$.  The existence of positive
solutions $\psi_{1,\pm}$
satisfying \eqref{7.27} has been investigated in
\cite{38} and \cite{42}.
In our application of Theorem~\ref{t7.2} in
Chapter~\ref{s9},
$\psi_{1,\pm}$ will be the branches of the
BA-function $\psi(P)$ and
positivity of $\psi_{1,\pm}$ can be verified
directly.\\
(ii). Depending on whether or not $H_1(0)$ (and
hence $H_1(t)$ for all $t
\in\bbR$, see \cite{42}) is critical or subcritical,
that is, whether or not
$H_1(0)\psi=0$ has a unique positive solution (up to
constant multiples)
or two linearly independent positive solutions,
Theorem~\ref{t7.2} yields
a unique solution, $\rho_0$ (respectively $a_{2,0}$,
$b_{2,0}$) or a
one-parameter family $\rho_\sig$ (respectively $a_{2,\sig}$,
$b_{2,\sig}$)
indexed by $\sig\in [-1, 1]$ ($\sig$ being
$t$-independent) of $\widetilde{KM}_r$
(respectively $\widetilde{TL}_r$) solutions.  Since $H_1(0) \geq 0$
implies the
existence of at least one weak positive solution
$\psi$ of $H_1 (0)
\psi=0$ (cf., e.g., \cite{42} and the references
therein) this case
distinction is exhaustive.  In addition,
$(a_1, b_1)$ and $\rho_\sig$,
$(a_{2,\sig}, b_{2,\sig})$ are all related by
the Miura-type
transformation \eqref{7.8}, \eqref{7.9}.\\
(iii). By Remark~\ref{r2.1} and Lemma~\ref{l4.1} we
assumed $a_1 < 0$
without loss of generality in Theorem~\ref{t7.2}.
The existence of weak
solutions $\psi_{1, \pm} > 0$ satisfying
$H_1 \psi_{1,\pm} =0$ then
necessarily yields $b_1 < 0$.\\
(iv). If $(a_1, b_1)$ are periodic (respectively quasi-periodic
finite-gap in
the sense of Chapter~\ref{s6}) then $\rho_\sig$ and $(a_{2,\sig},
b_{2,\sig})$ are periodic (respectively quasi-periodic
finite-gap) if and only
if $\sig =\pm 1$ (or if $H_1 (0)$ is critical).
This will be the case in
our final Chapter~\ref{s9} where we construct the
algebro-geometric
quasi-periodic finite-gap solutions of the $\km$
 hierarchy.
\end{rem}

The stationary $\km$ hierarchy is
characterized by $\dot
\rho=0$ in \eqref{7.13} (respectively \eqref{7.17}),
or more precisely, by
commuting matrix difference expressions of
the type
\begin{equation}
[Q_{2g+2}, M]=0.
\lb{7.38}
\end{equation}
In the special case where $L_1$ and $L_2$ are isospectral in
the sense
that the corresponding Burchnall-Chaundy polynomials
\eqref{2.37} coincide,
the analogs of \eqref{2.36} and \eqref{2.37}
then read
\begin{align}
Q_{2g+2}^2 & =
\prod_{m=0}^{2g+1} (M-E_m^{1/2}) (M+E_m^{1/2})
=\prod_{m=0}^{2g+1} (M^2 -E_m)
\lb{7.39}\\
\intertext{and}
y^2 & = \prod_{m=0}^{2g+1} (w-E_m^{1/2})
 (w+E_m^{1/2})
=\prod_{m=0}^{2g+1} (w^2 -E_m).
\lb{7.40}
\end{align}
We note that the curve \eqref{7.40}
becomes singular if
and only if $E_m=0$ for some $0 \leq m \leq 2g+1$.
In the self-adjoint
case where $0 \leq E_0 < E_1 < \cdots < E_{2g+1}$
this happens if and
only if $E_0 =0$ (i.e., if and only if $H_1$ and
hence $H_2$ are
critical).


\chapter{Spectral Theory for Finite-Gap
Dirac-Type Difference Operators}
\lb{s8}
\setcounter{prop}{0}
\setcounter{equation}{0}

In this chapter we briefly study spectral
properties of self-adjoint
$\ell^2 (\bbZ) \otimes \bbC^2$ realizations
associated with finite-gap
Dirac-type difference expressions.

Assuming $\rho\in\ell_\bbR^\infty (\bbZ)$,
$\rho (n) \neq 0$, $n\in\bbZ$
we start by introducing the general matrix-valued
difference expression
$M$ by
\begin{align}
\begin{split}
& M=\begin{pmatrix}
0 & A^*\\
A & 0
\end{pmatrix}, \quad A=\rho_o S^+ +\rho_e,
\quad A^* =\rho_o^- S^- +
\rho_e,\\
& \rho_e(n) =\rho(2n), \quad \rho_o(n)
=\rho(2n+1), \quad n\in\bbZ.
\lb{8.1}
\end{split}
\end{align}
We denote by $D$ the unique self-adjoint
realization associated with $M$
in $\ell^2 (\bbZ) \otimes \bbC^2$,
\begin{equation}
Df=Mf,\quad f\in\calD(D) =\ell^2 (\bbZ) \otimes \bbC^2.
\lb{8.2}
\end{equation}
The analog of Lemma~\ref{l4.1} then reads

\begin{lem} \lb{l8.1}
Let $\rho\in\ell_\bbR^\infty (\bbZ)$,
$\rho(n) \neq 0$, $n\in\bbZ$ and
introduce $\rho_\eps \in\ell_\bbR^\infty (\bbZ)$ by
\begin{equation}
\rho_\eps =\{ \eps (n)\rho(n) \}_{n\in\bbZ},
\; \eps (n) \in \{+1,
-1\}, \quad n\in\bbZ.
\lb{8.3}
\end{equation}
Define $D_\eps$ in $\ell^2 (\bbZ) \otimes \bbC^2$
as in \eqref{8.2} with
$M$ replaced by $M_\eps =\left(\begin{smallmatrix}
0 & A^*_\eps\\
A_\eps & 0
\end{smallmatrix}\right)$ and $\rho$ by
$\rho_\eps$.  Then $D$ and
$D_\eps$ are unitarily equivalent, that is, there
exists a unitary operator
$U_\eps$ in $\ell^2 (\bbZ) \otimes \bbC^2$ such
that
\begin{equation}
D=U_\eps D_\eps U_\eps^{-1}.
\lb{8.4}
\end{equation}
\end{lem}

\begin{proof}
$U_\eps$ is explicitly represented by
\begin{equation}
U_\eps =\begin{pmatrix}
U_{1,\eps} & 0\\
0 & U_{2,\eps} \end{pmatrix}, \; U_{k,\eps}
=(\ti \eps_k (n)
\del_{m,n})_{m,n\in\bbZ}, \quad k=1,2,
\lb{8.5}
\end{equation}
$\ti \eps_1 (n+1) \ti \eps_2 (n)
=\eps (2n+1)$, $\ti \eps_1(n) \ti
\eps_2(n) =\eps (2n)$,
$\ti \eps_1 (n) \ti \eps_2 (n-1) =\eps (2n-1)$,
$n\in\bbZ$.
\end{proof}

Next, we summarize the spectral properties of
$H_1$, $H_{2, \sig}$ and
$D_\sig$.  Since in this exposition we are interested
in quasi-periodic
finite-gap operators we shall restrict ourselves to
the case $\sig =\pm
1$ (this includes the case where $H_1$ (and hence
$H_{2, \sig}$) is critical
and hence $D_+ =D_- \equiv D_0$ as a limiting case).
To fix our notation
assume $a_1$, $b_1\in\ell_\bbR^\infty (\bbZ)$,
$a_1(n) < 0$, $b_1(n) <
0$, $n\in\bbZ$ and define $L_1$, $H_1$ as in
\eqref{4.1}, \eqref{4.2}.
Assuming $H_1 \geq 0$ let $\psi_1 > 0$ be a weak
solution of $L_1 \psi_1
=0$ and define
\begin{align}
\rho_e (n) & =
-[-a_1(n) \psi_1(n+1) / \psi_1 (n) ]^{1/2},
\lb{8.6}\\
\rho_o (n) &
= [-a_1(n) \psi_1(n) / \psi_1 (n+1) ]^{1/2},
\lb{8.7}\\
\rho (n) & = \begin{cases}
\rho_e(m), & n=2m\\
\rho_o(m), & n=2m+1
\end{cases},
\lb{8.8}\\
a_2(n) & = \rho_e(n+1) \rho_o (n),
\lb{8.9}\\
b_2(n) & = -\rho_e (n)^2 -\rho_o (n)^2.
\lb{8.10}
\end{align}
Given \eqref{8.6}--\eqref{8.10} one defines
$L_2$, $H_2$, $A$, $M$, and
$D$ as in \eqref{4.1}, \eqref{4.2}, \eqref{7.3},
\eqref{8.1}, and
\eqref{8.2}.

\begin{thm} \lb{t8.2}
Suppose $a_1$, $b_1\in\ell_\bbR^\infty (\bbZ)$ are
 $g$-gap sequences
satisfying \eqref{3.18}, \eqref{3.19}, and
$a_1 (n) < 0$, $b_1(n) < 0$,
$n\in\bbZ$.  Define $L_1$, $H_1$ as in \eqref{4.1},
\eqref{4.2}, suppose
$H_1 \geq 0$, and let $\psi_{1,\pm }(n)
=\psi_{1,\pm} (0,n,n_0)$ be the
branches of the stationary BA-function $\psi_1 (Q_0, n, n_0)$,
$Q_0 =(0,
R_{2g+2}(0)^{1/2})$ of $L_1$ in \eqref{3.21}
(respectively \eqref{3.66}).  Then
\begin{equation}
\psi_{1, \pm} (n) > 0, \quad n\in\bbZ
\lb{8.11}
\end{equation}
and we may define $\rho_{e,\pm}$, $\rho_{o,\pm}$,
$\rho_\pm$,
$a_{2,\pm}$, $b_{2,\pm}$, $L_{2,\pm}$, $H_{2,\pm}$,
$A_\pm$, $M_\pm$, and
$D_\pm$ as in \eqref{8.6}--\eqref{8.10},
\eqref{4.1}, \eqref{4.2},
\eqref{7.3}, \eqref{8.1}, and \eqref{8.2}.
Then $a_{2,\pm},\, b_{2,\pm},
\, \rho_\pm \in\ell_\bbR^\infty (\bbZ)$,
$a_{2\pm} (n) < 0,\, \rho_\pm
(n) \neq 0$, $n\in\bbZ$ and
\begin{align}
\sig (H_1) & = \sig(H_{2, \pm})
= \bigcup_{j=0}^g [E_{2j}, E_{2j+1}],
\quad E_0 \geq 0,
\lb{8.12}\\
\begin{split}
\sig (D_\pm) &
= \bigcup^{g+1}_{\substack{ j=-g-1\\ j\neq 0 }}
\sum\nolimits_j \; ,\\
\sum\nolimits_j & = [E_{2(j-1)}^{1/2}, E_{2j-1}^{1/2}],
\; \sum\nolimits_{-j}
=-\sum\nolimits_j \; ,
\quad 1 \leq j \leq g+1,
\lb{8.13}
\end{split}
\end{align}
\begin{equation}
\sig_{sc} (H_1) =\sig_{sc} (H_{2,\pm})
=\sig_{sc} (D_\pm) =\sig_p (H_1)
=\sig_p (H_{2,\pm}) =\sig_p (D_\pm) =\emptyset.
\lb{8.14}
\end{equation}
In addition, $H_1,\,
H_{2,\pm}$, and $D_\pm$ all have uniform
spectral multiplicity two.
\end{thm}

\begin{proof}
By Lemma~\ref{l3.3}~(i) and (iv),
$\psi_1 (P,n,n_0)$ given by
\eqref{3.66} satisfies
\begin{equation}
\psi_1(P,n,n_0) > 0 \text{ for } \ti\pi (P) \leq E_0.
\lb{8.15}
\end{equation}
Hence one infers $a_{2,\pm}, \, b_{2,\pm},\,
\rho_\pm\in\ell_\bbR^\infty
(\bbZ)$.  Theorem~\ref{t4.2}, the spectral
theorem, and the identities
\begin{align}
& D^2_\pm = H_1 \oplus H_{2,\pm},
\lb{8.16}\\
& \sig_3 D_\pm \sig_3^{-1}
=-D_\pm,\quad \sig_3 =\begin{pmatrix}
1 & 0\\
0 & -1
\end{pmatrix}
\lb{8.17}
\end{align}
then prove \eqref{8.12}--\eqref{8.14}.
\end{proof}

We note that the spectral gap
$(-E_0^{1/2}, E_0^{1/2})$ of $D_\pm$
``closes'' if and only if $H_1$ is critical,
that is, if and only if $E_0
=0$.

For $\sig \in (-1,1)$ one can show in contrast
to \eqref{8.14} that
\begin{equation}
\sig_p (H_{2,\sig}) =\sig_p (D_\sig) =\{0\},
\quad \sig \in (-1,1).
\lb{8.18}
\end{equation}
In fact, using identities of the type
\begin{equation}
(D-w)^{-1} =\begin{pmatrix}
w(H_1-w^2)^{-1} & A^* (H_2 -w^2)^{-1}\\
A(H_1-w^2)^{-1} & w(H_2-w^2)^{-1}
\end{pmatrix},\;
w^2 \in\bbC \bs \{ \sig (H_1) \cup \sig (H_2)\},
\lb{8.19}
\end{equation}
one can reduce the spectral analysis of general
 Dirac-type operators $D$
to that of $H_1$ and $H_2$.  Moreover, noting
that $H_1$ and $H_2$ are
essentially isospectral, that is,
\begin{equation}
\sig (H_1) \bs \{0\} =\sig (H_2) \bs \{0\}
\lb{8.20}
\end{equation}
and that $H_1 |_{\ker (H_1)^\perp}$ and
$H_2 |_{\ker (H_2)^\perp}$ are
unitarily equivalent,
a complete spectral analysis of $D$ in terms
of that of $H_1$ and $\ker
(H_2)$ can be given.  Since here we are mainly
concerned with finite-gap
operators $H_k$, $k=1,2$ and $D$, these
considerations are beyond the
scope of this exposition.  The interested reader
may find an exhaustive
discussion of such topics, for instance, in \cite{20},
\cite{35}, \cite{40},
\cite{41}, \cite{79}.


\chapter[Quasi-Periodic Solutions of
the Kac-van Moerbeke Hierarchy]{Quasi-Periodic
Finite-Gap Solutions of
the\\ Kac-van Moerbeke Hierarchy} \lb{s9}
\setcounter{prop}{0}
\setcounter{equation}{0}

In this final chapter we shall complete our main
goal and construct the
algebro-geometric quasi-periodic finite-gap
solutions of the Kac-van
Moerbeke hierarchy.  Given the extensive
preparations in
Chapters~\ref{s3}--\ref{s8}, our final task
will be relatively
straightforward.

We start with some notations.  Let $(a_1, b_1)$,
$a_1(n)<0$, $b_1(n) <
0$, $n\in\bbZ$ be the stationary $g$-gap
solution \eqref{3.18},
\eqref{3.19} and denote the corresponding
Dirichlet eigenvalues and
divisor by $\{\mu_{1,j}(n)\}_{1\leq j \leq g}$
and $\calD_{\humu_1 (n)}$
etc.  Given $(a_1, b_1)$,
$\{\hat\mu_j (n)\}_{1\leq j\leq g}$ we define
$L_1$, $H_1$, $\phi_1 (P,n)$,
$\psi_1 (P,n,n_0)$, $\uz_1 (P,n)$, and
$\uz_1(n)$ as in \eqref{4.1}, \eqref{4.2},
\eqref{3.20}, \eqref{3.21}
(respectively \eqref{3.65}--\eqref{3.68}), \eqref{3.34},
and \eqref{3.35}).
Next, identifying the branches
$\psi_{1,\pm} (0,n,n_0)$ of $\psi_1 (Q_0,
n,n_0)$, $Q_0 =(0,R^{1/2}_{2g+2}(Q_0))$ with
$\psi_{1,\pm}(n)$ in
Theorem~\ref{t7.2}, and noticing
\begin{equation}
\psi_1 (P,n,n_0) > 0,\quad \ti\pi (P) \leq E_0
\lb{9.1}
\end{equation}
as a consequence of Lemma~\ref{l3.3}~(i), (iv)
and
Theorem~\ref{t3.4}~(i), enables one to construct
$(a_{2,\pm},
b_{2,\pm})$, $\rho_\pm$ as in Theorem~\ref{t8.2}.
For convenience we
list these formulas below.
\begin{align}
\begin{split}
\rho_{e,\pm}(n) & =
 -[-a_1 (n) \psi_{1,\pm}(n+1) / \psi_{1,\pm} (n)
]^{1/2},\\
\rho_{o,\pm} (n) &
= [-a_1(n) \psi_{1,\pm}(n) / \psi_{1,\pm}
(n+1)]^{1/2},
\lb{9.2}
\end{split}\\
\rho_\pm (n) & = \begin{cases}
\rho_{e,\pm}(m), & n=2m\\
\rho_{o,\pm}(m), & n=2m+1
\end{cases},
\lb{9.3}\\
a_{2,\pm} (n) &
= \rho_{e,\pm} (n+1) \rho_{o,\pm} (n),
\lb{9.4}\\
b_{2,\pm} (n) & =
-\rho_{e,\pm}(n)^2 -\rho_{o,\pm}(n)^2.
\lb{9.5}
\end{align}
Given $(a_{2,\pm}, b_{2,\pm})$, $\rho_\pm$ one
then defines $L_{2,\pm}$,
$H_{2,\pm}$, $\phi_{2,\pm}(P,n),\,
\psi_{2,\pm}(P,n,n_0)$, $A_\pm$, and
finally $M_\pm,\, D_\pm$ as in the context of
Theorem~\ref{t8.2} using
\eqref{4.1}, \eqref{4.2}, \eqref{3.20},
\eqref{3.21} (respectively
\eqref{3.65}--\eqref{3.68}), \eqref{7.3}, and
\eqref{8.1}, \eqref{8.2}.
Moreover, defining
\begin{equation}
\phi_{1,\pm} (n) =-\rho_{e,\pm} (n) / \rho_{o,\pm}(n),
\quad n \in\bbZ,
\lb{9.6}
\end{equation}
one verifies
\begin{equation}
a_1 \phi_{1,\pm} +(a_1^- / \phi_{1,\pm}^-) =b_1,
\quad \phi_{1,\pm} > 0
\lb{9.7}
\end{equation}
and a comparison with the Riccati-type equation
\eqref{3.22} then yields
that
\begin{equation}
\phi_{1,\pm}(n) =\phi_{1,\pm} (0,n)
\lb{9.8}
\end{equation}
are the branches of $\phi_1 (Q_0, n)$,
$Q_0 =(0, R^{1/2}_{2g+2} (Q_0))$.
In particular, \eqref{3.21} implies
\begin{equation}
\psi_{1,\pm} (n) =\begin{cases}
\prod_{m=n_0}^{n-1} \phi_{1,\pm} (n),
& n\geq n_0 +1\\
1, & n=n_0\\
\prod_{m=n}^{n_0-1} \phi_{1,\pm} (n)^{-1},
& n\leq n_0 -1
\end{cases},
\lb{9.9}
\end{equation}
and
\begin{align}
L_1\psi_{1,\pm} & = 0, \quad \psi_{1,\pm} > 0
\lb{9.10}\\
\intertext{since}
\psi_{1,\pm} (n) & = \psi_{1,\pm} (0,n,n_0)
\lb{9.11}
\end{align}
are  the branches of $\psi_1 (Q_0, n, n_0)$,
$Q_0 =(0, R^{1/2}_{2g+2} (Q_0))$.  Next, defining
\begin{align}
\phi_{2,\pm,\mp} (n) & =
-\rho_{0,\pm}(n) / \rho_{e,\pm} (n+1), \quad
n\in\bbZ
\lb{9.12}\\
\intertext{and}
\psi_{2,\pm,\mp} (n) & = \begin{cases}
\prod_{m=n_0}^{n-1} \phi_{2,\pm,\mp} (n),
 & n \geq n_0 +1\\
1, & n=n_0\\
\prod_{m=n}^{n_0-1} \phi_{2,\pm,\mp} (n)^{-1},
 & n\leq n_0 -1
\end{cases},
\lb{9.13}
\end{align}
one verifies
\begin{equation}
a_{2,\pm} \phi_{2,\pm,\mp}
+(a_{2,\pm}^- / \phi_{2,\pm,\mp}) =b_{2,\pm}
\lb{9.14}
\end{equation}
and
\begin{equation}
L_{2,\pm} \psi_{2,\pm, \mp}=0.
\lb{9.15}
\end{equation}

In order to derive the $\theta$-function representation
for $(a_{2,\pm},
b_{2,\pm})$ and especially for $\rho_\pm$, we first
recall that for
$(a_1, b_1)$, $\phi_1(Q_0)$, $\psi_1(Q_0)$ from
Theorems~\ref{t3.4} and
\ref{t5.2},
\begin{align}
& a_1(n)  = \ti a [\theta (\uz_1 (n+1) )
\theta(\uz_1(n-1))
/ \theta (\uz_1(n))^2]^{1/2}, \lb{9.16}\\
& b_1(n)  = \sum_{j=1}^g \lam_j
-\frac12 \sum_{m=0}^{2g+1} E_m
-\sum_{j=1}^g c_j (g) \dfrac{\pa}{\pa w_j} \ln\left.
\left[ \dfrac{\theta
(\uw +\uz_1(n))}{\theta (\uw
+\uz_1 (n-1))} \right] \right|_{\uw=\underline{0}},
\lb{9.17}\\
& \phi_1 (Q_0, n)  = \left[ \dfrac{\theta (\uz_1 (n-1))}
{\theta (\uz_1
(n+1))} \right]^{1/2} \dfrac{\theta (\uz_1(Q_0, n+1))}
{\theta (\uz_1
(Q_0, n))} \exp \left( \int_{P_0}^{Q_0} \ome_{\infty_+,
\infty_-}^\bth
\right),
\lb{9.18}\\
\begin{split}
& \psi_1 (Q_0, n,n_0) =
C(n,n_0) \dfrac{\theta (\uz_1 (Q_0, n))}{\theta
(\uz_1 (Q_0, n_0))}
\exp \left[ (n-n_0) \int_{P_0}^{Q_0} \ome_{\infty_+,
\infty_-}^\bth \right],\\
& \uz_1 (P,n)  = \underline{\hat A}_{P_0}(P) -\hual_{P_0}
(\calD_{\humu_1(n)}
)
-\huxi_{P_0}, \quad \uz_1 (n) =\uz_1 (\infty_+, n),\\
& \hspace*{6cm} Q_0  =(0, R^{1/2}_{2g+2}(Q_0)).
\lb{9.19}
\end{split}
\end{align}
Here $C(n,n_0)$ is defined in \eqref{3.67},
\eqref{3.68} and the square
roots in \eqref{9.16}, \eqref{9.18} and hence in
\begin{equation}
C(n) = C(n+1, n) =\left[ \dfrac{\theta (\uz_1(n-1))}
{\theta (\uz_1(n+1))}
\right]^{1/2}
\lb{9.20}
\end{equation}
are all positive (cf.~\eqref{3.57}, \eqref{3.69},
\eqref{3.75}, and $\ti a < 0$).

In the following we explicitly need the branches
$\phi_{1,\pm} (z,n)$ of
$\phi_1 (P,n)$.  To fix notations we abbreviate
\begin{align}
\begin{split}
A_{E_0}^+ (z) & = \int_{P_0}^Q \ul{\ome},
\; \int_{E_0}^z \ome_{\infty_+,
\infty_-}^{\bth +} = \int_{P_0}^Q \ome_{\infty_+,
\infty_-}^\bth < 0,\\ Q & =
(z, R_{2g+2} (z)^{1/2} ) \in\Pi_+, \; \ti\pi (Q)
=z\leq E_0,
\lb{9.21}
\end{split}
\end{align}
where the path of integration from $P_0$ to $Q$ is
along the lift of the
straight line segment from $E_0$ to $z(\leq E_0)$.
(As in the proof of
Lemma~\ref{l3.3}, whenever the integration path meets
the cycle $b_k$ we
first move along $b_k$ until we hit the intersection
point with $a_k$.
Then we follow $a_k$ and return on the other side of
$b_k$ before we
continue the straight line path.  The contributions
on $b_k$ cancel and
the contribution from $a_k$ is irrelevant in
\eqref{9.24} below due to
the $\bbZ^g$-periodicity of the Riemann theta function
in \eqref{a.28}
and the normalization \eqref{3.41} of
$\ome_{\infty_+, \infty_-}^\bth$.)
Moreover, we use the notation
\begin{align}
\uz_{1,\pm} (z,n) & = \pm \ua_{E_0}^+ (z)
-\hual_{P_0}
(\calD_{\humu_1(n)}) -\huxi_{P_0},
\lb{9.22}\\
\uz_{1,\pm} (n) & = \uz_{1,\pm} (-\infty,n),
\; \uz_{1,+} (n) =\uz_1(n)
\lb{9.23}
\end{align}
for the branches of $\uz_1(P,n)$, $\ti\pi (P)
=z\leq E_0$.  The branches
$\phi_{1,\pm}(z,n)$ of $\phi_1(P,n)$ then
read explicitly,
\begin{align}
\phi_{1,\pm} (z,n) & =
\left[ \dfrac{\theta(\uz_1(n-1))}{\theta(\uz_1
(n+1))} \right]^{1/2} \dfrac{\theta(\uz_{1,\pm} (z,n+1))}
{\theta
(\uz_{1,\pm}(z,n))} e^{\pm \int_{E_0}^z \ome_{\infty_+,
\infty_-}^{\bth
+}}.
\lb{9.24}\\
\intertext{Together with \eqref{9.2},
\eqref{9.12}, and}
\phi_{1,\pm}(z,n) & =
\psi_{1,\pm}(z,n+1)/ \psi_{1,\pm} (z,n)
\lb{9.25}
\end{align}
this allows one to compute
\begin{align}
\begin{split}
\rho_{e,\pm} (n) & =
-[-a_1(n) \phi_{1,\pm}(0,n)]^{1/2}\\
& = -\left[ - \ti a \dfrac{\theta (\uz_1 (n-1))
\theta (\uz_{1,\pm}
(0,n+1))}{\theta (\uz_1(n))
\theta (\uz_{1,\pm} (0,n))} \right]^{1/2}
e^{\pm \frac12 \int_{E_0}^0 \ome_{\infty_+,
\infty_-}^{\bth +}},
\lb{9.26}\end{split}\\
\begin{split}
\rho_{o,\pm} (n) &
= [-a_1(n) \phi_{1,\pm} (0,n)^{-1}]^{1/2}\\
& = \left[-\ti a \dfrac{\theta (\uz_1 (n+1))
\theta(\uz_{1,\pm}
(0,n))}{\theta (\uz_1(n))
\theta (\uz_{1,\pm} (0, n+1))} \right]^{1/2}
e^{\mp \frac12 \int_{E_0}^0 \ome_{\infty_+,
\infty_-}^{\bth+}},
\lb{9.27}
\end{split}\\
\begin{split}
\phi_{2,\pm,\mp}(n) & =
-\rho_{o,\pm}(n) / \rho_{e,\pm} (n+1)\\
& = \left[ \dfrac{\theta (\uz_{1,\pm} (0,n))}
{\theta (\uz_{1,\pm}
(0,n+2))} \right]^{1/2} \dfrac{\theta(\uz_1 (n+1))}
{\theta (\uz_1(n))}
e^{\mp \int_{E_0}^0 \ome_{\infty_+,
\infty_-}^{\bth +}}.
\lb{9.28}
\end{split}
\end{align}
Next we define
\begin{align}
\uz_{2,\pm} (P,n) & = \hua_{P_0} (P)
-\hual_{P_0} (\calD_{\humu_{2,\pm}
(n)}) -\huxi_{P_0},
\lb{9.29}\\
\uz_{2,\pm}(n) & = \uz_{2,\pm} (\infty_+, n),
\lb{9.30}
\end{align}
where
\begin{align}
\begin{split}
& \ual_{P_0} (\calD_{\humu_{2,\pm}(n)})
=\ual_{P_0} (\calD_{\humu_1(n)})
-\eps_\pm \ua_{P_0} (Q_0) -\ua_{P_0} (\infty_+),\\
& \eps_+ =-\eps_- =\pm1 \text{ for } Q_0 \in\Pi_\pm,
\; Q_0 =(0,
R^{1/2}_{2g+2} (Q_0))
\lb{9.31}
\end{split}
\end{align}
describes the connection between the Dirichlet
divisors
$\calD_{\humu_{2,\pm}(n)}$ of $H_{2,\pm}
=A_\pm A_\pm^*$ and
$\calD_{\humu_1(n)}$ of $H_1=A_\pm^* A_\pm$.
(In the special case where
$H_1$ is critical and hence $E_0 =0$, that is,
$P_0 =Q_0$, one obtains $\hat
\mu_{2,+,j} (n) =\hat \mu_{2,-,j}(n) \equiv
\hat \mu_{2,0,j}(n)$, $1\leq j
\leq g$ and the sign ambiguity in \eqref{9.31}
vanishes.)  The branches
of $z_{2,\pm} (P,n)$ are then denoted by
\ba
\uz_{2,\eps, \eps'} (z,n) &=&
 \eps' \ua_{E_0}^+ (z) -\hual_{P_0}
(\calD_{\humu_{2,\eps}(n)}) -\huxi_{P_0},
\lb{9.32}\\
\uz_{2,\eps,\eps'} (n) &=&
 \uz_{2,\eps,\eps'} (-\infty, n), \;
\uz_{2,\eps, +} (n) =\uz_{2,\eps}(n),
\:\: \eps,\eps' \in \{+,-\},
\lb{9.33}
\ea
and $\phi_{2,+,-}$, $\phi_{2,-,+}$ in \eqref{9.28}
are seen to be
branches of the following meromorphic function
$\phi_{2,\pm} (P,n)$ on
$K_g$,
\begin{equation}
\phi_{2,\pm} (P,n)
=\left[ \dfrac{\theta (\uz_{2,\pm} (n-1))}{\theta
(\uz_{2,\pm} (n+1))} \right]^{1/2}
 \dfrac{\theta (\uz_{2,\pm}
(P,n+1))}{\theta (\uz_{2,\pm} (P,n))}
 e^{\int_{P_0}^P \ome_{\infty_+,
\infty_-}^\bth},
\lb{9.34}
\end{equation}
by noticing that
\begin{equation}
\uz_{2,\pm,\mp} (0,n) =\uz_{1,+} (n)
=\uz_1 (n), \; z_{2,\pm,+} (n-1)
=\uz_{2,\pm}(n-1) =\uz_{1,\pm}(0,n).
\lb{9.35}
\end{equation}
In particular, we may rewrite and extend
\eqref{9.28} in the form
\begin{equation}
\phi_{2,\eps,\eps'} (n)
=\left[ \dfrac{\theta (\uz_{2,\eps}(n-1))}{\theta
(\uz_{2,\eps}(n+1))} \right]^{1/2} \dfrac{\theta
(\uz_{2,\eps,-\eps'}(n+1))}
{\theta (\uz_{2,\eps,-\eps'}(n))} e^{\eps'
\int_{E_0}^0 \ome_{\infty_+,
\infty_-}^{\bth +}},\;
\eps, \eps' \in \{+,-\}.
\lb{9.36}
\end{equation}
The divisor of $\phi_{2,\pm} (P,n)$ thus reads
\begin{equation}
(\phi_{2,\pm} (.,n))=\calD_{\humu_{2,\pm}(n+1)}
-\calD_{\humu_{2,\pm}(n)}
+\calD_{\infty_+} -\calD_{\infty_-}
\lb{9.37}
\end{equation}
in analogy to that of $\phi_1(P,n)$
(cf.~\eqref{3.30})
\begin{equation}
(\phi_1(.,n))=\calD_{\humu_1(n+1)}
-\calD_{\humu_1(n)} +\calD_{\infty_+}
-\calD_{\infty_-}.
\lb{9.38}
\end{equation}

Given \eqref{9.36} respectively \eqref{9.26}--\eqref{9.33}
we can now express
$(a_{2,\pm}, b_{2,\pm})$ and $\rho_\pm$ in terms of
$\theta$-functions as
follows.

\begin{thm} \lb{t9.1}
Let $(a_1, b_1)$ in \eqref{9.16}, \eqref{9.17} be
the $g$-gap sequences
associated with $H_1=A^*_\pm A_\pm$.  Then the
sequences $(a_{2,\pm},
b_{2,\pm})$ associated with $H_{2,\pm}
=A_\pm A_\pm^*$ are explicitly
given by
\begin{align}
a_{2,\pm}(n) & =
\ti a [\theta (\uz_{2,\pm} (n+1) )\theta (\uz_{2,\pm}
(n-1)) / \theta (\uz_{2,\pm}(n))^2 ]^{1/2},
\lb{9.39}\\
b_{2,\pm}(n) & =
\sum_{j=1}^g \lam_j -\frac12 \sum_{m=0}^{2g+1} E_m
-\sum_{j=1}^g c_j (g) \dfrac{\pa}{\pa w_j} \ln \left.
\left[ \dfrac{
\theta
(\uw
+\uz_{2,\pm}(n))}{\theta (\uw
+\uz_{2,\pm} (n-1))} \right] \right|_{\uw
=\uzero}.
\lb{9.40}
\end{align}
In particular, $H_{2,\pm}$ are isospectral to
$H_1$ and hence
$(a_{2,\pm}, b_{2,\pm})$ are $g$-gap sequences
associated with the same
hyperelliptic curve $K_g$ as $(a_1, b_1)$ and
with nonspecial Dirichlet
divisors $\calD_{\humu_{2,\pm}(n)}$ satisfying
\begin{equation}
\mu_{2,\pm,j}(n)
=\ti\pi (\hat\mu_{2,\pm,j} (n))\in [E_{2j-1},
E_{2j}],\quad 1\leq j \leq g,\; n\in\bbZ
\lb{9.41}
\end{equation}
and
\begin{align}
\begin{split}
 \ual_{P_0} (\calD_{\humu_{2,\pm}(n)}) &
=\ual_{P_0} (\calD_{\humu_1(n)})
\mp \ua_{E_0}^+ (0) -\ua_{P_0} (\infty_+)\\
& = \ual_{P_0} (\calD_{\humu_1(n_0)})-2
 (n-n_0) \ua_{P_0}(\infty_+) \mp
\ua_{E_0}^+ (0) -\ua_{P_0} (\infty_+).
\lb{9.42}
\end{split}
\end{align}
Moreover, we have $H_{2,+} =H_{2,-}$, $a_{2,+}
=a_{2,-}$, $b_{2,+}
=b_{2,-}$, $\hat\mu_{2,+,j} =\hat\mu_{2,-,j}$,
$1\leq j \leq g$, etc.\ if
and only if $H_1$ is critical, that is, if and only
if $E_0 =0$ (i.e., $P_0
=Q_0$).
\end{thm}

\begin{proof}
Equation \eqref{9.39} is clear from \eqref{9.4},
\eqref{9.26}, \eqref{9.27},
and \eqref{9.35}.  That $H_{2,\pm}$ are isospectral
to $H_1$ has been
proven in Theorem~\ref{t8.2} and the nonspeciality of
$\calD_{\humu_{2,\pm}(n)}$ (cf.~Lemma~\ref{la.2})
together with
\eqref{9.41} is a consequence of
Lemma~\ref{l3.3}~(ii) and the fact that
\begin{equation}
i \Im [\hua_{P_0} (Q_0)] = i \Im [\hua_{P_0}(\infty_+)]
=\uzero \mod (L_g).
\lb{9.43}
\end{equation}
Equation \eqref{9.42} directly follows from \eqref{9.31}
and \eqref{9.21}.
Equation \eqref{9.40} for $b_{2,\pm}$ finally can be
derived from an expansion
of $\phi_{2,\pm}(P,n)$ near $P=\infty_+$ exactly as
in the proof of
Theorem~\ref{t5.2}.
\end{proof}

\begin{rem} \lb{r9.2}
Equations \eqref{9.31} respectively \eqref{9.42} illustrate the
effect of
commutation (i.e., $H_1 =A_\pm^* A_\pm \to H_{2,\pm}
= A_\pm A_\pm^*$) as
translations by $\mp \ua_{E_0}^+ (0)
-\ua_{P_0} (\infty_+) =-\eps_\pm
\ua_{P_0} (Q_0)-\ua_{P_0} (\infty_+)$,
$\eps_+=-\eps_- =\pm 1$ for
$Q\in\Pi_\pm$ on the Jacobi variety.  This
clearly resembles the
differential operator case pioneered by Burchnall
and Chaundy \cite{16},
\cite{17} and put into the context of B\" acklund
transformations for the
KdV equation in \cite{27}, \cite{29}, \cite{36},
\cite{41} and discussed
in connection with the spectral theory of Hill's
equation in \cite{60}--\cite{62}.
\end{rem}

Depending on whether or not $H_1$ (and hence
$H_{2,\pm}$) is critical,
that is, whether or not $E_0 =0$, the corresponding
Dirac-type operator
$D_\pm =\left( \begin{smallmatrix} 0 & A^*_\pm\\
A_\pm & 0
\end{smallmatrix} \right)$ in Theorem~\ref{t8.2}
has a spectral gap
containing 0 and hence altogether $2g+1$ spectral
gaps (if $H_1$,
$H_{2,\pm}$ are subcritical, that is, if $E_0 > 0$) or
 precisely $2g$
spectral gaps (if $H_1$, $H_{2,\pm}$ are critical,
i.e., if $E_0 =0$).
Accordingly we call the corresponding sequence
$\rho_\pm$
(respectively~$\rho_0$) a $(2g+1)$-gap (respectively~$2g$-gap)
sequence associated
with
$D_\pm$ (respectively $D_0$).  The explicit
$\theta$-function characterization
of $\rho_\pm$ (respectively $\rho_0$) then can be
summarized as follows.

\begin{thm}\lb{t9.3}
The $(2g+1)$-gap and $2g$-gap sequences $\rho_\pm$
and $\rho_0$
associated with $D_\pm
=\left(\begin{smallmatrix} 0 & A_\pm^*\\ A_\pm & 0
\end{smallmatrix}\right)$ and $D_0
= \left( \begin{smallmatrix} 0 &
A_0^*\\ A_0 & 0\end{smallmatrix}\right)$ are
given by
\begin{equation}
\rho_\pm (n) =\begin{cases}
\rho_{e,\pm}(m), & n=2m\\
\rho_{o,\pm} (m),
& n=2m+1\end{cases}  , \; \rho_0(n)
=\begin{cases}
\rho_{e,0}(m), & n=2m\\
\rho_{o,0} (m), & n =2m+1
\end{cases}, \:\: n \in\bbZ,
\lb{9.44}
\end{equation}
where
\begin{align}
\begin{split}
\rho_{e,\pm}(n) & =
-\left[ -\ti a \dfrac{\theta (\uz_1
(n-1))\theta(\uz_{1,\pm}(0, n+1))}
{\theta (\uz_1 (n))\theta(\uz_{1,\pm}
(0,n))} \right]^{1/2}
 e^{\pm \frac12 \int_{E_0}^0 \ome_{\infty_+,
\infty_-}^{\bth +}}\\
& = - \left[ -\ti a \dfrac{\theta (\uz_{2,\pm}(n))
\theta (\uz_{2,\pm,\mp}
(0,n-1))}{\theta (\uz_{2,\pm}(n-1) )
\theta(\uz_{2,\pm,\mp} (0,n))}
\right]^{1/2} e^{\pm \frac12 \int_{E_0}^0 \ome_{\infty_+,
\infty_-}^{\bth +}},
\lb{9.45}
\end{split}\\
\begin{split}
\rho_{o,\pm}(n) & =
\left[ -\ti a \dfrac{\theta(\uz_1(n+1))
\theta(\uz_{1,\pm} (0,n))}
{\theta (\uz_1 (n)) \theta (\uz_{1,\pm} (0,
n+1))} \right]^{1/2}
 e^{\mp \frac12 \int_{E_0}^0 \ome_{\infty_+,
\infty_-}^{\bth +}}\\
& = \left[ -\ti a \dfrac{\theta (\uz_{2,\pm}(n-1))
\theta(\uz_{2,\pm,\mp}
(0, n+1))}{\theta (\uz_{2,\pm}(n))
\theta(\uz_{2,\pm,\mp}(0,n))}
\right]^{1/2} e^{\mp \frac12 \int_{E_0}^0 \ome_{\infty_+,
\infty_-}^{\bth
+}} \\
&\hspace*{7cm} n\in\bbZ
\lb{9.46}
\end{split}
\end{align}
and $\rho_{e,0}(n)$, $\rho_{o,0}(n)$ are obtained
from \eqref{9.45},
\eqref{9.46} by taking $E_0 =0$.
\end{thm}

\begin{proof}
It suffices to combine \eqref{9.3}, \eqref{9.26},
\eqref{9.27}, and
\eqref{9.35}.
\end{proof}

Isospectral manifolds in connection with Toda
flows (including
non-Abelian generalizations) have attracted a
lot of interest (see, e.g.,
\cite{9}, \cite{14}, \cite{32}, \cite{65},
\cite{66}, \cite{67} and the
references therein).  In the present finite-gap
case the situation is
analogous to the (m)KdV case and briefly
summarized below.

\begin{rem}\lb{r9.4}
For the fixed hyperelliptic curve $K_g$
(cf.~\eqref{3.1}),
Lemma~\ref{l3.1} shows that all $g$-gap sequences
$(a,b)$ associated with
the Jacobi operator $H$ are parameterized by the
initial conditions
\begin{equation}
\hat\mu_j (n_0) =(\mu_j (n_0),
R_{2g+2} (\hat \mu_j (n_0))^{1/2}), \;
\mu_j (n_0) \in [E_{2j-1}, E_{2j}],
\quad 1 \leq j \leq g,
\lb{9.47}
\end{equation}
or equivalently, by the pairs
\begin{multline}
\{ (\mu_j (n_0), \sig_j(n_0))\}_{1\leq j \leq g},
\; \mu_j (n_0) \in
[E_{2j-1}, E_{2j}], \; \sig_j (n_0)
=\pm \text{ for } \hat \mu_j (n_0)
\in\Pi_\pm,\\ 1\leq j \leq g.
\lb{9.48}
\end{multline}
(Here we omit $\sig_j (n_0)$ in the special case
where $\mu_j (n_0)\in\{
E_{2j-1}, E_{2j}\}$.)  With this restriction in
mind, \eqref{9.48}
represents the product of $g$ circles $S^1$
when varying $\mu_j (n_0)$
(independently from $\mu_\ell (n_0)$,
$\ell \neq j$) in $[E_{2j-1},
E_{2j}]$, $1\leq j \leq g$.  In other words,
the isospectral set of all
$g$-gap sequences $(a,b)$ associated with $H$
can be identified with the
$g$-dimensional torus $T^g =\times_{j=1}^g S^1$.
Theorem~\ref{t5.2} then
provides a concrete realization of $T^g$.  By
Theorem~\ref{t9.3} the same
applies to the set of all $(2g+1)$-gap (respectively
$2g$-gap) sequences $\rho$
associated with the Dirac-type operator $D$.
More precisely, assuming
$H_1$ (and hence $H_{2,\pm}$) to be subcritical
(and thus $0\in\bbR \bs
\sig (D_\pm))$, the isospectral set of all
$(2g+1)$-gap sequences $\rho$
(in connection with the nonsingular hyperelliptic
curve $K_{2g+1}$ of
genus $2g+1$, cf.~\eqref{7.40}) is again
parameterized bijectively by the
Dirichlet divisor $\calD_{\humu_1(n_0)}$ (respectively
by the analog of
\eqref{9.48}) as is demonstrated in \eqref{9.42}.
In particular,
$\rho_+$ and $\rho_-$ in \eqref{9.44}--\eqref{9.46}
represent two
independent (yet equivalent) concrete realizations
of the isospectral
manifold $T^g$ of all $(2g+1)$-gap sequences $\rho$
 associated with $D$.
In the case where $H_1$ (and hence $H_{2,\pm}$) is
critical (i.e., $E_0
=0$ and thus $0\in \sig (D_0)$) the (fixed) curve
 $K_{2g+1}$ of
(arithmetic) genus $2g+1$ (cf.~\eqref{7.40}) is
singular, yet
$\calD_{\humu_1 (n_0)}$ still parameterizes the
 corresponding isospectral
set of $2g$-gap sequences $\rho$ in a one-to-one
and onto fashion. In
particular, $\rho_0$ in \eqref{9.44}--\eqref{9.46}
then
represents a concrete realization of the isospectral
torus $T^g$ of all
$2g$-gap sequences $\rho$ associated with $D$.
\end{rem}

Finally we briefly treat the $t$-dependent case.  Our
starting point is a
solution $(a_1(t), b_1(t))$, $a_1(n,t)<0$,
$b_1(n,t)<0$, $(n,t)\in\bbZ
\times \bbR$ of the $\widetilde{TL}_r$ equations
\eqref{6.6}
 with $g$-gap initial
conditions $(a_1^\bze, b_1^\bze)$ at $t=t_0$ and
$\theta$-function
representation as described in Theorem~\ref{t6.3}.
Next we note that as
in \eqref{9.1} the $t$-dependent BA-function
$\psi_1 (P, n,n_0, t,t_0)$
in \eqref{6.8}, respectively \eqref{6.34}, shares the
positivity condition
\begin{equation}
\psi_1 (P,n,n_0, t,t_0) > 0,
\quad \ti\pi (P) \leq E_0
\lb{9.49}
\end{equation}
and by \eqref{6.12}, \eqref{6.13} satisfies
\begin{gather}
L_1(t) \psi_1(P,t) =0, \quad t\in\bbR,\lb{9.50}\\
\dfrac{d}{dt} \psi_1 (P,t)
= \tilde P_{1,2r+2} \psi_1 (P,t), \quad r\in\bbN_0, \;
t\in\bbR.
\lb{9.51}
\end{gather}
At this point one can follow the stationary
considerations in
\eqref{9.2}--\eqref{9.15},
\eqref{9.21}--\eqref{9.38} step by step.
Especially, $\uz_{1,\eps} (z,n),\,
\uz_{2,\eps,\eps'} (z,n)$ are now
replaced by
\begin{align}
\begin{split}
\uz_{1,\eps} (z,n,t) &
= \eps \ua_{E_0}^+ (z) -\hual_{P_0}
(\calD_{\humu_1(n,t)} ) -\huxi_{P_0},\\
\uz_1 (n,t) & = \uz_{1,+} (-\infty, n,t),
\lb{9.52}
\end{split}\\
\begin{split}
\uz_{2,\eps,\eps'} (z,n,t) &
= \eps' \ua^+_{E_0} (z) -\hual_{P_0}
(\calD_{\humu_{2,\eps}(n,t)}) -\huxi_{P_0},\\
\uz_{2,\eps} (n,t) & =
 \uz_{2,\eps, +} (-\infty, n,t), \quad \eps, \eps'
\in \{+,-\},
\lb{9.53}
\end{split}
\end{align}
with (cf.~\eqref{3.42}, \eqref{6.27},
\eqref{6.55})
\begin{align}
\begin{split}
& \hual_{P_0} (\calD_{\humu_{2,\pm}(n,t)})
=\hual_{P_0}
(\calD_{\humu_1(n,t)}) \mp \ua_{E_0}^+ (0)
-\hua_{P_0} (\infty_+)\\
& = \hual_{P_0} (\calD_{\humu_1(n_0, t_0)})
-(n-n_0) \uU^\bth -(t-t_0)
\ti{\uU}_r^\btwo\mp \ua_{E_0}^+ (0)
-\hua_{P_0} (\infty_+).
\lb{9.54}
\end{split}
\end{align}
The divisors $\calD_{\humu_{2,\pm (n,t)}}$ are
all nonspecial, that is,
\begin{equation}
i(\calD_{\humu_{2,\pm} (n,t)})=0,
\quad (n,t) \in\bbZ\times\bbR
\lb{9.55}
\end{equation}
by exactly the same argument as in the proof
of Theorem~\ref{t9.1}.

Introducing $L_1(t)$, $H_1(t)$, $\rho_\pm (t)$,
$A_\pm (t)$, $H_{2,\pm}
(t)$, $M_\pm (t)$, $D_\pm (t)$, $a_{2,\pm} (t)$,
$b_{2,\pm} (t)$
according to \eqref{7.3}, \eqref{7.4},
\eqref{7.8}--\eqref{7.11},
\eqref{7.29}--\eqref{7.33}, \eqref{8.1},
\eqref{8.2} we may briefly
summarize our $t$-dependent results applying
Theorem~\ref{t7.2} as
follows.

\begin{thm} \lb{t9.5}
The $\theta$-function representation of the
$g$-gap solutions $(a_{2,\pm}
(t)$, $b_{2,\pm}(t))$ of the $\widetilde{TL}_r$ equations
 read
\begin{align}
\begin{split}
a_{2,\pm}(n,t) & =
\ti a [\theta (\uz_{2,\pm} (n+1, t))\theta(\uz_{2,\pm}
(n-1,t))/ \theta (\uz_{2,\pm}(n,t))^2 ]^{1/2},\\
& \hspace*{7cm} (n,t) \in\bbZ\times
\bbR,
\lb{9.56}\end{split}\\
\begin{split}
b_{2,\pm} (n,t) & =
\sum_{j=1}^g \lam_j -\frac12 \sum_{m=0}^{2g+1} E_m
-\sum_{j=1}^g c_j (g) \dfrac{\pa}{\pa w_j} \ln \left. \left[
\dfrac{\theta (\uw
+\uz_{2,\pm} (n,t))}{\theta (\uw +\uz_{2,\pm} (n-1,t))}
\right]\right|_{\uw = \uzero},\\
&\hspace*{7cm} (n,t) \in\bbZ\times \bbR.
\lb{9.57}
\end{split}
\end{align}
Similarly, the $(2g+1)$-gap and $2g$-gap solutions
$\rho_\pm$ and $\rho_0$ of the
$\widetilde{KM}_r$ equations
are given by
\begin{equation}
\rho_\pm (n,t) = \begin{cases}
\rho_{e,\pm} (m,t), & \!\! n=2m\\
\rho_{o,\pm}(m,t), & \!\! n=2m+1
\end{cases}, \:\:
\rho_0 (n,t) = \begin{cases}
\rho_{e,0} (m,t), & \!\! n=2m\\
\rho_{o,0} (m,t), & \!\! n=2m+1
\end{cases},
\lb{9.58}
\end{equation}
where
\begin{align}
\begin{split}
\rho_{e,\pm} (n,t) & =
-\left[ -\ti a \dfrac{\theta(\uz_1 (n-1, t))\theta
(\uz_{1,\pm}(0,n+1,t))}
{\theta (\uz_1(n,t) ) \theta (\uz_{1,\pm}
(0,n,t))} \right]^{1/2}
 e^{\pm \frac12 \int_{E_0}^0 \ome_{\infty_+,
\infty_-}^{\bth +}}\\
& = -\left[ -\ti a \dfrac{\theta
 (\uz_{2,\pm} (n,t))\theta
(\uz_{2,\pm,\mp}
(0,n-1,t))}{\theta (\uz_{2,\pm} (n-1,t))
\theta (\uz_{2,\pm,\mp} (0,n,t))}
\right]^{1/2} e^{\pm \frac12 \int_{E_0}^0 \ome_{\infty_+,
\infty_-}^{\bth
+}},
\lb{9.59}
\end{split}\\
\begin{split}
\rho_{o,\pm} (n,t) & = \left[ -\ti a \dfrac{\theta
(\uz_1(n+1,t))\theta(\uz_{1,\pm}(0,n,t))}
{\theta (\uz_1 (n,t)) \theta
(\uz_{1,\pm} (0,n+1,t))}\right]^{1/2}
 e^{\pm \frac12 \int_{E_0}^0
\ome_{\infty_+, \infty_-}^{\bth +}}\\
& = \left[ -\ti a \dfrac{\theta(\uz_{2,\pm}(n-1,t))
\theta
(\uz_{2,\pm,\mp} (0,n+1, t))}
{\theta (\uz_{2,\pm}(n,t))\theta
(\uz_{2,\pm,\mp} (0, n,t))} \right]^{1/2}
 e^{\pm \frac12 \int_{E_0}^0
\ome_{\infty_+, \infty_-}^{\bth +}}
\lb{9.60}
\end{split}
\end{align}
and $\rho_{e,0} (n,t),\, \rho_{o,0}(n,t)$ are
 obtained from \eqref{9.59},
\eqref{9.60} by taking $E_0 =0$.
\end{thm}

We end up with a brief outlook at possible
applications of the results of
this chapter.  In many respects the construction
of all real-valued
algebro-geometric quasi-periodic finite-gap solutions
of the $\km$ hierarchy
is by no means the end of the story but rather the
beginning of the next
chapter in view of possible applications of this
material. For instance, in
addition to the applications described in \cite{84},
it appears tempting
to transfer results on the Toda shock problem (see,
 e.g., \cite{56a}, \cite{44},
\cite{56}, \cite{82} and the references therein)
to that of the $\km$ system
and to search for connections between algorithms
for eigenvalue
computation of real matrices with the Toda flows
(see, e.g., \cite{21},
\cite{23}, \cite{33} and the references therein).
 Similarly the solution
of certain discrete Peierls models for
quasi-one-dimensional conducting
polymers in connection with finite-gap Toda
solutions (see, e.g.,
\cite{54}, \cite{55} and the references
therein and in Ch.~8 of
\cite{10}) and especially the phenomenon of
soliton excitations in
conducting polymers (such as polyacetylen) and
Fermion number
fractionization (see, e.g., the reviews \cite{43},
\cite{71}), where the
underlying model Hamiltonians are related to the
Dirac-type expression
\eqref{7.4}, offer a variety of applications for
finite-gap solutions of
the $\km$ hierarchy.

\appendix


\chapter{Hyperelliptic Curves of the Toda-Type
and Theta Functions}
\lb{app-a}

We briefly summarize our basic notation in connection
with hyperelliptic
Toda curves and their theta functions as employed in
Chapters~\ref{s3},
\ref{s5}, \ref{s6}, and \ref{s9}.  For background
information on this
standard material we refer, for instance, to \cite{30},
\cite{31}, \cite{52},
\cite{70}.

Consider the points
\begin{equation}
\{E_m\}_{0\leq m \leq 2g+1} \subset \bbR,
\; E_0 < E_1 < \cdots <
E_{2g+1}, \quad g\in\bbN_0
\lb{a.1}
\end{equation}
and define the cut plane
\begin{equation}
\Pi =\bbC \bs \bigcup_{j=0}^g [E_{2j}, E_{2j+1}]
\lb{a.2}
\end{equation}
with the holomorphic function
\begin{equation}
R_{2g+2} (.)^{1/2} : \left\{\begin{array}{l}
\Pi \to \bbC\\
z\mapsto [\prod_{m=0}^{2g+1} (z-E_m)]^{1/2}
\lb{a.3}
\end{array}\right.
\end{equation}
on it.  We extend $R_{2g+2}^{1/2}$ to all of
$\bbC$ by
\begin{equation}
R_{2g+2} (\lam)^{1/2}
=\lim\limits_{\eps \downarrow 0} R_{2g+2} (\lam +
i\eps)^{1/2}, \quad \lam \in\bbC\bs \Pi,
\lb{a.4}
\end{equation}
with the sign of the square root chosen
according to
\begin{equation}
R_{2g+2} (\lam)^{1/2} = \begin{cases}
-| R_{2g+2} (\lam)^{1/2}|,
& \lam \in (E_{2g+1}, \infty)\\
(-1)^{g+j+1} |R_{2g+2} (\lam)^{1/2} |,
& \lam \in (E_{2j+1}, E_{2j+2}),
\; 0 \leq j \leq g-1\\
(-1)^g |R_{2g+2}(\lam)^{1/2}|,
& \lam \in (-\infty, E_0)\\
(-1)^{g+j+1} i |R_{2g+2} (\lam)^{1/2}|,
& \lam \in (E_{2j}, E_{2j+1}), \;
0\leq j \leq g
\end{cases}.
\lb{a.5}
\end{equation}
Next we define the set
\begin{equation}
M=\{(z,\sig R_{2g+2} (z)^{1/2})| z\in\bbC,
\, \sig\in \{-,+\}\} \cup \{
\infty_+, \infty_-\}
\lb{a.6}
\end{equation}
and
\begin{equation}
B=\{(E_m, 0)\}_{0 \leq m \leq 2g+1} \; ,
\lb{a.7}
\end{equation}
the set of branch points.  $M$ becomes a Riemann
surface upon introducing
the charts $(U_{P_0}, \zeta_{P_0})$ defined as
follows:
\begin{align}
\begin{split}
P_0 &  = (z_0, \sig_0 R_{2g+2} (z_0)^{1/2})
\text{ or } P_0 =\infty_\pm, \;
P=(z,\sig R_{2g+2} (z)^{1/2})
 \in U_{P_0} \subset M,\\
V_{P_0} & = \zeta_{P_0} (U_{P_0})
 \subset \bbC.
\lb{a.8}
\end{split}
\end{align}
\underline{$P_0  \notin \{B\cup \{\infty_+,
\infty_-\}\}$.}
\begin{align*}
U_{P_0} = & \{P\in M \big| |z-z_0| < C,
\; \sig R_{2g+2} (z)^{1/2} \text{
the branch obtained by straight line}\\ & \text{analytic
continuation starting from } z_0\}, \quad C =
\underset{m}{\min} | z_0 - E_m|,\\
V_{P_0} = & \{\zeta\in\bbC \big| | \zeta| < C\},
\end{align*}
\begin{align}
\begin{split}
& \zeta_{P_0} : \left\{
\begin{array}{l} U_{P_0} \to V_{P_0}\\
P \mapsto z-z_0
\end{array} \right.,
\quad \zeta_{P_0}^{-1}: \left\{ \begin{array}{l}
V_{P_0} \to U_{P_0}\\
\zeta \mapsto (z_0 +\zeta, \, \sig R_{2g+2} (z_0
+\zeta)^{1/2})\end{array} \right. .
\lb{a.9}
\end{split}
\end{align}
\underline{$P_0 = (E_{m_0}, 0)$.}
\begin{align*}
& U_{P_0} = \big\{P\in M\big| | z-E_{m_0}|
< C_{m_0}\big\}, \; C_{m_0}
=\underset{m\neq m_0}{\min} | E_m - E_{m_0}|,\\
& V_{P_0} =\{\zeta\in\bbC
\big| | \zeta| < C^{1/2}_{m_0} \},
\end{align*}
\begin{align}
\begin{split}
& \zeta_{P_0} : \left\{ \begin{array}{l}
U_{P_0} \to V_{P_0}\\
P\mapsto \sig (z-E_{m_0})^{1/2}\end{array} \right.,\quad
\begin{array}{l} (z-E_{m_0})^{1/2}
=|(z-E_{m_0})^{1/2}|
 e^{(i/2)\arg (z-E_{m_0})},\\
\arg(z-E_{m_0}) \in \begin{cases} [0, 2\pi),
& m_0 \text{ even}\\
(-\pi, \pi], & m_0 \text{ odd }
\end{cases}, \end{array} \\
& \zeta_{P_0}^{-1} : \left\{ \begin{array}{l}
V_{P_0} \to U_{P_0}\\
\zeta \mapsto (E_{m_0} +\zeta^2, \;
\zeta[\prod_{m\neq m_0} (E_{m_0} -E_m
+\zeta^2)]^{1/2})\end{array} \right.,\\
& \left[\prod_{m\neq m_0} (E_{m_0} -E_m
+\zeta^2)\right]^{1/2} =(-1)^g
i^{-m_0 -1}
\Big| \Big[ \prod_{m\neq m_0}
 (E_{m_0} -E_m)\Big]^{1/2} \Big|\times\\
& \hskip 1truein \times \Big[1
+\frac12 \zeta^2 \sum_{m\neq m_0} (E_{m_0}
-E_m)^{-1} +O(\zeta^4)\Big].
\lb{a.10}
\end{split}
\end{align}
\underline{$P_0 = \infty_\pm$.}
\begin{align*}
& U_{P_0} = \{P\in M\big| |z|> C_{\infty}\},\: C_{\infty}
=\underset{m}{\max} |E_m|, \:
V_{P_0} =\big\{ \zeta\in\bbC \big|
| \zeta| < C^{-1}_{\infty}\big\},
\end{align*}
\begin{align}
\begin{split}
& \zeta_{P_0}: \left\{ \begin{array}{l}
U_{P_0} \to V_{P_0}\\
P\mapsto z^{-1}\\
\infty_\pm \mapsto 0
\end{array}\right., \quad
\zeta_{P_0}^{-1} : \left\{
\begin{array}{l}
V_{P_0} \to U_{P_0}\\
\zeta \mapsto (\zeta^{-1} ,
\pm \zeta^{-g-1} [\Pi_m (1-\zeta
E_m)]^{1/2})\\
0 \mapsto \infty_\pm
\end{array}\right.,\\
& \Big[ \Pi_{m} (1-\zeta E_m)\Big]^{1/2}
=-1 +\frac12 \zeta \sum_m E_m
+0
(\zeta^2).
\lb{a.11}
\end{split}
\end{align}
It will also be useful to introduce the subsets
$\Pi_\pm \subset M$ (upper and lower
sheets)
\begin{equation}
\Pi_\pm = \{ (z, \pm R_{2g+2} (z)^{1/2})
\in M |z\in\Pi\}
\lb{a.12}
\end{equation}
and the associated charts
\begin{equation}
\zeta_\pm : \left\{ \begin{array}{l}
\Pi_\pm \to \Pi\\
P \mapsto z
\end{array} \right..
\lb{a.13}
\end{equation}
The topology introduced by the charts
\eqref{a.9}--\eqref{a.11} is
Hausdorff and second countable (finitely many
of them cover $M$).  In
addition, $\Pi_\pm$ are connected (being
homeomorphic to $\Pi$) and so
are their closures $\opi_\pm$.  Moreover, since
$M= \opi_+ \cup \opi_-$
and $\opi_+$ and $\opi_-$ have points in common,
$M$ is connected and
\eqref{a.9}--\eqref{a.11} define a complex structure
on $M$.  We shall
denote the resulting Riemann surface (curve)
by $K_g$.  Topologically,
$K_g$ is a sphere with $g$ handles and hence has
genus $g$.

Next, consider the holomorphic sheet exchange
map (involution)
\begin{equation}
* : \left\{ \begin{array}{l}
K_g \to K_g\\
(z,\sig R_{2g+2} (z)^{1/2})
\mapsto (z, \sig R_{2g+2} (z)^{1/2})^* =(z,-\sig
R_{2g+2} (z)^{1/2})\\
\infty_\pm \mapsto \infty_\pm^* =
 \infty_\mp \end{array} \right.
\lb{a.14}
\end{equation}
and the two meromorphic projection maps
\begin{equation}
\ti \pi : \begin{cases}
K_g \to \bbC \cup \{\infty\}\\
(z, \sig R_{2g+2} (z)^{1/2}) \mapsto z\\
\infty_\pm \mapsto \infty
\end{cases}, \quad
R_{2g+2}^{1/2} : \begin{cases}
K_g \to \bbC\cup \{\infty\}\\
(z, \sig R_{2g+2} (z)^{1/2}) \mapsto
 \sig R_{2g+2}(z)^{1/2}.\\
\infty_\pm \mapsto \infty
\end{cases}
\lb{a.16}
\end{equation}
$\ti \pi$ has poles of order 1 at $\infty_\pm$
and two simple zeros at
$(0, \pm R_{2g+2}(0)^{1/2})$ if
$R_{2g+2} (0) \neq 0$ or a double zero at
$(0,0)$ if $R_{2g+2} (0) =0$ (i.e., if
$0\in \{E_m\}_{0 \leq m \leq
2g+1}$) and $R_{2g+2}^{1/2}$ has poles of order
$g+1$ at $\infty_\pm$ and
$2g+2$ simple zeros at $(E_m, 0)$,
$0 \leq m \leq 2g+1$.  Moreover,
\begin{equation}
\ti \pi (P^*) =\ti \pi (P), \; R_{2g+2}^{1/2} (P^*)
=-R^{1/2}_{2g+2} (P),
\quad P \in K_g.
\lb{a.17}
\end{equation}
Thus $K_g$ is a two-sheeted ramified covering of
the Riemann sphere
$\bbC_\infty (\cong \bbC \cup \{ \infty\}$),
$K_g$ is compact (since
$\ti\pi$ is open and $\bbC_\infty$ is compact), and $K_g$ is
hyperelliptic (since it admits the meromorphic
function $\ti\pi$ of
degree two).

Using our local charts one infers that for
$g\in \bbN$, $d\ti\pi /
R_{2g+2}^{1/2}$ is a holomorphic differential
on $K_g$ with zeros of
order $g-1$ at $\infty_\pm$ and hence
\begin{equation}
\eta_j =\dfrac{\ti\pi^{j-1} d\ti\pi}{R_{2g+2}^{1/2}},
\quad 1\leq j \leq
g
\lb{a.18}
\end{equation}
form a basis for the space of holomorphic
differentials on $K_g$.

Next we introduce a canonical homology basis
$\{a_j, b_j\}_{1\leq j \leq
g}$ for $K_g$ as follows. The cycle $a_\ell$
starts near $E_{2\ell-1}$
on $\Pi_+$ surrounds $E_{2\ell}$ counterclockwise
thereby changing to
$\Pi_-$, and returns to the starting point
 encircling $E_{2\ell -1}$
changing sheets again.  The cycle $b_\ell$
surrounds $E_0$, $E_{2\ell -1}$
counterclockwise (once) on $\Pi_+$.  The cycles
 are chosen so that their
intersection matrix reads
\begin{equation}
a_j \circ b_k =\del_{j,k}, \quad 1\leq j,k \leq g.
\lb{a.19}
\end{equation}
Introducing the invertible matrix $C$ in $\bbC^g$,
\begin{align}
\begin{split}
& C=(C_{j,k})_{1\leq j,k\leq g}, \; C_{j,k}
=\int_{a_k} \eta_j
=2\int_{E_{2k-1}}^{E_{2k}} \dfrac{z^{j-1} \, dz}
{R_{2g+2}(z)^{1/2}} \in
\bbR,\\
& \uc (k) =(c_1(k), \ldots, c_g(k)), \; c_j (k)
=C_{j,k}^{-1},
\lb{a.20}
\end{split}
\end{align}
the normalized differentials $\ome_j$,
$1\leq j \leq g$,
\begin{equation}
\ome_j =\sum_{\ell=1}^g c_j (\ell) \eta_\ell,
\; \int_{a_k} \ome_j
=\del_{j,k}, \quad 1\leq j,k\leq g
\lb{a.21}
\end{equation}
form a canonical basis for the space of
holomorphic differentials on
$K_g$.  The matrix $\tau$ in $\bbC^g$ of
$b$-periods,
\begin{equation}
\tau=(\tau_{j,k})_{1\leq j,k\leq g}, \quad \tau_{j,k}
 =\int_{b_k}\ome_j
\lb{a.22}
\end{equation}
satisfies
\begin{equation}
\tau_{j,k} =\tau_{k,j}, \quad 1\leq j,k\leq g,
\lb{a.23}
\end{equation}
\begin{equation}
\tau =iT,\quad T>0.
\lb{a.24}
\end{equation}

In the charts $(U_{\infty_\pm},
\zeta_{\infty_\pm} \equiv \zeta)$
induced by $1/ \ti\pi$ near $\infty_\pm$ one
infers
\begin{align}
\begin{split}
\uome & = \mp \sum_{j=1}^g \uc (j) \dfrac{\zeta^{g-j}\,
d\zeta}{[\Pi_m
(1-\zeta E_m)]^{1/2}}\\
& = \pm \Big\{ \uc (g)
+\zeta \Big[ \frac12 \uc (g) \sum_{m=0}^{2g+1} E_m
+\uc (g-1)\Big] +O(\zeta^2) \Big\} \, d\zeta.
\lb{a.25}
\end{split}
\end{align}

Associated with the homology basis
$\{a_j, b_j\}_{1\leq j \leq g}$ we
also recall the canonical dissection of $K_g$
along its cycles yielding
the simply connected interior $\hat K_g$ of the
fundamental polygon $\pa
\hat K_g$ given by
\begin{equation}
\pa \hat K_g =a_1 b_1 a_1^{-1} b_1^{-1}
a_2 b_2 a_2^{-1} b_2^{-1} \cdots
a_g^{-1} b_g^{-1}.
\lb{a.26}
\end{equation}

The Riemann theta function associated with
$K_g$ is defined by
\begin{equation}
\theta (\uz) =\sum_{\un \in\bbZ^g}
 \exp [2\pi i (\un,\uz) + \pi i (\un,
\tau \un)], \quad \uz =(z_1,
\ldots, z_g) \in\bbC^g,
\lb{a.27}
\end{equation}
where $(\uu, \uv)=\sum_{j=1}^g \overline{u}_j v_j$
denotes the
scalar product
in $\bbC^g$. It has the fundamental properties
\begin{align}
\begin{split}
& \theta(z_1, \ldots, z_{j-1}, -z_j, z_{j+1},
\ldots, z_g) =\theta
(\uz),\\
& \theta (\uz +\um +\tau \un)
=\exp [-2 \pi i (\un,\uz) -\pi i (\un, \tau
\un) ] \theta (\uz), \quad \um, \un \in\bbZ^g.
\lb{a.28}
\end{split}
\end{align}

A divisor $\calD$ on $K_g$ is a map
$\calD: K_g \to \bbZ$, where $\calD
(P) \neq 0$ for only finitely many
$P\in K_g$.  The set of all divisors
on $K_g$ will be denoted by $\Div (K_g)$.
With $L_g$ we denote the
period lattice
\begin{equation}
L_g : = \{ \uz \in\bbC^g | \uz = \um +\tau \un,
\; \um, \un \in\bbZ^g\}
\lb{a.29}
\end{equation}
and the Jacobi variety $J(K_g)$ is defined by
\begin{equation}
J(K_g) =\bbC^g / L_g.
\lb{a.30}
\end{equation}
The Abel maps $\ua_{P_0} (.)$ respectively
$\ual_{P_0} (.)$ are defined by
\begin{align}
\ua_{P_0} & : \begin{cases}
K_g \to J(K_g)\\
P\mapsto \ua_{P_0} (P) =\int_{P_0}^P \uome \mod (L_g)
\end{cases},
\lb{a.31}\\
\ual_\pze & : \begin{cases}
\Div (K_g) \to J(K_g)\\
\calD \mapsto \ual_\pze (\calD)
=\sum_{P \in K_g} \calD (P) \ua_\pze (P)
\end{cases},
\lb{a.32}
\end{align}
with $P_0 \in K_g$ a fixed base point.
(In the main text we agree
to fix $\pze =(E_0, 0)$ for convenience.)

In connection with \eqref{a.26} we shall also
need the maps
\begin{equation}
\hua_\pze: \begin{cases}
\hat K_g \to \bbC^g\\
P\mapsto \int_\pze^P \uome
\end{cases},
\quad \hat\ual_\pze\ : \begin{cases}
\Div (K_g) \to \bbC^g\\
\calD \mapsto \sum_{P\in\hat K_g} \calD (P)
 \hua_\pze (P)
\end{cases},
\lb{a.33}
\end{equation}
with path of integration lying in $\hat K_g$.

Let $\calM (K_g)$ and $\calM^1 (K_g)$ denote the
set of meromorphic
functions (0-forms) and meromorphic
differentials (1-forms)
on $K_g$. The residue of a meromorphic differential
$\nu\in \calM^1 (K_g)$ at a
point $Q_0 \in K_g$ is defined by
\begin{equation}
\res_{Q_0} (\nu)
=\frac{1}{2\pi i} \int_{\gam_{Q_0}} \nu,
\lb{a.34}
\end{equation}
where $\gam_{Q_0}$ is a counterclockwise oriented
smooth simple closed
contour encircling $Q_0$ but no other pole of
$\nu$.  Holomorphic
differentials are also called (Abelian) differentials
of the first kind (dfk),
(Abelian) differentials of the second kind
(dsk) $\ome^\btwo \in \calM^1 (K_g)$ are characterized
by the property
that all their residues vanish.  They are
normalized, for
instance, by demanding that all their $a$-periods
vanish, that is,
\begin{equation}
\int_{a_j} \ome^\btwo =0, \quad 1\leq j \leq g.
\lb{a.35}
\end{equation}
If $\ome_{P_1, n}^\btwo$ is a dsk on $K_g$ whose
only pole is $P_1 \in
\hat K_g$ with principal part $\zeta^{-n-2}\,
d\zeta$, $n\in\bbN_0$ near
$P_1$ and $\ome_j =
 (\sum_{m=0}^\infty d_{j,m} (P_1) \zeta^m)\, d\zeta$
near $P_1$, then
\begin{equation}
\int_{b_j} \ome_{P_1, n}^\btwo =
 \frac{2\pi i}{n+1} d_{j,n} (P_1).
\lb{a.36}
\end{equation}

Any meromorphic differential $\ome^\bth$ on
$K_g$ not of the first or
second kind is said to be of the third
kind (dtk).
A dtk $\ome^\bth \in \calM^1 (K_g)$
is usually normalized by the vanishing of its
$a$-periods, that is,
\begin{equation}
\int_{a_j} \ome^\bth =0, \quad 1\leq j\leq g.
\lb{a.38}
\end{equation}
A normal dtk $\ome_{P_1, P_2}^\bth$ associated
with two points $P_1$,
$P_2 \in \hat K_g$, $P_1 \neq P_2$ by definition
has simple poles at
$P_1$ and $P_2$ with residues $+1$ at $P_1$ and
$-1$ at $P_2$ and
vanishing $a$-periods.  If $\ome_{P,Q}^\bth$ is a
normal dtk associated
with $P$, $Q\in\hat K_g$, holomorphic on
$K_g \bs \{ P,Q\}$, then
\begin{equation}
\int_{b_j} \ome_{P,Q}^\bth =2\pi i \int_{Q}^P \ome_j,
\quad 1\leq j \leq
g,
\lb{a.39}
\end{equation}
where the path from $Q$ to $P$ lies in
$\hat K_g$ (i.e.,
does not touch any of the cycles $a_j$, $b_j$).

We shall always assume (without loss of generality)
that all poles of
dsk's and dtk's on $K_g$ lie on $\hat K_g$ (i.e.,
not on $\pa \hat K_g$).

For $f\in \calM (K_g) \bs \{0\}$,
$\ome \in \calM^1 (K_g) \bs \{0\}$ the
divisors of $f$ and $\ome$ are denoted
by $(f)$ and
$(\ome)$, respectively.  Two
divisors $\calD$, $\calE\in \Div (K_g)$ are
called equivalent, denoted by
$\calD \sim \calE$, if and only if $\calD -\calE
=(f)$ for some
$f\in\calM (K_g) \bs \{0\}$.  The divisor class
$[\calD]$ of $\calD$ is
then given by $[\calD]
=\{\calE \in \Div (K_g) | \calE \sim \calD\}$.  We
recall that
\begin{equation}
\deg ((f))=0,\, \deg ((\ome)) =2(g-1),\,
f\in\calM (K_g) \bs
\{0\},\,  \ome\in \calM^1 (K_g) \bs \{0\},
\lb{a.40}
\end{equation}
where the degree $\deg (\calD)$ of $\calD$ is given
by $\deg (\calD)
=\sum_{P\in K_g} \calD (P)$.  It is custom to call
$(f)$ (respectively,
$(\ome)$) a principal (respectively, canonical)
divisor.

Introducing the complex linear spaces
\begin{align}
\calL (\calD) & =\{f\in \calM (K_g) | f=0
 \text{ or } (f) \geq \calD\}, \;
r(\calD) =\dim_\bbC \calL (\calD),
\lb{a.41}\\
\calL^1 (\calD) & =
 \{ \ome\in \calM^1 (K_g) | \ome=0
 \text{ or } (\ome) \geq
\calD\},\; i(\calD) =\dim_\bbC \calL^1 (\calD),
\lb{a.42}
\end{align}
($i(\calD)$ the index of specialty of $\calD$) one
infers that $\deg
(\calD)$, $r(\calD)$, and $i(\calD)$ only depend on
the divisor class
$[\calD]$ of $\calD$.  Moreover, we recall the
following fundamental
facts.

\begin{thm} \lb{ta.1}
Let $\calD \in \Div (K_g)$,
$\ome \in \calM^1 (K_g) \bs \{0\}$. Then\\
(i).
\begin{equation}
i(\calD) =r(\calD-(\ome)), \quad g\in\bbN_0.
\lb{a.43}
\end{equation}
(ii) (Riemann-Roch theorem).
\begin{equation}
r(-\calD) =\deg (\calD) + i (\calD) -g+1,
\quad g\in\bbN_0.
\lb{a.44}
\end{equation}
(iii) (Abel's theorem).  $\calD\in \Div (K_g)$,
$g\in\bbN$ is principal
if and only if
\begin{equation}
\deg (\calD) =0 \text{ and } \ual_\pze (\calD)
=\uzero.
\lb{a.45}
\end{equation}
(iv) (Jacobi's inversion theorem).  Assume
$g\in\bbN$, then $\ual_\pze
: \Div (K_g) \to J(K_g)$ is surjective.
\end{thm}

For notational convenience we agree to abbreviate
\begin{equation}
\calD_Q: \begin{cases}
K_g \to \{0,1\}\\
P \mapsto \begin{cases}
1, & P=Q\\
0, & P\neq Q
\end{cases}
\end{cases}
\lb{a.46}
\end{equation}
and, for $\uq =(Q_1, \ldots, Q_g) \in \sig^g K_g$
($\sig^n K_g$ the $n$-th symmetric power of $K_g$),
\begin{equation}
\calD_\uq : \begin{cases}
K_g \to \{0,1,\ldots, g\}\\
P \mapsto \begin{cases}
m & \text{if }
P \text{ occurs } m \text{ times in } \{Q_1, \ldots,
Q_g\}\\
0 & \text{if } P \notin \{Q_1,\ldots, Q_g\}
\end{cases}
\end{cases}.
\lb{a.47}
\end{equation}
Moreover, $\sig^n K_g$ can be identified with
the set of positive
divisors $0< \calD \in \Div (K_g)$ of degree $n$.

\begin{lem} \lb{la.2}
Let $\calD_\uq \in \sig^g K_g$,
$\uq=(Q_1, \ldots, Q_g)$.  Then
\begin{equation}
1 \leq i (\calD_\uq ) =s(\leq g/2)
\lb{a.48}
\end{equation}
if and only if there are $s$ pairs of the type
$(P, P^*)\in \{Q_1,
\ldots, Q_g\}$ (this includes, of course, branch
points for which
$P=P^*$).
\end{lem}

We emphasize that most results in this appendix
immediately extend to the
case where $\{E_m\}_{0 \leq m \leq 2g+1} \subset \bbC$.
(In this case
$\tau$ is no longer purely imaginary as stated in
\eqref{a.24} but has a
positive definite imaginary part.)


\chapter{Periodic Jacobi Operators}
\lb{app-b}

Due to the extensive attention paid in the literature
to the theory of
periodic Jacobi matrices (see, e.g., \cite{2}, \cite{9},
\cite{19},
\cite{22}, \cite{49}, \cite{50}, \cite{59},
\cite{65}--\cite{67}) we shall now summarize the highlights
of this special case.
Throughout this appendix we shall use (and extend) the
notation
established in Chapter~\ref{s4} in connection with
(bounded) Jacobi
operators.

In addition to the assumption $a,\,
b\in \ell_\bbR^\infty (\bbZ)$, $a(n)
\neq 0$, $n\in\bbZ$ in \eqref{4.1}, \eqref{4.2}
we now add the
periodicity condition
\begin{equation}
a(n+N) =a(n), \; b(n+N) =b(n), \quad n\in\bbZ
\lb{b.1}
\end{equation}
for some $N\in\bbN$.  (In most formulas below we
tacitly avoid the
trivial case $N=1$ (cf. Appendix C) but assume
$N\geq 2$ instead.)  We agree to
abbreviate
\begin{equation}
A=\prod_{n=1}^N a(n) = \prod_{n=1}^N a(n_0+n),
\; B=\sum_{n=1}^N b(n)
=\sum_{n=1}^N b(n_0+n), \quad n_0 \in\bbZ.
\lb{b.2}
\end{equation}
Given the fundamental system of solutions
$c(z,n,n_0)$, $s(z,n,n_0)$ (see
\eqref{4.16}) of \eqref{4.6} one defines the
fundamental matrix
\begin{align}
\begin{split}
\Phi (z,n,n_0) & =\begin{pmatrix}
c(z,n,n_0) & s(z,n,n_0)\\
c(z,n+1, n_0) & s(z,n+1, n_0)
\end{pmatrix}\\
& =\begin{cases}
U_n(z) \cdots U_{n_0+1}(z), & n\geq n_0 +1\\
1, & n=n_0\\
U_{n+1}^{-1}(z) \cdots U_{n_0}^{-1}(z), & n\leq n_0 -1
\end{cases},
\lb{b.3}
\end{split}
\end{align}
where
\begin{align}
\begin{split}
U_m (z) & = \frac{1}{a(m)} \begin{pmatrix}
0 & a(m)\\
-a(m-1) & z+b(m)
\end{pmatrix}, \\
U_m (z)^{-1} & = \frac{1}{a(m-1)} \begin{pmatrix}
z+b(m) & -a(m)\\
a(m-1) & 0
\end{pmatrix}.
\lb{b.4}
\end{split}
\end{align}
Since
\begin{equation}
W(c(z,.,n_0), \, s(z,.,n_0))=a(n_0),
\lb{b.5}
\end{equation}
an arbitrary solution $\psi (z)$ of \eqref{4.6} is
of the type
\begin{equation}
\psi (z,n) =\psi(z,n_0) c(z,n,n_0)
+ \psi(z,n_0+1) s(z,n,n_0),
\lb{b.6}
\end{equation}
or equivalently,
\begin{equation}
\binom{\psi (z,n)}{\psi (z,n+1)}
= \Phi (z,n,n_0) \binom{\psi
(z,n_0)}{\psi (z,n_0+1)}.
\lb{b.7}
\end{equation}
Moreover, one infers
\begin{align}
& \det [\Phi (z,n,n_0)] = \frac{a(n_0)}{a(n)},
\lb{b.8}\\
& \Phi (z,n,n_0) =\Phi (z,n,n_1) \Phi (z,n_1, n_0),
\lb{b.9}\\
& \Phi (z,n,n_0)^{-1} =\Phi (z,n_0, n).
\lb{b.10}
\end{align}
The monodromy matrix $M(z,n)$ is then defined by
\begin{equation}
M(z,n) =\Phi (z,n+N, n)
\lb{b.11}
\end{equation}
and hence
\begin{equation}
M(z,n) =\Phi (z,n,n_0) M(z,n_0) \Phi (z,n,n_0)^{-1}
\lb{b.12}
\end{equation}
and
\begin{equation}
\det [M(z,n)]=1.
\lb{b.13}
\end{equation}
The Floquet discriminant $\Del (z)$ defined by
\begin{equation}
\Del (z) = \tr [M(z,n)]/2
\lb{b.14}
\end{equation}
is independent of $n$ (cf.~\eqref{b.12}) and the
Floquet multipliers
$m_\pm (z)$ (the eigenvalues of $M(z,n)$) then read
\begin{equation}
m_\pm (z) =\Del (z) \pm [\Del (z)^2 -1]^{1/2}.
\lb{b.15}
\end{equation}
Again by \eqref{b.12} they are independent of $n$
and satisfy
\begin{equation}
m_+ (z) m_- (z) =1,\; m_+ (z) + m_- (z) =2\Del (z).
\lb{b.16}
\end{equation}
Let $\{\ti E_\ell\}_{0 \leq \ell \leq 2N-1}$ be the
zeros of $\Del (z)^2
-1$ and write
\begin{align}
\Del (z)^2 -1 & =
\dfrac1{4A^2} \prod_{\ell =0}^{2N-1} (z- \ti E_\ell)
\lb{b.17}\\
\intertext{and}
\Del (z) \mp 1 & = \dfrac1{2A} \prod_{j=1}^N (z-E_j^\pm).
\lb{b.18}
\end{align}
The zeros $\{E_j^\pm\}_{1 \leq j \leq N}$ turn out
to be the eigenvalues
of the following periodic respectively antiperiodic Jacobi
matrices $\ti
H_{n_0}^\pm$ in $\bbC^N$.  More generally, define
$\ti H_{n_0}^\theta$ in
$\bbC^N$ associated with the boundary conditions
\begin{multline}
a(n_0 +N) \psi (\nze +N)
=e^{i\theta} a(n_0) \psi(n_0), \, \psi(n_0
+N+1) =e^{i\theta} \psi (n_0 +1),
\\ \qquad 0 \leq \theta < 2\pi
\lb{b.19}
\end{multline}
by
\begin{multline}
\ti H_{n_0}^\theta =\begin{pmatrix}
-b({\scriptstyle n_0 +1}) & a({\scriptstyle n_0 +1}) &
0 & \cdots \:\: 0
& e^{-i\theta} a({\scriptstyle n_0 +N})\\
a({\scriptstyle n_0 +1}) & -b({\scriptstyle n_0 +2}) &
\ddots \; & & 0\\
0 & & \ddots & \ddots \;
& \vdots\\
\vdots & \ddots & \ddots \;
& \ddots & 0\\
0 & & \ddots &
-b({\scriptstyle n_0+N-1}) & a({\scriptstyle n_0 +N-1})\\
e^{i\theta } a({\scriptstyle n_0 +N}) & 0 \:\: \cdots
& 0 & a({\scriptstyle n_0 +N-1}) &
-b({\scriptstyle n_0+N})
\end{pmatrix},\\
0 \leq \theta < 2\pi.
\lb{b.20}
\end{multline}
One infers that $\ti H_\nze^\theta$ and
$\ti H_\nze^{2\pi -\theta}$ are
antiunitarily equivalent.  The periodic
respectively antiperiodic Jacobi
matrices $\ti H_\nze^\pm$ alluded to above
are then defined by
\begin{equation}
\ti H_\nze^+ = \ti H_\nze^0, \; \ti H_\nze^-
=\ti H_\nze^\pi.
\lb{b.21}
\end{equation}
The eigenvalues of $\ti H_\nze^\theta$ are then
given by
\begin{equation}
(m_+ (z) -e^{i\theta}) (m_- (z) -e^{i\theta}) =0,
\text{ that is, } \Del (z)
=\cos (\theta).
\lb{b.22}
\end{equation}
They are simple for $\theta\in (0,\pi)\cup (\pi, 2\pi)$
and at most twice
degenerate for $\theta =0,\pi$.  In the latter case
one infers
\begin{multline}
E_1^\pm < E_1^\mp \leq E_2^\mp < E_2^\pm
\leq E_3^\pm < \cdots <
E_{N-1}^{(-1)^{N-1}}
\leq E_N^{(-1)^{N-1}} < E_N^{(-1)^N},\\
\qquad \sgn (A) = \pm(-1)^N,
\lb{b.23}
\end{multline}
and $\{\ti E_\ell\}_{0 \leq \ell \leq 2N-1}$
coincides with the
corresponding sequence in \eqref{b.23}.
Another way to
express these facts is to invoke the theory
of direct integral
decompositions (see, e.g., \cite{75},
Sect.~XIII.~16)
\begin{equation}
\ell^2 (\bbZ) \cong \int_{[0,2 \pi)}^\oplus
\dfrac{d\theta}{2\pi} \ell^2
((n_0 +1, n_0 +N)), \; H \cong \int_{[0, 2\pi)}^\oplus
\dfrac{d\theta}{2\pi} \ti H_\nze^\theta
\lb{b.25}
\end{equation}
and $\cong$ denotes unitary equivalence.  In
particular, the spectrum
$\sig (H)$ of $H$ is characterized by
\begin{equation}
\sig (H)
=\{ \lam \in \bbR\big| |\Del (\lam )| \leq 1 \} =
\bigcup_{j=0}^{N-1} [\ti E_{2j}, \ti E_{2j-1}]
\lb{b.26}
\end{equation}
(see Theorem~\ref{t4.2} for additional information).

Returning to the square root $[\Del (z)^2 -1]^{1/2}$
in \eqref{b.15}, we
shall consider it as a fixed branch defined as follows,
\begin{align}
\begin{split}
[\Del (\lam)^2 -1]^{1/2} & =
-\sgn (A) | [ \Del(\lam)^2 -1]^{1/2}|, \quad
\lam > \ti E_{2N-1},\\
[\Del (\lam)^2 -1]^{1/2} &
= \lim\limits_{\eps\downarrow 0} [\Del (\lam +
i\eps)^2 -1]^{1/2}, \quad \lam \in\bbR,
\lb{b.26a}
\end{split}
\end{align}
assuming $[\Del (z)^2 -1]^{1/2}$ to be analytic
in $\bbC \bs
\bigcup_{j=0}^{N-1} [\ti E_{2j}, \ti E_{2j+1}]$.
As a consequence one
obtains
\begin{equation}
|m_+ (z)|\leq 1, \; |m_- (z) |\geq 1
\lb{b.27}
\end{equation}
and the (normalized) Floquet functions
$\psi_\pm (z,n,n_0)$ in
\eqref{4.17} then satisfy
\begin{equation}
\psi_\pm (z,n+N,\nze) =m_\pm (z) \psi_\pm (z,n,n_0)
\lb{b.28}
\end{equation}
and \eqref{4.18}.  In addition, one infers
(cf.~\eqref{4.4},
\eqref{4.17}--\eqref{4.19})
\begin{align}
\begin{split}
\phi_\pm (z,n_0) = \phi_\pm (z,n_0 +N) &
= \dfrac{m_\pm (z) -c(z,n_0+N,
n_0)}{s(z,n_0+N,n_0)}\\
& =\dfrac{c(z,n_0+N+1,n_0)}{m_\pm (z)
-s(z,n_0+N+1,n_0)},
\lb{b.29}\end{split}
\end{align}
\begin{equation}
W(\psi_- (z,.,n_0),\, \psi_+ (z,.,n_0))
=\dfrac{ 2a (\nze) [\Del(z)^2
-1]^{1/2}}{s(z,n_0 +N,n_0)},
\lb{b.30}
\end{equation}
\begin{equation}
G(z,n,n) = \dfrac{s(z,n+N,n)}
{2a(n) [\Del(z)^2 -1]^{1/2}}  =
\dfrac{\prod_{j=1}^{N-1} [z-\mu_j (n)]}
{\big\{ \prod_{\ell=0}^{2N-1}
[z-\ti E_\ell]\big\}^{1/2}},
\lb{b.31}
\end{equation}
\begin{equation}
\psi_+ (z,n,n_0) \psi_- (z,n,n_0)
=\dfrac{a(n_0) s(z,n+N,n)}{a(n)
s(z,n_0+N,n_0)} =\prod_{j=1}^{N-1}
\left[ \dfrac{z-\mu_j(n)}{z-\mu_j
(n_0)}\right].
\lb{b.32}
\end{equation}
If all spectral gaps of $H$ are ``open'',
that is, the spectra of $\ti
H_\nze^\pm$ are both simple, we have
\begin{equation}
g=N-1,\quad \left[ \prod_{\ell=0}^{2N-1}
(z-\ti E_\ell)\right]^{1/2} =R_{2g+2} (z)^{1/2}
=2A [\Del (z)^2 -1]^{1/2},
\lb{b.33}
\end{equation}
see \eqref{a.4}, \eqref{a.5}.  In the case where
some spectral gaps
``close'' we introduce the index sets
\begin{equation}
J' = \{1 \leq j' \leq N-1| \ti E_{2j'-1}
=\ti E_{2j'} \}, \;
J=\{0,1,\ldots, 2N-1\} \bs \{j', j'+1 | j' \in J'\}
\lb{b.34}
\end{equation}
and define
\begin{equation}
Q(z) =\dfrac1{2A} \prod_{j'\in J'} (z- \ti E_{2j'-1}),
\quad R_{2g+2}(z)
=\prod_{j\in J} (z-\ti E_j).
\lb{b.35}
\end{equation}
In order to establish the connection with the
notation employed in the
main text and in Appendix~\ref{app-a} we agree
to identify
\begin{equation}
\{\ti E_j \}_{j\in J} \text{ and }
\{E_m\}_{0 \leq m \leq 2g+1}.
\lb{b.36}
\end{equation}
Then one infers
\begin{align} \no
g=N-1 -|J'| =N-1-\deg (Q) = (|J|-2)/2, \\
[\Del(z)^2 -1]^{1/2}
=R_{2g+2}(z)^{1/2} Q(z),
\lb{b.37}
\end{align}
where $|J|,|J'|$ abbreviates the cardinality
of $J,J'$.

Next we shall give the

\begin{proof}[Proof of Theorem~\ref{t5.7}]
First we claim that in the periodic case
$\ome_{\infpm}^\bth$ is
explicitly given by
\begin{equation}
\ome_\infpm^\bth = \dfrac{\sgn (A) \Del'
\, d\ti \pi}{N R_{2g+2}^{1/2} Q}
=\dfrac{\sgn (A) \Del'
\, d\ti \pi}{N [\Del^2 -1]^{1/2}}.
\lb{b.38}
\end{equation}
For a proof of \eqref{b.38} we only need to
check that it is
appropriately normalized, that is, all its $a$-periods
vanish.  This is seen from
\begin{align}
\begin{split}
\left| \int_{a_j} \ome_\infpm^\bth\right| &
= \dfrac2N \left|
\int_{E_{2j-1}}^{E_{2j}} \dfrac{dz \Del' (z)}
{|[ \Del (z)^2 -1]^{1/2}|}
\right| \\ & =\dfrac2N \big| \ln \{ \Del (z)
+[\Del (z)^2 -1]^{1/2} \}
\big|^{E_{2j}}_{z=E_{2j-1}} \big|\\ & = 0,
\quad 1 \leq j \leq g.
\lb{b.39}
\end{split}
\end{align}
For the $b$-periods, one computes
\begin{align}
\begin{split}
\int_{b_j} \ome_\infpm^\bth & =
 \dfrac{2i}{N} \sum_{k=1}^j \left|
\int_{E_{2k}}^{E_{2k+1}} \dfrac{dz \Del' (z)}
{[ \Del (z)^2 -1]^{1/2}}
\right| =\dfrac{2i}{N} \sum_{k=1}^j
\big| \arcsin [\Del (z)]
\big|_{z=E_{2k}}^{E_{2k+1}} \big|\\
& = 2\pi i (j/N), \quad 1\leq j \leq g.
\lb{b.40}
\end{split}
\end{align}
By \eqref{3.42} this implies
\begin{equation}
2N\hua_\pze (\infty_+) =\uzero \mod (L_g)
\lb{b.41}
\end{equation}
which completes the proof.
\end{proof}

Next, we indicate a systematic approach to
high-energy expansions of
$c(z,n,n_0)$ and $s(z,n,n_0)$.  First we note
that \eqref{b.10} yields
\begin{align}
\begin{split}
s(z,n+1,n_0) & = a(n_0) a(n)^{-1} c(z,n_0,n),\\
s(z,n,n_0) & = -a(n_0) a(n)^{-1} s(z,n_0,n),\\
c(z,n,n_0) & = a(n_0) a(n)^{-1} s(z,n_0+1,n),\\
c(z,n+1,n_0) & =-a(n_0) a(n)^{-1} c(z,n_0+1, n)
\lb{b.42}
\end{split}
\end{align}
and \eqref{b.7} implies
\begin{align}
\begin{split}
s(z,n,n_0+1) & = -a(n_0+1) a(n_0)^{-1} c(z,n,n_0),\\
c(z,n,n_0-1) & =-a(n_0-1) a(n_0)^{-1} s(z,n,n_0),\\
s(z,n,n_0-1) & = c(z,n,n_0) +[b(n_0)
+z] a(n_0)^{-1} s(z,n,n_0),\\
c(z,n,n_0+1) & = s(z,n,n_0) +[b(n_0 +1)
+z] a(n_0)^{-1} c(z,n,n_0).
\lb{b.43}
\end{split}
\end{align}
Next we define the Jacobi matrix $J_\nze (k)$
in $\bbC^k$
\begin{equation}
J_\nze (k) = \begin{pmatrix}
-b({\scriptstyle n_0 +1}) & a({\scriptstyle n_0 +1}) & 0 &
\cdots & 0\\
a({\scriptstyle n_0 +1}) & -b({\scriptstyle n_0+2}) &
\ddots &\ddots & \vdots\\
0 & \ddots \; & \; \ddots & \ddots & 0 \\
\vdots & \ddots \; & \ddots \; & -b({\scriptstyle n_0 +k-1})
& a({\scriptstyle n_0 +k-1})\\
0 & \cdots & 0 & a({\scriptstyle n_0 +k-1}) &
-b({\scriptstyle n_0 +k})
\end{pmatrix}
\lb{b.44}
\end{equation}
and introduce
\begin{equation}
P_\nze (n,k) = \frac{1}{n}
\{ \tr [J_\nze (k)^n] -\sum_{j=1}^{n-1} P_\nze (j,k)
\tr[J_\nze (k)^{n-j} ] \}.
\lb{b.45}
\end{equation}
One then obtains
\begin{equation}
s(z,n_0+k+1, n_0) =\dfrac{\det [z-J_\nze (k)]}
{\prod_{n=1}^k a(n_0+n)}
=\dfrac{z^k -\sum_{\ell=1}^k P_\nze (\ell,
k) z^{k-\ell}}{\prod_{n=1}^k
a(n_0 +n)},\;
k\in\bbN.
\lb{b.46}
\end{equation}
Explicitly, one computes
\begin{align}
\begin{split}
\tr [J_\nze (k)] & = -\sum_{n=n_0 +1}^{n_0
+k} b(n),\\
\tr [J_\nze (k)^2] & = \sum_{n=n_0 +1}^{n_0 +
k} b(n)^2 +2 \sum_{n=n_0
+1}^{ n_0 +k-1} a(n)^2,\\
\tr [J_\nze (k)^3 ] & = -\sum_{n=n_0 +1}^{n_0
+k} b(n)^3 -3\sum_{n=n_0
+1}^{n_0 +k -1} a(n)^2 [b(n) + b(n+1)],\\
& \text{etc.}
\lb{b.47}
\end{split}
\end{align}
Using \eqref{b.42} and \eqref{b.43} one can extend
\eqref{b.46} to $k\leq
-1$ and to corresponding results for $c(z,n,n_0)$.
A direct calculation
yields for $k\in\bbN$
\begin{align}
\begin{split}
c(z,n_0+k+1,n_0) & = -\dfrac{a(n_0) z^{k-1}}
{\prod_{n=1}^k a(n_0 +n)}
\left[ 1+z^{-1} \sum_{n=2}^k b(n_0+n)
+O(z^{-2})\right],\\
c(z,n_0-k,n_0) & = \dfrac{z^k}
{\prod_{n=1}^k a(n_0 -n)} \left[ 1+z^{-1}
\sum_{n=0}^{k-1} b(n_0 -n)
+O(z^{-2})\right],\\
s(z,n_0 +k +1, n_0) &
= \dfrac{z^k}{\prod_{n=1}^k a(n_0 +n)} \left[
1+z^{-1} \sum_{n=1}^k b(n_0+n)
+O(z^{-2})\right],\\
s(z,n_0-k,n_0) & = -\dfrac{a(n_0) z^{k-1}}
{\prod_{n=1}^k a(n_0 -n)}
\left[ 1+z^{-1} \sum_{n=1}^{k-1} b(n_0 -n)
+O(z^{-2}) \right].
\lb{b.48}
\end{split}
\end{align}
We emphasize that \eqref{b.42}--\eqref{b.48} hold
for general (not
necessarily periodic or finite-gap) Jacobi operators.
In the following
we shall apply \eqref{b.48} to the periodic case.
Equations \eqref{b.15},
\eqref{b.27} yield the expansion
\begin{align}
\begin{split}
m_\pm (z) & \underset{|z|\to\infty}{=} (1\mp 1)
\Del (z) \pm
\dfrac1{2\Del (z)} +O(\Del(z)^{-3})\\
& \underset{|z|\to\infty}{=} (z^N / A)^{\mp 1}
 [1+O(z^{-1})]
\lb{b.49}
\end{split}
\end{align}
and \eqref{b.29} and \eqref{b.48} then imply
\begin{equation}
\phi_\pm (z,n) \underset{|z|
\to\infty}{=} [a(n) / z]^{\pm 1} \left[1\mp
z^{-1} b\binom{\scriptstyle n+1}{\scriptstyle n}
+ O(z^{-2})\right].
\lb{b.50}
\end{equation}
The relation
\begin{equation}
\psi_\pm (z,n,n_0) = \begin{cases}
\prod_{m=n_0}^{n-1} \phi_\pm (z,m),
& n\geq n_0 +1\\
1, & n=n_0\\
\prod_{m=n}^{n_0-1} \phi_\pm (z,m)^{-1},
& n\leq n_0 -1
\end{cases}
\lb{b.51}
\end{equation}
then yields
\begin{align} \no
\psi_\ppm (z,n_0+k,n_0) & = \left[ z^{-k}
\prod_{n=0}^{k-1} a({\scriptstyle n_0+n})
\right]^{\ppm 1} \left[ 1 \underset{(+)}{-} z^{-1}
\sum_{n=1(0)}^{k(k-1)} b({\scriptstyle n_0+n})
+O(z^{-2})\right],\\ \no
\psi_\ppm (z,n_0-k,n_0) & = \left[ z^{-k}
\prod_{n=1}^k a({\scriptstyle n_0-n})
\right]^{\pmp 1} \left[ 1\pmp z^{-1}
\sum_{n=1(0)}^{k(k-1)} b({\scriptstyle n_0 -n}) +0
(z^{-2})\right],\\
& \hspace*{8cm} k\in\bbN.
\lb{b.52}
\end{align}
Expansions \eqref{b.50}--\eqref{b.52} also hold
in the general case if
$\psi_\pm$ are the solutions of \eqref{4.6} which
are in $\ell^2((0,\pm\infty))$.

These expansions can now be employed to explicitly
compute $\ti a$, $\ti
b_1$, in Lemma~\ref{l5.1}~(i).

\begin{lem} \lb{lb.1}
In the periodic case one obtains
\begin{align}
\ti a & = -|A|^{1/N},
\lb{b.53}\\
\ti b_1 & = B/N,
\lb{b.54}\\
\intertext{and}
B & = \frac12 \sum_{\ell=0}^{2N-1} \ti E_\ell.
\lb{b.55}
\end{align}
\end{lem}

\begin{proof}
Combining \eqref{b.28}, \eqref{b.52}, and \eqref{5.1}
yields
\begin{align}
\begin{split}
m_\pm (z,n_0) & = \psi_\pm (z,n_0 +N, \nze)
=(A/ z^N)^{\pm 1} [1\mp
z^{-1} B+O(z^{-2})]\\
& = \sgn (A) (-\ti a / z)^{\pm N}
 [1\mp z^{-1} N \ti b_1 +O(z^{-2})]
\lb{b.56}
\end{split}
\end{align}
and hence \eqref{b.53} (noting $\ti a < 0$) and
\eqref{b.54}.  Combining
\eqref{b.54} and \eqref{5.3} (accounting for the
possibility of closing
spectral gaps) then yields \eqref{b.55}.
\end{proof}


\chapter{Examples, $g=0,1$}
\lb{app-c}

In this Appendix we illustrate the two simplest
examples in connection
with genus $g=0$ and $1$.

We start with \boldmath $g=0$\unboldmath:\\[3mm]
Let $N\in\bbN$ be fixed and consider
\begin{equation}
a(n) =a, \; b(n) =b,\quad n\in\bbZ.
\lb{c.1}
\end{equation}
One then verifies the following series of formulas,
\begin{align}
\phi_\pm (z,n) & =
 (2a)^{-1} (z+b) \pm [(2a)^{-2} (z+b)^2 -1]^{1/2},
\lb{c.2}\\
\psi_\pm (z,n,n_0) & =
 \{ (2a)^{-1} (z+b) \pm [(2a)^{-2} (z+b)^2
-1]^{1/2} \}^{(n-n_0)},
\lb{c.3}\\
\begin{split}
s(z,n,n_0) & =  \{ \{ (2a)^{-1} (z+b)
+[(2a)^{-2}
(z+b)^2
-1]^{1/2}\}^{(n-n_0)}\\
& \quad - \{ (2a)^{-1} (z+b) -[(2a)^{-2} (z+b)^2
-1]^{1/2}\}^{(n-n_0)}\} \times\\
& \times \{ 2[(2a)^{-2} (z+b)^2 -1]^{1/2} \}^{-1},
\lb{c.4}\end{split}\\
c(z,n,n_0) & = -s(z,n-1,n_0),
\lb{c.5}
\end{align}
\begin{align}
\begin{split}
\Del (z) & = \frac12 \{ (2a)^{-1} (z+b)
+[ (2a)^{-2} (z+b)^2
-1]^{1/2}\}^N\\
& \quad + \frac12 \{(2a)^{-1} (z+b)
- [(2a)^{-2} (z+b)^2 -1]^{1/2} \}^N,
\lb{c.6}
\end{split}\\
m_\pm (z) & = \{(2a)^{-1} (z+b)
 \pm [(2a)^{-2} (z+b)^2 -1]^{1/2} \}^N,
\lb{c.7}
\end{align}
\begin{align}
& A=a^N, \; B=Nb,
\lb{c.8}\\
\begin{split}
& \ti E_0 =-2 |a|-b, \; \ti E_{2j+1}
=\ti E_{2j+2} =\mu_j (n) =-2 |a|
\cos (j\pi / N) -b,\\
& \qquad \qquad 0 \leq j \leq N-2, \; n\in\bbZ,
\; \ti E_{2N-1} =2|a| -b,
\lb{c.9}
\end{split}\\
& J' = \{1,2,\ldots, N-1\},\; J=\{0,2N-1\},
\lb{c.10}\\
\begin{split}
& E_0 =-2|a|-b,\; E_1 =2|a| -b,\\
& |a| =(E_1 -E_0)/ 4, \; b=-(E_0+E_1)/2,
\lb{c.11}
\end{split}\\
& H=a(S^+ + S^-) -b, \; \calD (H) =\ell^2 (\bbZ),
\lb{c.12}\\
& \sig (H) =[E_0, E_1] =[-2|a| -b, \, 2|a|-b],
\lb{c.13}\\
& R_2(z) = (z-E_0) (z-E_1),
\lb{c.14}\\
& \ti a =-|a|,\; \ti b_1 =-b.
\lb{c.15}
\end{align}
Concerning the $t$-dependence of the branches of
the BA-function $\psi$
in the simplest case where $r=0$, that is, for the
original Toda system, one
obtains
\begin{multline}
\psi_\pm (z,n,n_0, t,t_0) =\{ (2a)^{-1} (z+b)
\pm [(2a)^{-2} (z+b)^2
-1]^{1/2} \}^{(n-n_0)}\times\\
\times \exp [\pm (t-t_0) R_2 (z)^{1/2}
].
\lb{c.16}
\end{multline}

Finally, assuming $a_1(n) =a< 0$,
$b_1(n) =b< 0$, $n\in\bbZ$ and $H_1
\geq 0$, that is, $|b| \geq 2|a|$, one computes,
\begin{align}
\begin{split}
\rho_{e,\pm} (n) & =
 -\{\frac12 |b| \pm \frac12 [b^2 -4a^2]^{1/2}
\}^{1/2},\\
\rho_{o,\pm}(n) & = -\rho_{e,\mp}(n),
\lb{c.17}
\end{split}\\
\rho_\pm (n) & = \begin{cases}
-\{\frac12 |b|
\pm \frac12 [b^2 -4a^2]^{1/2}\}^{1/2}, & n =2m\\
\{ \frac12 |b|
\mp \frac12 [b^2 -4a^2 ]^{1/2} \}^{1/2}, & n=2m+1
\end{cases},
\lb{c.18}\\
a_{2,\pm}(n) & = a,\; b_{2,\pm}(n) =b,
\lb{c.19}\\ \no
& \text{etc.}
\end{align}

Next we turn to the case \boldmath $g=1$\unboldmath:\\[3mm]
We suppose
\begin{equation}
E_0 < E_1 < E_2 < E_3, \; R_4(z)
=\prod_{m=0}^3 (z-E_m)
\lb{c.20}
\end{equation}
and introduce the following notations.
\begin{align}
k& =\left[ \dfrac{(E_2 -E_1)(E_3 -E_0)}
{(E_3 -E_1)(E_2 - E_0)}
\right]^{1/2} \in (0,1), \\ k'
& =\left[ \dfrac{(E_3
-E_2)(E_1-E_0)}{(E_3-E_1)(E_2 - E_0)} \right]^{1/2} \in (0,1),
\lb{c.21}
\end{align}
such that $k^2 +{k'}^2 =1$,
\begin{equation}
\bar u (z)
=\left[ \dfrac{(E_3-E_1)(E_0-z)}{(E_3-E_0)(E_1
-z)}\right]^{1/2},\;
C=\dfrac{2}{[(E_3-E_1)(E_2-E_0)]^{1/2}},
\lb{c.22}
\end{equation}
and Jacobi's integral of the first
\begin{align}
F(z,k) & = \int_0^z \dfrac{dx}
{[(1-x^2)(1-k^2 x^2)]^{1/2}},
\lb{c.23}\\
\intertext{second}
E(z,k) & = \int_0^z \, dx \left[ \dfrac{1-x^2}{1-k^2 x^2
} \right]^{1/2},
\lb{c.24}
\end{align}
and third kind
\begin{equation}
\Pi (z,\al^2, k) =\int_0^z \dfrac{dx}
{(1-\al^2 x^2)[(1-x^2)(1-k^2
x^2)]^{1/2}} \, , \quad \al^2 \in\bbR,
\lb{c.25}
\end{equation}
respectively.  (We refer, e.g., to \cite{18}
for details on Jacobi
elliptic integrals.)  We define
\begin{equation}
K(k) =F(1,k), \; E(k) =E(1,k), \; \Pi (\al^2, k)
=\Pi (1,\al^2, k)
\lb{c.26}
\end{equation}
and note that all square roots are assumed to be
positive for $x\in
(0,1)$.  We observe that $E(z,k)$ has a simple
pole at $\infty$ while
$\Pi (z,\al^2,k)$ has simple poles at
 $z =\pm \al^{-1}$.

Given these concepts we can now express the
basic objects in connection
with the elliptic curve $K_1$ in terms of the
quantities
\eqref{c.21}--\eqref{c.26}.  We list a series
of results below.

The Abelian dfk $\ome_1$ reads in the charts
$(\Pi_\pm, z)$,
\begin{align}
\ome_1 & = \dfrac{dz}{\pm 2C K(k) R_4(z)^{1/2}}
\lb{c.27}\\
\intertext{and one computes}
\tau_{1,1} & = \int_{b_1} \ome_1 = iK (k') / K (k).
\lb{c.28}
\end{align}
The Abel map $A_\pze$ reads
\begin{align}
A_\pze (P) & = \pm \dfrac{F(\bar u (z), k)}{2K(k)}
 \mod (L_1), \; P =(z,\pm
R_4 (z)^{1/2})
\lb{c.29}\\
\intertext{and hence}
A_\pze (\infty_+) & = \dfrac{F\left( \left(
\frac{E_3-E_1}{E_3-E_0}\right)^{1/2},k\right)}{2K(k)}
 \mod (L_1).
\lb{c.30}
\end{align}
The Riemann constant is base point independent
and given by
\begin{equation}
\Xi =\dfrac{1-\tau_{1,1}}{2} \mod (L_1).
\lb{c.31}
\end{equation}
Moreover, one computes
\begin{align}
\ome_\infpm^\bth & = \dfrac{(z-\lam_1)\, dz}
{\pm R_4(z)^{1/2}},\; \lam_1
=E_0 +\dfrac{E_1-E_0}{K(k)}\Pi \left( \dfrac{E_2 -E_1}
{E_2-E_0},
k\right),
\lb{c.32}\\
\int_{b_1} \ome_\infpm^\bth & =
2\pi i \left[ K(k)^{-1} F\left( \left(
\dfrac{E_3 - E_1}{E_3 - E_0} \right)^{1/2},
k \right) +1 \right],
\lb{c.33}\\
\begin{split}
\int_{P_0}^P \ome_\infpm^\bth & =
 \pm C (E_1 -E_0) \Bigg\{ \left[
1 - K(k)^{-1} \Pi \left( \dfrac{E_2
-E_1}{E_2-E_0}, k\right) \right] F(\bar u (z),k)\\
& \qquad - \Pi \left( \bar u (z), \dfrac{E_3-E_0}
{E_3 - E_1}, k \right)
\Bigg\}, \quad P=(z,\pm R_4 (z)^{1/2}).
\lb{c.34}
\end{split}
\end{align}
Restricting ourselves to the case $r=0$ (i.e.,
the original Toda system)
one obtains for
\begin{equation}
\Ome_0^\btwo =\ome_{\infty_+,0}^\btwo -
\ome_{\infty_-,0}^\btwo
\lb{c.35}
\end{equation}
(cf.~\eqref{6.24}) the explicit relations
\begin{equation}
2\pi i U_{0,1}^\btwo
=\int_{b_1} \Ome_0^\btwo =4 \pi i c_1(1)
=\dfrac{2\pi i}{CK(k)}, \quad
r=0,
\lb{c.36}
\end{equation}
\begin{align}
\begin{split}
& \int_\pze^P \Ome_0^\btwo =\frac12
C(E_2-E_0) \Bigg\{ (E_3-E_1) K(k)^{-1} E(k)
 F(\bar u (z), k)\\
& -(E_3 -E_1) E(\bar u (z), k)
-(E_3-E_0) \left[ 1- \dfrac{E_3-E_0}{E_3-E_1}
\bar u (z)^2 \right]^{-1} \times \\
& \times \bar u (z) [1-\bar u
(z)^2]^{1/2} [1-k^2 \bar u (z)^2]^{-1/2}\Bigg\},\\
& \hspace*{6cm} r=0, \; P=(z,\pm R_4 (z)^{1/2}).
\lb{c.37}
\end{split}
\end{align}
The relation
\begin{equation}
A_\pze (\hmu_1 (n,t)) =A_\pze (\hat \mu_1 (n_0, t_0))
-2(n-n_0) A_\pze
(\infty_+) -2(t-t_0) c_1(1)
\lb{c.38}
\end{equation}
then yields
\begin{align}
\mu_1 (n,t) = &{\scriptstyle E_1 \left\{1-
\big( \frac{E_2-E_1}{E_2-E_0} \big)
\frac{E_0}{E_1} \sn^2 \big[2K(k) \del_1+
2 (n-n_0) F\big(\big(
\frac{E_3-E_1}{E_3-E_0} \big)^{1/2}, k\big)
+2C^{-1} (t-t_0)\big]
\right\} }\times \notag\\
& \times {\scriptstyle \left\{1-\big( \frac{E_2-E_1}
{E_2-E_0}\big) \sn^2
\big[ 2K (k) \del_1
+2(n-n_0) F\big( \big( \frac{E_3-E_1}{E_3-E_0}
\big)^{1/2},k\big)+2C^{-1} (t-t_0)\big] \right\}^{-1} },
\lb{c.39}
\end{align}
where we abbreviated
\begin{equation}
A_\pze (\hmu_1(n_0, t_0)) =\left(-\del_1
+\dfrac{\tau_{1,1}}{2}\right)
 \mod (L_1)
\lb{c.40}
\end{equation}
and
\begin{equation}
\sn(w) =z,\; w= \int_0^z \dfrac{dx}
{[(1-x^2)(1-k^2x^2)]^{1/2}} =F(z,k).
\lb{c.41}
\end{equation}
The corresponding sheet of $\hmu_1(n,t)$ can
 be read off the sign of $\sn
\scriptstyle \big[2K(k) \del_1 +\dots \big]$.

Finally we recall that
\begin{equation}
\theta (z) = \vartheta_3 (z) =\sum_{n\in\bbZ}
 \exp [2\pi i nz +\pi i
\tau_{1,1} n^2].
\lb{c.42}
\end{equation}
The results \eqref{c.27}--\eqref{c.41} now enable
one to express all
objects like $a_1 (n,t)$, $b_1(n,t)$,
$a_{2,\pm}(n,t)$, $b_{2,\pm}
(n,t)$, $\rho_\pm (n,t)$, (for $r=0$) in
terms of the quantities
\eqref{c.21}--\eqref{c.26}, \eqref{c.41}, and
\eqref{c.42}.  We omit
further details at this point.

\chapter*{Acknowledgments}
W.~B.~is indebted to the Department of Mathematics
of the University of Missouri, Columbia for the
hospitality extended to him during stays in
the springs of 1990 and 1993. Furthermore, he gratefully
acknowledges financial supports for both stays by the
Amt der Steierm\"arkischen Landesregierung
and the Technical University of Graz, Austria in 1993.

F.~G.~is indebted to the Department of Mathematical
Sciences of the University of Trondheim, NTH, Norway for the
extraordinary hospitality
extended to him during one month stays in the summers
of 1993 and 1994.

Support by the Norwegian Research Council
(F.~G. and H.~H.) is gratefully acknowledged.

G.~T.~gratefully acknowledges a kind invitation from
the Department of Mathematics of the University of Missouri,
Columbia (December 1, 1993 to March 31, 1994) where parts
of this research were performed. This stay was supported
by a Fellowship of the Austrian
Ministry of Science and a Fellowship of the
Technical University of Graz, Austria.

\bibliographystyle{amsplain}

\end{document}